\documentclass[12pt]{article}
\pdfoutput=1

\usepackage{xcolor}

\usepackage{hyperref}
\usepackage{subcaption}
\usepackage[utf8]{inputenc}

\usepackage{setspace}
\usepackage{amsmath, amssymb, amsthm, float, graphicx}
\numberwithin{equation}{section}

\hypersetup{colorlinks=false, linkcolor=blue, citecolor=red}

\def\beq{\begin{eqnarray}}\def\eeq{\end{eqnarray}}
\def\be{\begin{equation}}\def\ee{\end{equation}}
\def\g{\gamma}
\def\r{\rho}
\def\s{\sigma}
\def\m{\mu}
\def\n{\nu}
\def\a{\alpha}
\def\e{\epsilon}
\def\k{\kappa}
\def\b{\beta}
\def\d{\delta}

\def\D{\Delta}
\def\G{\Gamma}
\def\l{\lambda}

\def\ta{\tau}

\def\la{\langle}
\def\ra{\rangle}

\def\G{\Gamma}

\begin{document}

\title{\bf A Mellin Space Approach \\ to the Conformal Bootstrap}
\date{}

\author{\!\!\!\! Rajesh Gopakumar$^{a}$\footnote{rajesh.gopakumar@icts.res.in},  Apratim Kaviraj$^{b}$\footnote{apratim@cts.iisc.ernet.in}, \\~Kallol Sen$^{b,c}$\footnote{kallolmax@gmail.com} ~and Aninda Sinha$^{b}$\footnote{asinha@chep.iisc.ernet.in}\\ ~~~~\\
\it ${^a}$International Centre for Theoretical Sciences (ICTS-TIFR),\\
\it Survey No. 151, Shivakote, Hesaraghatta Hobli,
Bangalore North, India 560 089\\
\it ${^b}$Centre for High Energy Physics,
\it Indian Institute of Science,\\ \it C.V. Raman Avenue, Bangalore 560012, India. \\
\it $^c$Kavli Institute for the Physics and Mathematics of the Universe (WPI),\\ \it The University of Tokyo Institutes for Advanced Study,  Kashiwa, Chiba 277-8583, Japan}
\maketitle
\maketitle
\vskip 1.5cm
\abstract{We describe in more detail our approach to the conformal bootstrap which uses the Mellin representation of $CFT_d$ four point functions and expands them in terms of crossing symmetric combinations of $AdS_{d+1}$ Witten exchange functions. We consider arbitrary external scalar operators and set up the conditions for consistency with the operator product expansion. Namely, we demand cancellation of spurious powers (of the cross ratios, in position space) which translate into spurious poles in Mellin space.  We discuss two contexts in which we can immediately apply this method by imposing the simplest set of constraint equations. The first is the epsilon expansion. We mostly focus on the Wilson-Fisher fixed point as studied in an epsilon expansion about $d=4$. We reproduce Feynman diagram results for operator dimensions to $O(\e^3)$ rather straightforwardly. This approach also yields new analytic predictions for OPE coefficients to the same order which fit nicely with recent numerical estimates for the Ising model (at $\e =1$). We will also mention some leading order results for scalar theories near three and six dimensions. The second context is a large spin expansion, in any dimension, where we are able to reproduce and go a bit beyond some of the results  recently obtained using the (double) light cone expansion. We also have a preliminary discussion about numerical implementation of the above bootstrap scheme in the absence of a small parameter.}

\vskip 1cm
\tableofcontents

\onehalfspacing

\section{Introduction}

Quantum Field Theory (QFT) is one of the most robust frameworks we have in theoretical physics. Its versatility is attested by the fact that it plays a central role in many contexts in high energy physics, condensed matter physics and statistical physics. Thanks to the work of Wilson and others \cite{wilson,vicari}, QFT was understood beyond a perturbative 
Feynman diagram expansion. The central role in this modern understanding is played by scale invariant fixed points of the Renormalisation Group (RG) flow.  When combined with $d$ dimensional Poincare invariance, these fixed points are believed to have an enhanced $SO(d,2)$ conformal invariance \cite{Polyakov1970xd}. The resulting CFTs while being dynamically nontrivial are also strongly constrained by the conformal symmetry. 

The conformal bootstrap is the philosophy that these constraints are strong enough to largely determine the dynamical content of the CFT viz. the spectrum of operator dimensions of primaries and their three point functions. The presence of a convergent OPE then implies that all other correlators can be fixed in terms of this data \cite{Ferrara:1973vz,Polyakov}. The conventional approach to the bootstrap, which proved to be very successful in $d=2$ \cite{bpz}, employs the associativity of the four point function, as we describe below. Recently, making use of the progress in finding efficient expressions for conformal blocks \cite{dolanosborn}, this approach was revived for $d>2$ \cite{rrtv} where associativity constraints, often together with positivity on the squares of OPE coefficients,  were implemented numerically through linear programming and semi-definite programming, together with judicious truncation of the operator spectrum \cite{reviews, bootstrap}. This has led to remarkably precise bounds on low-lying operator dimensions in a number of nontrivial CFTs. This includes, famously, the 3d Ising model \cite{3dising,mostprecise} which is in the same universality class as the critical point of the liquid-vapour transition of water. There are also very strong indications of such theories living at special points (``kinks") in the numerically allowed regions of parameter space. This suggests that these theories are special in some way and perhaps amenable to analytic treatment. These numerical methods have also been extended to supersymmetric theories \cite{susyboot}. Furthermore, there are also certain analytic results available at large spin \cite{dlce,others,alday,kss}. However, the existing approaches do not appear to be well suited for extracting analytic results in general. Also limited progress has been made in the case where external operators carry spin, see e.g. \cite{spins, Iliesiu:2015akf}.


Recently, using the conformal invariance of the three point function, the leading order (in $\epsilon$) anomalous dimensions of certain operators in $d=(4-\epsilon)$ dimensions were calculated for the Wilson-Fisher fixed point  \cite{rychkovtan}. This approach was further generalized to extract leading order anomalous dimensions for other theories in \cite{rtothers}. Results have also been obtained at leading order (both for the $\epsilon$-expansion as well as $1/N$ expansion) for anomalous dimensions of almost conserved higher spin currents \cite{giombi,skvortt, gio-shirm}. These results crucially rely on the use of the equations of motion or a higher spin symmetry, that is present when the coupling constant goes to zero. It is not immediately obvious how to systematize these approaches to subleading orders. In \cite{sensinha}, a dispersion relation based method of Polyakov \cite{Polyakov} (which had built in crossing symmetry)  was re-visited and it was found that this approach could be extended to get the subleading order anomalous dimension for the $\phi^2$ operator\footnote{ It was also shown how the leading order anomalous dimension at $O(\e^2)$ for large spin operators could be extracted using large spin bootstrap arguments based on \cite{dlce}.}. In spite of this encouraging result (though it took more than 40 years to reach here!), it was again not clear how to extend this dispersion relation based approach to operators with spin or to make it a starting point for a systematic algorithm. A major stumbling block was the reliance on momentum space where the underlying conformal symmetry is not fully manifest. 


In this paper, we will describe in more detail a novel approach to the conformal bootstrap that was recently outlined in \cite{usprl}. This approach is calculationally effective and at the same time conceptually quite suggestive. It combines two important ingredients. The first goes back to an alternative approach to the above dispersion based one, also attempted by Polyakov in his original bootstrap paper \cite{Polyakov}. He outlined a general way in 
which demanding consistency of the operator product expansion with crossing symmetry gave rise to constraints on operator dimensions and OPE coefficients. This was then implemented in position space which made the symmetries more manifest compared to momentum space. The idea behind this approach was to expand the $CFT_d$ four point function not in terms of the conventional conformal blocks but rather in terms of a new set of building blocks with built-in crossing symmetry from the beginning. We will see, in our modern incarnation, that these new building blocks can be chosen to be essentially tree level Witten exchange diagrams in $AdS_{d+1}$. This is very suggestive of a reorganisation of the CFT in terms of a dual AdS description though this will not be the main thrust of the present work. 

The second ingredient we introduce is to implement the above bootstrapping procedure in Mellin space rather than position space as used in \cite{Polyakov}. The position space approach made the equations in \cite{Polyakov} quite cumbersome and not explicit, especially for exchanges involving spin. We are familiar with this from the complicated form that Witten diagrams take in position space. The technology of the Mellin representation has been developed quite a bit in recent years starting from the work of Mack \cite{mack,pene,mig,joao, joaoreview, aldaybissi2,rastelli}. As has been amply stressed in these works, Mellin space is very natural for a CFT and plays a role analogous to momentum space in usual QFTs. This enables one to exploit properties such as meromorphy and more generally, features of scattering amplitudes (to which Mellin space amplitudes naturally transition to, in an appropriate flat space limit).  This, we will see, brings us big calculational gains. We will be able to reproduce many of the analytic results available in the literature for the conformal bootstrap in a fairly straightforward manner. In addition, we will be able to derive new results which we subject to various cross checks. We also give some preliminary evidence that this approach might also be workable into a useful computational scheme, complementary to existing ones. 

In the rest of this section we give a broad sketch of the new philosophy that we adopt and state some of the new results obtained with this approach. We first describe the ideas in position space and only later translate them into Mellin space. 

\subsection{The philosophy outlined}

Consider a four point function (of four identical scalars, for definiteness - we will consider the general case in Sec.2). In essence, we expand this amplitude in a new basis of building blocks as follows
\begin{eqnarray}\label{polyapp}
{\cal A}(u,v)  &=& \langle {\cal O}(1){\cal O}(2){\cal O}(3){\cal O}(4) \rangle \nonumber \\
&=& \sum_{\Delta, \ell} c_{\Delta, \ell}\bigg( W_{\Delta,\ell}^{(s)}(u,v)+W_{\Delta,\ell}^{(t)}(u,v)+W_{\Delta,\ell}^{(u)}(u,v) \bigg) \ .
\end{eqnarray}
Here $(u,v)$ are the usual conformally invariant cross ratios, whose dependence captures the nontrivial dynamical information of the four point function (we have suppressed a trivial additional dependence on positions which is predetermined).  
In the second line we sum over the {\it entire} physical spectrum of primary operators generically characterised by the operator dimensions $(\Delta)$ together with the spin ($\ell$) quantum numbers. The building block $W_{\Delta,\ell}^{(s)}$ can, for the moment, be viewed as the Witten exchange function in $AdS_{d+1}$  -- it will be defined more precisely later. This is diagrammatically represented in Fig.1. It involves the four identical scalars with an exchange in the $s-$channel of a field of spin $\ell$ and corresponding to a dimension $\Delta$. Similarly, for the $t$ and $u$-channels. The to-be determined coefficients $ c_{\Delta, \ell}$ will turn out to be proportional to the (square) of the three point OPE coefficients 
$C_{{\cal O}{\cal O}{\cal O}_{\Delta,\ell}}\equiv C_{\D,\ell}$. 

\begin{figure}[H]
\begin{center}
\vskip 2pt
\resizebox{450pt}{120pt}{\includegraphics{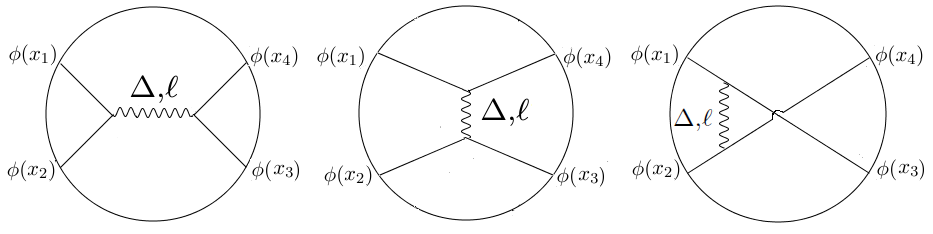}}
\end{center}
\caption{ Witten Diagrams in $AdS_{d+1}$ for identical scalars with exchange of a field corresponding to an operator of dimension $\D$ and spin $\ell$.}
\end{figure}

The idea behind this expansion, which we will contrast below to the usual conformal block expansion, is that we 
are expanding in a basis which
\begin{enumerate}
\item Is conformally invariant, as Witten exchange diagrams are;
\item Is consistent with factorisation, in that the individual blocks factorise on the physical operators with the right residues corresponding to three point functions;
\item Is {\it crossing symmetric} by construction since we are summing over all three channels. 
\end{enumerate}

The Witten exchange diagrams satisfy the second criterion since they arise from a local field theory 
in $AdS$. This will be much more explicitly seen in the Mellin representation. 
The last criterion ensures that we don't need to check channel duality since that is built in. But what is not obvious now is that expanding the resulting amplitude in any one channel, say the $s$-channel, is consistent with the operator product expansion. In other words, if we expand ${\cal A}(u,v)$ in powers of $u$, it is not guaranteed that all the powers that appear are those of the physical primary operators together with their descendants.  

In fact, generically, such an expansion will have spurious power law dependence. For instance, with identical external scalars (of dimension $\Delta_{\phi}$), we will see that there are pieces which go like $u^{\Delta_{\phi}}$ and $u^{\Delta_{\phi}}\ln{(u)}$. The  $u^{\Delta_{\phi}}$ would indicate the presence of an operator with dimension $2\Delta_{\phi}$, which generically does not exist in the theory\footnote{There could be special operators in interacting superconformal theories - ``chiral primaries" - for which there indeed are physical operators with dimension $2\D_\phi$. Such cases would have to be treated specially, perhaps using mixed correlators or by focussing on other spurious powers.}. These are often called ``double-trace operator" (``${\cal O}^2$") contributions in the AdS/CFT literature since these are there interpreted as contributions from two particle states whose energy is almost the same (in a large $N$ limit) as the two external (single) particle states\footnote{The logarithmic dependence is a consequence of having identical scalars. If we had generic dimensions $\Delta_i$ for the external operators, the spurious powers would take the form $u^{\frac{\Delta_1+\Delta_2}{2}}$ and $u^{\frac{\Delta_3+\Delta_4}{2}}$ corresponding to the two sets of double trace operators associated with the external states in the $s$-channel. The logarithm arises in the coincident limit $\Delta_i\rightarrow \Delta_{\phi}$. We also emphasise that the logarithmic dependence has nothing to do with anomalous dimensions since we are not making any expansion in a small parameter (yet).}. {\it We will then obtain constraints on operator dimensions as well as the coefficients $c_{\Delta, \ell}$ (and thus the OPE coefficients) from requiring that such spurious powers vanish.} Note that these are strong constraints implying an infinite number of relations since there is a full function (of $v$) multiplying these powers. Though we will not make use of them in this work, there are additional spurious powers (and logs) of the form $u^{\Delta_{\phi}+n}$ and 
$u^{\Delta_{\phi}+n}\ln{(u)}$. These can viewed as contributions from descendants as well as other double-trace primaries (what would have been ``${\cal O}\partial^{2n}{\cal O}$" in a weakly coupled theory). One would obtain additional constraints from requiring their vanishing but we will not explore the consequences of this in this paper (see \cite{rajanind}). 

We should stress that the Witten exchange diagrams are being employed as a {\it convenient kinematical basis} for this expansion, for an arbitrary $CFT_d$. We are not assuming (and it does not have to be) that the theory has an $AdS_{d+1}$ gravity dual. We could have alternatively used conformal blocks as a basis of expansion. But as will become clearer in Mellin space these are not very well behaved at infinity\footnote{Polyakov made a similar observation in terms of the behaviour of these blocks in the spectral parameter space \cite{Polyakov}.}. In contrast, Witten exchange diagrams will be polynomially bounded and thus a better basis for expansion. We note that since each Witten exchange diagram contains the conformal block contribution of the exchanged operator and since we are summing over the full primary operator spectrum we are not undercounting in this basis. In particular, what would have been double trace operators are included separately in the sum -- this is different from what we do in AdS/CFT where we only sum over single trace primaries. In this context note also that contact four point Witten diagrams  make no appearance in our approach.  We do not have to include them since it is known that they are decomposable into the double trace conformal blocks and thus, in our context, have purely spurious power law contributions.

Another important point to note is that we are implicitly assuming that the sums over $(\Delta,\ell)$ in the spurious pole cancellation conditions are convergent or can be analytically continued. In the examples we have considered  in this paper, the spurious poles have gotten contributions from only a small set of operators and hence we did not have to worry about convergence. It would be good to investigate the issue more generally. As for the physical contributions, once the spurious pole cancellation has been achieved, the remaining sum is just the usual sum over the physical conformal blocks which is believed to be convergent in a finite domain. 
 
Let us contrast this approach to the more ``conventional" bootstrap approach to CFTs \cite{Ferrara:1973vz, bpz} where we expand the four point function
\begin{eqnarray}\label{convapp}
{\cal A}(u,v)  &=& \sum_{\Delta, \ell} C_{\Delta, \ell} G_{\Delta,\ell}^{(s)}(u,v)  \nonumber \\
&=& \sum_{\Delta, \ell} C_{\Delta, \ell} G_{\Delta,\ell}^{(t)}(u,v) =\sum_{\Delta, \ell} C_{\Delta, \ell} G_{\Delta,\ell}^{(u)}(u,v)\ .
\end{eqnarray}
In this expansion, we choose to expand in terms of the conformal block in a particular channel, say, $s$-channel, as in the first line. The conformal blocks are
\begin{enumerate}
\item Are conformally invariant by construction;
\item Are consistent with factorisation, in that the individual blocks give the factorised contribution on physical operators with the right residues;
\item Are consistent with the OPE by construction since we are summing over all physical operators in any given channel. 
\end{enumerate}

The last criterion now ensures compatibility with the operator product expansion in terms of the powers that appear, but it is now not guaranteed that the resulting amplitude is crossing symmetric. In other words, the equality of the first line with the second line does not automatically follow. Demanding this associativity of the OPE is the nontrivial requirement which constrains operator dimensions and the OPE coefficients $C_{\Delta, \ell}$. Recent progress has come from efficient ways to translate the constraints of associativity and positivity (which follows from unitarity) into inequalities which can be numerically implemented.

Coming back to our approach, to convert Polyakov's scheme into a calculationally effective tool we mix in our second ingredient which is the Mellin representation of CFT amplitudes. The position space amplitude ${\cal A}(u,v)$ has the Mellin representation 
\be\label{idmelldef}
{\cal A}(u,v)= \int_{-i\infty}^{i\infty}\frac{ds}{2\pi i} \, \frac{dt}{2\pi i} \, u^{s}v^t \Gamma^2(-t)\Gamma^2(s+t)\Gamma^2(\Delta_{\phi}-s){\cal M}(s,t)
\ee
The additional kinematic $\Gamma$ factors in the measure are defined for convenience \cite{mack}. ${\cal M}(s,t)$ is the (reduced) Mellin amplitude for the original conformal amplitude ${\cal A}(u,v)$.  

Mellin amplitudes are ideally suited to our present purpose since they share many of the features of momentum space for standard S-matrix amplitudes. In particular, the contributions of different operators show up as poles with a factorisation of the residues into lower point amplitudes. Moreover, our building blocks, the Witten exchange diagrams, are complicated in position space but are analytically easier to deal with in the Mellin representation. In fact, they can be viewed as the meromorphic piece of the conformal blocks in Mellin space, which are also known explicitly since the work of Mack. They therefore have the same poles as the corresponding conformal blocks together with the same residues thus exhibiting the needed factorisation. In fact, as mentioned above, the Witten blocks are better behaved in Mellin space compared to conformal blocks: the latter have exponential dependence on the Mellin variables compared to the former which are polynomially bounded. 

We can now translate the presence of spurious power law (as well as log) dependence in position space into Mellin space. The $u^{\D_\phi}$ (and $u^{\D_\phi}\ln{u}$) behaviour arise from spurious poles (and double poles) at $s=\Delta_{\phi}$ where $s$ is the Mellin variable conjugate to the cross ratio $u$. Therefore, we now demand that these residues vanish identically, i.e. as a function of the other Mellin variable, $t$. This gives an infinite set of constraints on operator dimensions and OPE coefficients\footnote{As mentioned above, there are additional spurious powers which lead to subsidiary spurious poles (double as well as single) at 
$s=\Delta_{\phi}+m$ (with $m=1,2, \ldots$).}. Here another advantage of the Mellin representation makes its appearance. In analogy with partial wave expansions for momentum space scattering amplitudes, there is a natural decomposition of the residues into a sum over a basis of orthogonal polynomials in the $t$-variable. These polynomials (known as the continuous Hahn polynomials in the mathematics literature) go over to the generalised Legendre (or Gegenbauer) polynomials in an appropriate flat space limit. This decomposition makes the imposition of our infinite set of conditions more tractable analytically. One big simplification is that operators of spin-$\ell$ contribute (in the $s$-channel) only to the orthogonal polynomial of degree $\ell$, as one might expect in analogy with flat space scattering. In the $t$-channel an infinite number of spins do contribute but this happens in a relatively controlled way.  This makes it natural to impose the vanishing residue conditions independently for each partial wave $\ell$.\footnote{In the discussion section we will mention an alternative procedure for imposing the constraints which maybe more efficient numerically -- by expanding in a power series around a special point $t=t_0$ and setting each of the terms to zero. This set of conditions is linearly related to the set of conditions from  the partial waves.} This feature of the Mellin space approach to the conformal bootstrap makes it very close in spirit to the flat space S-matrix bootstrap (see \cite{penedsmat} for one recently proposed way of connecting the two).

\subsection{Results}

It turns out to be simplest to implement this schema when there is a small parameter to expand in. 
We will focus here on two such examples. The first is the canonical  $\epsilon$ expansion in $d=(4-\e)$ dimensions for a single real scalar at the Wilson-Fisher fixed point. The second is the large spin limit (in any dimension) in scalar theories with a twist gap. 

In the former case, we will see that, in the Mellin partial wave decomposition, there are some significant simplifications when we impose the vanishing constraints. By assuming the existence of a stress tensor of dimension $d$ and examining the $\ell=(0,2)$ partial wave contributions we find that the lowest couple of orders in 
$\e$ get contribution only from the $\phi^2$ primary exchange, in addition to the identity operator. This enables us to recover known results for the anomalous scaling dimensions ($\D_\phi$ and $\D_0$, respectively) of both 
$\phi$ and $\phi^2$. The anomalous dimensions of these operators are known upto $O(\e^5)$ \cite{kleinert}. 
\be\label{Dphi0}
\Delta_\phi =1-\frac{\epsilon }{2}+\frac{1}{108} \epsilon^2+\frac{109}{11664} \epsilon^3+O(\e^4) ;\, \, \,\,\ \,  \D_0=2-\frac{2}{3}\e+\frac{19}{162}\e^2+O(\e^3).
\ee
We also find the OPE coefficient $C_{\phi\phi\phi^2}=C_0$ with a new result for the $O(\e^2)$ piece
\be\label{opewf}
C_{0}=2-\frac{2}{3}\e-\frac{34}{81}\e^2+O(\e^3)\,.
\ee
In fact, if we take the input from Feynman diagram calculations of the $O(\e^3)$ contribution to $\D_0$ then we can make a new prediction \eqref{C0new} for the corresponding $O(\e^3)$ contribution to $C_0$ as well.
By moving onto the partial wave $\ell$ we again find that, in the $s$-channel, it is only the leading twist operators of spin $\ell$ (of the schematic form $J^\ell= \phi \partial^{\ell}\phi$) that contribute to the first two non-vanishing orders in $\e$. Once again, to this same order in the $t$-channel it is only the $\phi^2$ (and identity) contribution that is needed. This enables one to obtain, in a fairly easy way, the nontrivial results for the anomalous dimensions of these operators
\be\label{genspin}
\D_\ell=d-2+\ell+\left(1-\frac{6}{\ell (\ell +1)}\right)\frac{\epsilon^2}{54}++\d^{(3)}_{\ell}\e^3+O(\e^4)\,,
\ee
with $\d^{(3)}_{\ell}$ being given in \eqref{dl3}. The $O(\e^3)$ term matches with the nontrivial Feynman diagram computations of \cite{gracey}. 
Our approach gives the OPE coefficients too with equal ease unlike other methods. We thus obtain 
$C_{\phi\phi J^\ell}=C_{\ell}$ to $O(\e^3)$ as given in \eqref{clcfr}, \eqref{Cl3} which are both new results. In the case of $\ell=2$ we can compare with previous results on the central charge $c_T$ which is related to $C_2$ (by \eqref{cTformula}). Our result
\be\label{ct30}
\frac{c_T}{c_{free}}=1-\frac{5 \epsilon ^2}{324}-\frac{233 \epsilon ^3}{8748}+O(\e^4)\, ,
\ee
agrees with previous calculations to $O(\e^2)$ \cite{hathrell, jack-osborn, petkou} and gives a new prediction at $O(\e^3)$. 

As we describe in section \ref{ising} these results, after setting $\e=1$, compare very well with some of the numerical results obtained for the 3d Ising model.   

The second context is of the large spin asymptotics (in a general dimension $d$, for Wilson-Fisher like fixed points) we consider the two regimes that have been analysed in the literature through the (double) lightcone expansion. Our techniques here allow us to reproduce results in both the large and small twist gap regimes. Thus we reproduce the results of \cite{anboot,Komargodski} for the anomalous dimensions of the operators $J^\ell$ in \eqref{gammastrng} and the leading corrections to the OPE coefficients in \eqref{opestrng}. In an opposite ``weakly coupled" regime \cite{alday} we reproduce again the  anomalous dimensions of the operators $J^\ell$ in
\eqref{gammaellweak} together with a new determination of the coefficients in \eqref{alphai}. 

The plan of the paper is as follows. In section 2 we discuss both Witten diagrams and the usual conformal blocks in Mellin space. We also discuss the spectral function representation of Witten diagrams that we employ in the rest of the paper. In section 3 we explain how to implement the bootstrap in Mellin space. In section 4 we turn to the identical scalar case which sets up the explicit 
$\e$-expansion calculation in section 5. Section 6 deals with large spin asymptotics both for strongly coupled theories and weakly coupled theories. We conclude in section 7 with a preliminary discussion on numerics and future directions. There are several appendices containing useful identities and intermediate results.

\section{Witten diagrams \& conformal blocks in Mellin space}

In this section we begin the process of migrating to Mellin space by carrying over the familiar conformal blocks and the associated Witten exchange diagrams from position space. 

We will consider the somewhat more general case of arbitrary scalar external operators and define our amplitudes, setting notation in the process.  
Let ${\cal A}(x_1,x_2,x_3,x_4)$ denote the four point function of four scalar operators in a CFT (the scalar ${\cal O}_i$ has dimension $\Delta_i$). 
\begin{eqnarray}\label{overall}
{\cal A}(x_1,x_2,x_3,x_4) &=& \langle {\cal O}_1(x_1){\cal O}_2(x_2){\cal O}_3(x_3){\cal O}_4(x_4)\rangle\,  \nonumber \\
&=& \frac{1}{(x_{12}^2)^{\frac{1}{2}(\D_1+\D_2)}(x_{34}^2)^{\frac{1}{2}(\D_3+\D_4)}}\bigg(\frac{x_{14}^2}{x_{24}^2}\bigg)^{a_s}\bigg(\frac{x_{14}^2}{x_{13}^2}\bigg)^{b_s} {\cal A}(u,v) \,.
\end{eqnarray}
Here we have pulled out overall factors in the four point function appropriate for an $s$-channel decomposition and defined
\be\label{asbs}
a_s=-\frac{1}{2}(\Delta_1-\Delta_2)\,, \ \ b_s=\frac{1}{2}(\D_{3}-\Delta_4)\, .
\ee 
The cross ratios $(u,v)$ are defined in the standard way
\be
u=\frac{x_{12}^2x_{34}^2}{x_{13}^2x_{24}^2} ; \, \, \, \, \, v=\frac{x_{14}^2x_{23}^2}{x_{13}^2x_{24}^2} \,.
\ee
The corresponding Mellin amplitude reads as\footnote{This is related to the conventional Mellin variables \cite{mack} $\delta_{ij}$ by some shifts: $s= \frac{1}{2}(\Delta_1+\Delta_2)-\delta_{12}; t= \delta_{12}+\delta_{13}-\frac{1}{2}(\D_1+\D_2+\D_3-\D_4)$. See also \cite{dolanosborn2}.} 
\begin{align}
\begin{split}\label{nonidmelldef}
{\cal A}(u,v) =&\int_{-i\infty}^{+i\infty} \frac{ds}{2\pi i}\ \frac{dt}{2\pi i}\ u^s v^t \G(\frac{\D_1+\D_2}{2}-s)\G(\frac{\D_3+\D_4}{2}-s)\G(-t)\G(-a_s-b_s-t)\\
&\times\G(s+t+a_s)\G(s+t+b_s){\cal M}(s,t)  \\
\equiv &\int_{-i\infty}^{+i\infty} \frac{ds}{2\pi i}\ \frac{dt}{2\pi i}\ u^s v^t \rho_{\{\D_i\}}(s,t) {\cal M}(s,t) . 
\end{split}
\end{align}
Setting $\Delta_i=\Delta_{\phi}$, we recover the previous expression Eq.(\ref{idmelldef}). 
Note that we are making a particular choice here so that even when we consider $t,u$-channel exchange diagrams, we will still be using the convention (\ref{nonidmelldef}) with the overall factors as in (\ref{overall}).  

Though we will not be utilising them very much, let us discuss how the conformal blocks look in Mellin space \cite{mack, fitzpatrick}. Under the transform of  Eq.(\ref{nonidmelldef}), 
the conformal blocks 
\be
G_{\Delta,\ell}^{(s)}(u,v) \rightarrow B_{\Delta,\ell}^{(s)}(s,t).
\ee
The Mellin space conformal blocks $B_{\Delta,\ell}^{(s)}(s,t)$ take the form \cite{fitzpatrick}
\begin{eqnarray}\label{mellbl}
B_{\Delta,\ell}^{(s)}(s,t) &\equiv & e^{i\pi(h-\Delta)}(e^{i\pi(2s+\Delta+\ell-2h)}-1)
\Omega_{\Delta-h,\ell}^{(s)}(s) P^{(s)}_{\Delta-h,\ell}(s,t) \nonumber \\
&=&e^{i\pi(h-\Delta)}(e^{i\pi(2s+\Delta+\ell-2h)}-1)
\frac{\G(\frac{\Delta-\ell}{2}-s)\G(\frac{2h-\Delta-\ell}{2}-s)}{\G(\frac{\Delta_1+\Delta_2}{2}-s) \G(\frac{\Delta_3+\Delta_4}{2}-s)} P^{(s)}_{\Delta-h,\ell}(s,t) \nonumber \\
&=& e^{i\frac{\pi}{2}(2s-\Delta+\ell)}\frac{(-2\pi i)\G(\frac{\Delta-\ell}{2}-s)}{\G(s+1-h+\frac{\Delta-\ell}{2})\G(\frac{\Delta_1+\Delta_2}{2}-s) \G(\frac{\Delta_3+\Delta_4}{2}-s)} P^{(s)}_{\Delta-h,\ell}(s,t)   \, .
\end{eqnarray}
Here $\Omega_{\Delta-h,\ell}^{(s)}(s) $ is defined, for later use, by the equality between the first and second lines. 
The Gamma functions in the numerator of  $\Omega_{\Delta-h,\ell}^{(s)}(s)$ exhibit poles at both $2s=\Delta-\ell+2m$ ($m=0,1,2\ldots$), which are physical, as well as at the so-called ``shadow" values $2s=2h-\Delta-\ell+2m$. (Here and below we use the conventional notation $h=\frac{d}{2}$, with $d$ the spacetime dimension). Since we would like to project out the contribution of the shadow poles the prefactor in brackets was introduced in \cite{fitzpatrick} so that it has zeroes precisely at these unphysical values. This cancellation of poles is made manifest in the third line. The projection, however, leads to an exponential dependence on $s$ at large values of this Mellin variable. 

The crucial piece of the conformal blocks in Mellin space are the $P^{(s)}_{\Delta-h,\ell}(s,t)$ -- the so-called Mack Polynomials which are of degree $\ell$ in the Mellin variables $(s,t)$. In addition to the dependence on $\Delta$, they also depend on the external scalars through 
$a_s, b_s$, but we suppress this dependence, so as not to clutter notation\footnote{Both the Mack Polynomials as well as $\Omega$ have a parametric dependence which is naturally in the combination $(\Delta-h)$ and this is reflected in their subscript.}. We merely signal this dependence through the superscript which indicates that we are considering parameters relevant to an $s$-channel. The explicit form of these polynomials is given in Appendix \ref{mack}.\footnote{Our normalisation of the Mack Polynomials agrees with that of Mack and differs from that of \cite{joao} by a factor of $(\Delta-1)_{\ell}(2h-\Delta-1)_{\ell}$.} 

The conformal blocks factorise on the physical poles $2s=\Delta-\ell+2m$ giving residues which are kinematic polynomials in the variable $t$ determined by the spin $\ell$ of the intermediate state and the level $m$ of the conformal descendants \cite{joao}. 
\be\label{mellblexp}
B_{\Delta,\ell}^{(s)}(s,t) = \sum_{m=0}^{\infty} \frac{(-1)^m\sin{\pi(\Delta-h)}\G(h-\Delta-m)}{m!\G(\frac{\Delta_1+\Delta_2-\Delta+\ell-2m}{2}) \G(\frac{\Delta_3+\Delta_4-\Delta+\ell-2m}{2})}\frac{Q^{\Delta}_{\ell,m}(t)}{2s-\Delta+\ell-2m}+\ldots
\ee
The dots refer to the entire function piece of the block in Eq.(\ref{mellbl}). That is the part which has an exponential behaviour at infinity. 
The $Q^{\Delta}_{\ell,m}(t)$ polynomials are single variable specialisations of the Mack Polynomials. 
\be
Q^{\Delta}_{\ell,m}(t) \equiv \frac{4^\ell }{(\Delta-1)_{\ell}(2h-\Delta-1)_{\ell}}P^{(s)}_{\Delta-h,\ell}(s=\frac{\Delta-\ell}{2}+m,t). 
\ee
In particular, the case with $m=0$ is special and will play an important role in the following i.e. 
\be\label{hahndef}
Q^{\Delta}_{\ell,0}(t)= \frac{4^\ell }{(\Delta-1)_{\ell}(2h-\Delta-1)_{\ell}}P^{(s)}_{\Delta-h,\ell}(s=\frac{\Delta-\ell}{2},t). 
\ee
The $Q^{\Delta}_{\ell,0}(t)$ turn out to be a family of orthogonal polynomials (continuous Hahn Polynomials) whose properties are given in appendix \ref{A}. These can be viewed as the generalisations of the Legendre/Gegenbauer polynomials that accompany the partial wave decomposition for scattering amplitudes. 

Just as for the conformal blocks, we can consider the Mellin version of the contribution from Witten exchange diagrams under the transform \eqref{nonidmelldef}
\be
W_{\Delta,\ell}^{(s)}(u,v) \rightarrow M_{\Delta,\ell}^{(s)}(s,t).  
\ee
Witten exchange diagrams in Mellin space have been investigated in the literature \cite{pene, mig, joao}. It is known that they have the {\it same poles and residues} as the corresponding conformal blocks. However, they are polynomially bounded for large $(s,t)$, in contrast to the exponential dependence of conformal blocks 
$B_{\Delta,\ell}^{(s)}(s,t)$. They therefore take the form 
\be\label{mellwit}
M_{\Delta,\ell}^{(s)}(s,t)=\sum_{m=0}^{\infty} \frac{(-1)^m\sin{\pi(\Delta-h)}\G(h-\Delta-m)}{m!\G(\frac{\Delta_1+\Delta_2-\Delta+\ell-2m}{2}) \G(\frac{\Delta_3+\Delta_4-\Delta+\ell-2m}{2})}\frac{Q^{\Delta}_{\ell,m}(t)}{2s-\Delta+\ell-2m}+R_{\ell-1}(s,t) 
 \ee
where $R_{\ell-1}(s,t)$ is a polynomial of degree at most $(\ell-1)$ in $(s,t)$. Note that the first term is identical to that in \eqref{mellblexp}. The second term is an additional polynomial ambiguity coming from freedom in the choice of three point vertices in the bulk $AdS$ in defining the exchange diagram. The meromorphic piece is however fixed to be the same as that of the conformal blocks (\ref{mellblexp}). Since our interest is to use an appropriate basis, we will choose the ambiguity to our convenience. A particularly simple choice of basis would, for instance, be to only use the meromorphic piece of the conformal block  i.e. just the first term in (\ref{mellwit}). We can write this sum (for any $\ell$) in terms of a finite sum of hypergeometric functions. Our choice will actually involve the additional polynomial piece $R_{\ell-1}(s,t)$ as well. In the case of a scalar exchange, however, such terms don't enter and the answer for the corresponding sum in \eqref{mellwit} is particularly simple \cite{pene, mig}
\begin{align}
M_{\Delta,\ell=0}^{(s)}(s,t)= \frac{1}{2s-\Delta}\frac{\Gamma^2(\Delta_{\phi}+\frac{\Delta-2h}{2})}{\Gamma(1+\Delta-h)} {} _3F_2\bigg[\begin{matrix} &1-\Delta_{\phi}+\frac{\Delta}{2}, 1-\Delta_{\phi}+\frac{\Delta}{2},\frac{\Delta}{2}-s \\
& 1+\frac{\Delta}{2}-s, 1+\Delta-h\end{matrix};1\bigg] .
\end{align}

In forthcoming work \cite{rajanind} we will employ this direct method to explicitly write down the Witten exchange function. In the current paper, we use an alternative approach to writing the exchange diagram in terms of a spectral function representation. While this introduces some additional terminology, it will have some advantages for implementing our bootstrap philosophy.

\subsection{The spectral function representation}

Our starting point will be the spectral representation of the Witten exchange function in position space (in, say, the $s$-channel).  Following (\ref{overall}) we define
\be\label{witoverall}
W_{\Delta,\ell}^{(s)}(x_i) =  \frac{1}{(x_{12}^2)^{\frac{1}{2}(\D_1+\D_2)}(x_{34}^2)^{\frac{1}{2}(\D_3+\D_4)}}\bigg(\frac{x_{14}^2}{x_{24}^2}\bigg)^{a_s}\bigg(\frac{x_{14}^2}{x_{13}^2}\bigg)^{b_s}W_{\Delta,\ell}^{(s)}(u,v)\,.
\ee
The spectral representation is then a decomposition in terms of conformal partial waves (see for e.g. \cite{costa}, Sec. 6). This follows from a ``split" representation of the bulk-to-bulk propagator in terms of two bulk-to-boundary propagators with a spectral parameter $\n$ that is integrated over. The latter can be expressed in terms of conformal partial waves
\be\label{hf}
W_{\Delta,\ell}^{(s)}(u,v) = \int_{-i\infty}^{i\infty} d\nu\, \mu^{(s)}_{\Delta, \ell}(\nu) F^{(s)}_{\nu,\ell}(u,v)\,.
\ee
The conformal partial waves $F^{(s)}_{\nu,\ell}(u,v)$ are closely related \cite{dolanosborn2, joao} to the conformal blocks being just linear combinations of 
a block of fictitious dimension $\Delta=h+ \nu$ and its shadow with dimension $d-\Delta= h-\nu$. 
\be\label{fgrel}
F_{\nu,\ell}^{(s)}(u,v)=\left(n(\nu,\ell) G^{(s)}_{h+\nu,\ell}(u,v)+n(-\nu,\ell)G^{(s)}_{h-\nu,\ell}(u,v)\right)\,,
\ee
where 
\be\label{nnorm}
n(\nu,\ell)=\frac{2^{-\ell}\G(-\nu)(h-\nu-1)_\ell}{\G(h+\nu+\ell)}\gamma_{\lambda_1,a_s}\gamma_{\lambda_1,b_s}\,.
\ee
We also follow \cite{dolanosborn2} in introducing the notation $\lambda_1=(h+\nu+\ell)/2$, $\bar\lambda_1=(h-\nu+\ell)/2$ and 
\be\label{gammadef}
\g_{x,y}=\G(x+y)\G(x-y)\,.
\ee

The spectral function $\mu^{(s)}_{\Delta, \ell}(\nu)$ itself is the dynamical piece that contains information about the exchanged operator with dimension $\D$.
We can further break it as $\mu^{(s)}_{\Delta,\ell}(\nu)= \sigma_\ell(\nu)\rho^{(s)}_{\Delta, \ell}(\nu)$ to exhibit a piece which is group theoretical (the Plancherel measure for the conformal group, see for e.g. Appendix B of \cite{GGL}))
\be
\sigma_{\ell}(\nu)= \frac{\G(h+\ell)}{2(2\pi)^h\ell !}\frac{\G(h+\nu-1)\G(h-\nu-1)}{\G(\nu)\G(-\nu)}[(h+\ell-1)^2-\nu^2]
\ee
and a piece which is dynamical (i.e. $\rho^{(s)}_{\Delta, \ell}(\nu)$) and thus knows about $\Delta$. The explicit expression for $\rho^{(s)}_{\Delta, \ell}(\nu)$ is
\be\label{specfn}
\rho^{(s)}_{\Delta, \ell}(\nu) = \frac{1}{2\pi i ((\D-h)^2-\nu^2)}\frac{\G(\frac{\D_1+\D_2-h+\ell+\nu}{2})\G(\frac{\D_1+\D_2-h+\ell-\nu}{2})\G(\frac{\D_3+\D_4-h+\ell+\nu}{2})\G(\frac{\D_3+\D_4-h+\ell-\nu}{2})}{\G(h+\nu+\ell)\G(h-\nu+\ell)}\,.
\ee
The interpretation as a spectral function comes from the fact that we can evaluate the integral along the imaginary axis by closing the contour (when the integral is well behaved at infinity) on, say, the right half plane and picking up the residues at the simple poles. These then correspond to the primary operators which are exchanged whose contribution, along with their conformal descendants, is captured by the conformal block in \eqref{fgrel}. The contour is chosen to enclose either an operator or its shadow, but not both. The superscript in $\rho^{(s)}$ (and $\mu^{(s)}$) signifies the channel and is reflected in the dependence on the $\D_i$ in \eqref{specfn}.

Let us see which primaries contribute.
The spectral function in (\ref{specfn}) has simple poles at $\pm \nu =\Delta -h$ corresponding to the operator $(\Delta, \ell)$ (and its shadow). But it also has simple poles at $h\pm \nu = \Delta_1+\Delta_2+\ell+2n$
and $h\pm \nu = \Delta_3+\Delta_4+\ell+2m$ where  $n$ and $m$ are non-negative integers. In a generic theory there are no operators of this dimension. These are dubbed as ``double-trace" operator contributions in the $AdS/CFT$ literature. This is because in a weakly coupled (``generalised free field") theory, like in the large N limit,  these would be the dimensions of double trace operators of the schematic form ${\cal O}_1\partial^{\ell} (\partial^2)^n {\cal O}_2$ (and similarly with ${\cal O}_3$ and ${\cal O}_4$). It is known \cite{liu, rastelli-etal} that precisely these double trace primary operators (of spin $\ell$) do contribute to the Witten diagram (in the $s$-channel) and the spectral function merely reproduces this fact.   Note that when we close the $\nu$ contour on the right half plane only the poles with the plus sign will (typically) contribute\footnote{Note that this would imply that there are no poles coming from $\gamma_{\lambda_1,a_s}\gamma_{\lambda_1,b_s}$ in the integrand. For general $a_s$, there may be a finite set of poles on the right but an infinite set of poles on the left. Our choice of contour is such that the entire infinite chain is on one side of the contour.}. 

The full spectral function (with the Plancherel measure) thus takes the form
\be\label{specunitry0}
\mu^{(s)}_{\Delta,\ell}(\nu)=\frac{\G(\frac{\D_1+\D_2-h+\ell+\nu}{2})\G(\frac{\D_1+\D_2-h+\ell-\nu}{2})\G(\frac{\D_3+\D_4-h+\ell+\nu}{2})\G(\frac{\D_3+\D_4-h+\ell-\nu}{2})}{2\pi i ((\D-h)^2-\nu^2)\G(\nu)\G(-\nu)(h+\nu-1)_\ell(h-\nu-1)_\ell}\,.
\ee
We have seen that this has the right behaviour to reproduce the known properties of the Witten exchange diagram. There is one subtlety though that we should mention in this form of the spectral function. There are a finite number of extra poles in the $\nu$ integral from the denominator factors of $(h\pm \nu-1)_{\ell}$ whose contributions need to be cancelled by adding lower order spin terms $(\ell^\prime < \ell)$. These have been explicitly studied in \cite{costa}. However, these additional terms will not contribute to the terms of interest to us which are the residues at the double trace poles which come purely from the above piece.  

In the particular case of identical scalars for $d=4$, the above spectral function $\mu^{(s)}_{\Delta,\ell}(\nu)$  agrees with what was constructed by Polyakov for what he called the `unitary' amplitude. Polyakov, of course, did not come to this from Witten exchange diagrams. He first constructed a spectral function for what he called the `algebraic amplitude'  which is nothing other than the conformal block itself. This turns out to have the form 
\be\label{specalg}
\mu^{Alg}_{\Delta,\ell}(\nu)=\frac{1}{2\pi i ((\D-h)^2-\nu^2)\G(\nu)\G(-\nu)(h+\nu-1)_\ell(h-\nu-1)_\ell}\,.
\ee
This algebraic spectral function $\mu^{Alg}_{\Delta,\ell}(\nu)$ is designed to reproduce the conformal block on the LHS  if we insert it in the RHS of \eqref{hf} instead of $\mu^{(s)}_{\Delta,\ell}(\nu)$. We see that now the $\nu$ contour integral over the right half plane only gets contributions from the single pole at $\nu=\D-h$ after using  (\ref{fgrel}),(\ref{nnorm}). Note that $\mu^{Alg}_{\Delta,\ell}(\nu)$ in \eqref{specalg} differs from  $\mu^{(s)}_{\Delta,\ell}(\nu)$ given in  (\ref{specunitry0}) by the four numerator $\G$-functions in the latter which were the double trace contributions. 
However, $\mu^{Alg}_{\Delta,\ell}(\nu)$ suffers from the problem that it diverges as one goes to 
$\pm i\infty$. As we will see later, this is related to the poor behaviour of the conformal block in Mellin space, at infinity along the imaginary axis. To cure this problem, Polyakov prescribes adding certain additional factors of 
$\G$-functions to the numerator. These turn out to be precisely (in his case, for identical scalars) the double trace ones which appear in the numerator of (\ref{specunitry0}) so that we indeed get the spectral function appropriate to the Witten exchange diagram!

We can now translate this spectral representation to Mellin space. We use the fact that the partial wave appearing in (\ref{hf}) has the Mellin representation \cite{dolanosborn2}
\begin{align}
\begin{split}\label{sparwavemell}
F_{\nu,\ell}^{(s)}(u,v)=& \int_{-i\infty}^{+i\infty} \frac{ds}{2\pi i}\ \frac{dt}{2\pi i}\ u^s v^t \rho_{\{\D_i\}}(s,t)\Omega_{\nu, \ell}^{(s)}(s)P^{(s)}_{\nu, \ell}(s,t) 
\end{split}
\end{align}
in terms of the Mack Polynomials as well as the $\Omega_{\nu,\ell}^{(s)}(s)$ defined in (\ref{mellbl}) and the standard Mellin measure $\rho_{\{\D_i\}}(s,t)$ in \eqref{nonidmelldef}. 

Then, combining (\ref{sparwavemell}) with (\ref{hf}) we have the spectral representation for the $s$-channel Witten exchange diagram in Mellin space to be 
\begin{align}
\begin{split}\label{sunitrymell}
M_{\Delta,\ell}^{(s)}(s,t) = \int_{-i\infty}^{i\infty} d\nu\, \mu^{(s)}_{\Delta,\ell}(\nu)
\Omega_{\nu, \ell}^{(s)}(s)P^{(s)}_{\nu, \ell}(s,t)\,.
\end{split}
\end{align}

\subsection{Adding in the $t,u$ channels}

Our discussion was for the $s$-channel contribution $W^{(s)}_{\Delta,\ell}(u,v)$ to (\ref{polyapp}) and the corresponding $M^{(s)}_{\Delta,\ell}(s,t)$ in Mellin space. It is not difficult to extend the discussion to the other two channels. The main point to keep in mind is that our conventions are chosen, for definiteness, for an $s$-channel expansion. Thus we pull out the {\it same} external factor as in (\ref{overall}) when we are considering the reduced amplitude in the $t$ and $u$-channels also. This is even though the natural definition for the Witten diagram in the $t$-channel would involve an interchange of subscripts $(2,4)$ (and similarly $(2,3)$ for the $u$-channel) of the $s$-channel which gives, for instance, 
answer (\ref{witoverall})
\be\label{prefactt}
W^{(t)}_{\Delta,\ell}(x_i) = \frac{1}{(x_{14}^2)^{\frac{1}{2}(\D_1+\D_4)}(x_{23}^2)^{\frac{1}{2}(\D_2+\D_3)}}\bigg(\frac{x_{12}^2}{x_{24}^2}\bigg)^{a_t}\bigg(\frac{x_{12}^2}{x_{13}^2}\bigg)^{b_t}
W_{\Delta,\ell}^{(t)}(u,v)\,,
\ee
where $a_t= -\frac{1}{2}(\D_1-\D_4)$ and $b_t= \frac{1}{2}(\D_3-\D_2)$. If we recast this in the form (\ref{overall}) by pulling out the same external factor as in that equation, then this corresponds to multiplying $W_{\Delta,\ell}^{(t)}(u,v)$ by an extra factor of $u^{\frac{1}{2}(\D_3+\D_4)}v^{-\frac{1}{2}(\D_2+\D_3)}$. In a similar way, an extra factor of $u^{\frac{1}{2}(\D_1+\D_4)}$ multiplies $W_{\Delta,\ell}^{(u)}(u,v)$. Here, both $W_{\Delta,\ell}^{(t,u)}(u,v)$ are obtained from $W_{\Delta,\ell}^{(s)}(u,v)$ by the interchange of labels $(2,4)$ (and $(2,3)$, respectively).


We can translate this to Mellin space in a straightforward manner. Thus the $t$-channel partial wave (the analogue of (\ref{sparwavemell}) reads, with the above prefactor, as
\begin{align}
\begin{split}\label{tparwavemell}
&u^{\frac{1}{2}(\D_3+\D_4)}v^{-\frac{1}{2}(\D_2+\D_3)}F_{\nu,\ell'}^{(t)}(u,v)=u^{\frac{1}{2}(\D_3+\D_4)}v^{-\frac{1}{2}(\D_2+\D_3)}\int \frac{ds}{2\pi i} \frac{dt}{2\pi i}\ u^s\ v^t\G(-s)\G(-a_t-b_t-s)\\ &\times 
\G(a_t+s+t)\G(b_t+s+t)\G(\frac{1}{2}(\Delta-\ell')-t)\G(\frac{1}{2}(2h-\Delta-\ell')-t)P^{(t)}_{\nu, \ell'}(s,t) \,.
\end{split}
\end{align}
Here the integrand on the RHS is obtained from the corresponding one of the $s$-channel (\ref{sparwavemell}),  with the interchange $(s \leftrightarrow t)$ i.e. of labels $(2,4)$. The superscript $t$ on the Mack polynomials also indicates this exchange -- we have $P^{(t)}_{\nu, \ell}(s,t)=P^{(s)}_{\nu, \ell}(t,s)|_{a_s\rightarrow a_t, b_s\rightarrow b_t}$. Here $\ell'$ denotes the spin in the $t$-channel.

But now observe that by shifting variables 
\be
s\rightarrow s-\frac{1}{2}(\D_3+\D_4)\,, \ \ \ t\rightarrow t+\frac{1}{2}(\D_2+\D_3)\, ,
\ee 
we can make the RHS of (\ref{tparwavemell}) now in the same form as  (\ref{sparwavemell}) i.e.
\begin{align}
\begin{split}\label{tparwavemell1}
\int \frac{ds}{2\pi i} \frac{dt}{2\pi i}\ & u^s v^t \rho_{\{\D_i\}}(s,t)\Omega_{\nu, \ell'}^{(t)}(t)
P^{(t)}_{\nu, \ell'}(s-\frac{1}{2}(\D_3+\D_4),t+\frac{1}{2}(\D_2+\D_3)) \,,
\end{split}
\end{align}
where 
\be
\Omega^{(t)}_{\nu, \ell'}(t)= \frac{\G(\frac{1}{2}(h+\nu-\ell')-t-\frac{1}{2}(\D_2+\D_3))\G(\frac{1}{2}(h-\nu-\ell')-t-\frac{1}{2}(\D_2+\D_3))}{\G(-t)\G(-a_s-b_s-t)} \, .
\ee

Similarly, in the $u$-channel, we have 
\begin{align}
\begin{split}\label{uparwavemell}
&u^{\frac{1}{2}(\D_1+\D_4)}F_{\nu,\ell'}^{(u)}(u,v)=\int \frac{ds}{2\pi i}\ \frac{dt}{2\pi i}\ u^s v^t
\rho_{\{\D_i\}}(s,t)
\Omega_{\nu, \ell'}^{(u)}(s+t)P^{(u)}_{\nu, \ell'}(s-\frac{1}{2}(\D_1+\D_4),t)
\end{split}
\end{align}
where
\be
\Omega^{(u)}_{\nu, \ell'}(s+t)= \frac{\G(\frac{1}{2}(h+\nu-\ell')+s+t-\frac{1}{2}(\D_1+\D_4))\G(\frac{1}{2}(h-\nu-\ell')+s+t-\frac{1}{2}(\D_1+\D_4))}
{\G(b_s+s+t)\G(a_s+s+t)} \,,
\ee
and  $P^{(u)}_{\nu, \ell}(s,t)=P^{(s)}_{\nu, \ell}(-s-t,t)|_{a_s\rightarrow a_u, b_s\rightarrow b_u}$.
Here $a_u= -\frac{1}{2}(\D_1-\D_3)$ and $b_u= \frac{1}{2}(\D_2-\D_4)$.

This was for the partial waves in the $t,u$-channels. We can now employ the corresponding versions of (\ref{hf}) to write the expressions for the corresponding Witten exchange diagrams in the spectral representation in  Mellin space i.e. the counterparts of (\ref{sunitrymell}).
Combining (\ref{tparwavemell}) with the analogue of (\ref{hf}), we find 
\begin{align}
\begin{split}\label{tunitrymell}
M_{\Delta,\ell'}^{(t)}(s,t) = \int_{-i\infty}^{i\infty} d\nu\, \mu^{(t)}_{\Delta,\ell'}(\nu)\Omega_{\nu, \ell'}^{(t)}(t)
P^{(t)}_{\nu, \ell'}(s-\frac{1}{2}(\D_3+\D_4),t+\frac{1}{2}(\D_2+\D_3)) \, .
\end{split}
\end{align}
And similarly in the $u$-channel with (\ref{uparwavemell})
\begin{align}
\begin{split}\label{uunitrymell}
M_{\Delta,\ell'}^{(u)}(s,t) = \int_{-i\infty}^{i\infty} d\nu\, \mu^{(u)}_{\Delta,\ell'}(\nu)\Omega_{\nu, \ell'}^{(u)}(s+t) P^{(u)}_{\nu, \ell'}(s-\frac{1}{2}(\D_1+\D_4),t ) \, .
\end{split}
\end{align}

Here the spectral weights, $\mu^{(t)}_{\Delta,\ell'}(\nu), \mu^{(u)}_{\Delta,\ell'}(\nu)$ are given by (\ref{specunitry0}) with the exchange of subscripts $(2 \leftrightarrow 4)$ and $(2 \leftrightarrow 3)$ respectively.

\section{The bootstrap strategy implemented}

With all this machinery in place, we are now ready to come to the crux of our strategy. As mentioned in the introduction, we write the four point function as a sum over a set of crossing symmetric Witten exchange diagrams as in (\ref{polyapp}). In position space this can be written, using the spectral representation (\ref{hf}), as
\begin{align}\label{totpos}
\begin{split}
{\cal A}(u,v)= &   \sum_{\Delta, \ell}  \int_{-i\infty}^{i\infty} d\nu\, 
\bigg(c^{(s)}_{\D, \ell}  \mu^{(s)}_{\Delta,\ell}(\nu)F_{\nu,\ell}^{(s)}(u,v)
+c^{(t)}_{\D, \ell} \mu^{(t)}_{\Delta,\ell}(\nu) u^{\frac{1}{2}(\D_3+\D_4)}v^{-\frac{1}{2}(\D_2+\D_3)}F_{\nu,\ell}^{(t)}(u,v)\\
+& c^{(u)}_{\D, \ell} \mu^{(u)}_{\Delta,\ell}(\nu) u^{\frac{1}{2}(\D_1+\D_4)}F_{\nu,\ell}^{(u)}(u,v) \bigg) \,. 
\end{split}
\end{align}
Here the sum over $\D, \ell$ is over the entire physical (primary) operator spectrum of the CFT. Note that we have, in general, to-be-determined coefficients $c^{(s,t,u)}_{\D, \ell}$ which are mutually related by exchanges of the labels (e.g. $(2 \leftrightarrow 4)$ or $(2 \leftrightarrow 3)$). This ensures that the full amplitude is crossing symmetric.

Since we are {\it not} making an expansion of the amplitude in terms of conformal blocks in a fixed channel, we are not guaranteed that this expansion will have the right power law dependences on the positions (or equivalently, cross-ratios) that is consistent with the OPE. For instance, in the case of identical scalars we see from (\ref{specunitry}) that the spectral function  $\mu^{(s)}_{\Delta,\ell}(\nu)$ has double poles (at $h+\nu=\Delta_\phi$, where $\Delta_\phi$ is the dimension of the common external scalar). When we perform the $\nu$ integral, this double pole gives rise to $u^{\Delta_\phi}\log{u}$ terms in the sum, as well as $u^{\Delta_\phi}$ terms. Both of these dependences would imply the presence of an operator with dimension $2\Delta_{\phi}$ in the spectrum which is generically not the case.  More generally, we will have spurious power laws of the form  $u^{\frac{\D_1+\D_2}{2}}$ and $u^{\frac{\D_3+\D_4}{2}}$ when we expand \eqref{totpos} in the $s$-channel. 
There are generically no operators corresponding to dimensions $(\D_1+\D_2)$ and $(\D_3+\D_4)$. 
Thus we have to demand that these terms identically vanish after including the contributions from the other channels and on summation over $(\D, \ell)$.

As discussed, it will be easier to implement this in Mellin space. In other words, we look at the total Mellin space amplitude corresponding to  (\ref{totpos}) which we obtain by putting together (\ref{sunitrymell}), (\ref{tunitrymell}) and  (\ref{uunitrymell})
\begin{align}\label{totmell}
\begin{split}
{\cal M}(s,t)= &   \sum_{\Delta, \ell}  \bigg(c^{(s)}_{\D, \ell}M_{\Delta,\ell}^{(s)}(s,t)+c^{(t)}_{\D, \ell}M_{\Delta,\ell}^{(t)}(s,t)
+c^{(u)}_{\D, \ell}M_{\Delta,\ell}^{(u)}(s,t) \bigg) \\
= \sum_{\Delta, \ell}  & \int_{-i\infty}^{i\infty} d\nu\, 
\bigg( c^{(s)}_{\D, \ell}\mu^{(s)}_{\Delta,\ell}(\nu)\Omega_{\nu, \ell}^{(s)}(s)P^{(s)}_{\nu, \ell}(s,t)
+c^{(t)}_{\D, \ell}\mu^{(t)}_{\Delta,\ell}(\nu)\Omega_{\nu, \ell}^{(t)}(t)
P^{(t)}_{\nu, \ell}(s-\frac{1}{2}(\D_3+\D_4),t+\frac{1}{2}(\D_2+\D_3))  \\
+&c^{(u)}_{\D, \ell} \mu^{(u)}_{\Delta,\ell}(\nu) \Omega_{\nu, \ell}^{(u)}(s+t) P^{(u)}_{\nu, \ell}(s-\frac{1}{2}(\D_1+\D_4),t ) \bigg) \,.
\end{split}
\end{align}

The definition of the Mellin transform in  (\ref{nonidmelldef}) imply that the spurious powers in position space mentioned in the previous para arise from spurious poles at $s=\frac{1}{2}(\Delta_1+\Delta_2)$ and $s=\frac{1}{2}(\Delta_3+\Delta_4)$. When the external scalars are identical, these two sets of spurious poles coalesce to give double as well as single poles at $s=\Delta_{\phi}$.  
It is important to note that these are statements about the full Mellin space amplitude and not just the reduced one, ${\cal M}(s,t)$. In other words, recalling the notation of  (\ref{nonidmelldef}) we need to examine the spurious poles of $\rho_{\{\D_i\}}(s,t){\cal M}(s,t)$. In particular, for identical scalars, the $\G^2(\Delta_{\phi}-s)$ piece of in (\ref{idmelldef}) already has double and single poles at  $s=\Delta_{\phi}$. So we will need to look at the constant as well as terms linear in $(s-\D_{\phi})$ of ${\cal M}(s,t)$ to isolate the poles of interest to us. 

In either case, the residues at these spurious poles will be a function of $t$ and we will obtain an infinite number of constraints on our CFT by setting these identically to zero. Below, we will individually look at the Mellin amplitudes in each channel, for non-identical scalars, and isolate the residues. We can then add them all up and find the conditions for consistency with the OPE. In the following section we will examine the special features that arise for identical scalars.
 




\subsection{The $s$-channel}

We start with the unitary block in the $s$-channel (i.e. the Mellin transform of the Witten exchange diagram) given in Eq. (\ref{sunitrymell}). 
\begin{align}
\begin{split}\label{sunitrymell1}
M_{\Delta,\ell}^{(s)}(s,t) = \int_{-i\infty}^{i\infty} d\nu\, \mu^{(s)}_{\Delta,\ell}(\nu)
\Omega_{\nu, \ell}^{(s)}(s)P^{(s)}_{\nu, \ell}(s,t)
\end{split}
\end{align}
where, as in (\ref{specunitry0}), we have the spectral function 
\be\label{specunitry2}
\mu^{(s)}_{\Delta,\ell}(\nu)=\frac{\G(\frac{\D_1+\D_2}{2}-\l_2)\G(\frac{\D_1+\D_2}{2}- \bar{\l}_2)\G(\frac{\D_3+\D_4}{2}-\l_2)\G(\frac{\D_3+\D_4}{2}- \bar{\l}_2)}{2\pi i ((\D-h)^2-\nu^2)\G(\nu)\G(-\nu)(h+\nu-1)_\ell(h-\nu-1)_\ell}\,
\ee
and
\be\label{oms0}
\Omega^{(s)}_{\nu, \ell}(s)= \frac{\G(\l_2-s)\G(\bar{\l}_2-s)}{\G(\frac{\D_1+\D_2}{2}-s)\G(\frac{\D_3+\D_4}{2}-s)} \, .
\ee 
Here we have introduced, for compactness, the notation  \cite{dolanosborn2}:
\be\label{lam2}
 \l_2=(h+\nu -\ell)/2, \,\, \bar{\l}_2  =(h- \nu -\ell)/2 \,
\ee
We are to carry out the $\nu$ integral by closing the contour on the right half plane. 

The ``physical pole" in the spectral function is the one at $\nu=(\D-h)$\footnote{We are assuming $\D-h\geq 0$. If not, we have to deform the contour so that we include this pole but {\it not} that of the shadow operator which would now lie on the right half plane \cite{mack}.}. In this case the factors of $\G(\frac{1}{2}(\D_1+\D_2)-s)$ and $\G(\frac{1}{2}(\D_3+\D_4)-s)$ in the denominator of $\Omega^{(s)}_{\nu, \ell}(s)$ cancel out with the corresponding  factors in $\rho_{\{\D_i\}}(s,t)$ of Eq.(\ref{nonidmelldef}).\footnote{For the scalar, this can be explicitly seen in the denominator $\G$ factors in (\ref{mellwit}) which cancel against $\rho_{\{\D_i\}}(s,t)$ at the physical pole in $s$.} The residue at this physical pole in $\nu$ has factors 
of $\G$-functions from the numerator of (\ref{oms0}) which give rise to the physical pole in the $s$-variable (as well as for the shadow) i.e.
\be
2s =\D-\ell +2n \, ; \,\,\,\,\,\,\, 2s =(2h-\D)-\ell +2n; \,\,\,\ (n=0,1, 2\ldots) \, .
\ee
When we do the $s$ integral (again closing the contour appropriately) of the Mellin amplitude, this gives rise to the physical contribution with a $u^{(\D-\ell)/2+n}$ dependence but does not pick up the shadow. 

However, there are other poles in $\nu$ which give rise to the spurious poles in $s$ that we described earlier. For instance, when we consider the poles from the numerator of the spectral function (\ref{specunitry2})
\be\label{spupole1}
h-\ell\pm \nu\ = \D_1+\D_2+2n
\ee
the residue will get a contribution from the numerator of (\ref{oms0}) 
$\propto \G(\frac{1}{2}(\D_1+\D_2)+n-s)$.\footnote{As well as a shadow piece $\G(h-\ell -\frac{1}{2}(\D_1+\D_2)-n -s)$ which will always be understood to be present but which we will ignore since we will choose to close the Mellin contour so as to exclude this set of poles.}
The denominator of  (\ref{oms0}) then cancels with the Mellin measure but the above piece gives spurious poles at
$s=\frac{1}{2}(\D_1+\D_2)+n$. By a similar argument there are spurious poles also at 
$s=\frac{1}{2}(\D_3+\D_4)+n$. Instead, if we had taken the poles from the numerator of (\ref{oms0}) i.e.
\be\label{spupole2}
h-\ell\pm \nu =2s-2n
\ee
then the residue contribution from the numerator of (\ref{specunitry2}) would be again $\propto \G(\frac{1}{2}(\D_1+\D_2)+n-s)$. Thus the residues of the Mellin amplitude evaluated on these second set of poles gives a factor of two to the previous contribution\footnote{The poles at $s=\frac{1}{2}(\D_1+\D_2)+n$ with $(n>0)$ actually come with a multiplicity. But since we will be focussing on the $n=0$ case in this paper, we will not worry about this factor.}.  We have already discussed (see around \eqref{specunitry0} and footnote 10) the absence of any role from all the other poles of the spectral function. 

Thus we will focus on the residue contribution on the poles  in \eqref{spupole2} (for $n=0$) that correspond to spurious poles at $s=\frac{\D_1+\D_2}{2}$ i.e. for $h-\ell+\nu=\D_1+\D_2$. The two variable Mack Polynomial simplifies in this case, when we further put $2s=\D_1+\D_2$. We use the general relation (\ref{hahndef}) to obtain 
\be\label{mackres}
P^{(s)}_{(\D_1+\D_2 +\ell-h),\ell}(s, t)\bigg|_{s=\frac{\D_1+\D_2}{2}}=4^{-\ell}(\D_1+\D_2 +\ell-1)_\ell(2h-\D_1-\D_2-\ell-1)_\ell\ Q^{\D_1+\D_2+\ell}_{\ell,0}(t)\,.
\ee

As a result the nett residue of the Mellin amplitude at the unphysical pole is 
\begin{align}\label{Mst_nuint}
\begin{split}
M^{(s)}(s,t)\bigg|_{s=\frac{1}{2}(\D_1+\D_2)}=&\sum_{\D,\ell}c^{(s)}_{\D,\ell}4^{1-\ell} (\D_1+\D_2+\ell-1)_\ell(2h-\D_1-\D_2-\ell-1)_\ell\G(h-\ell-\D_1-\D_2)\\
\times & \frac{2\pi i\mu^{(s)}_{\D,\ell}(2s-h+\ell)}{\G(\frac{1}{2}(\D_1+\D_2)-s)\G(\frac{1}{2}(\D_3+\D_4)-s)}\bigg|_{s=\frac{1}{2}(\D_1+\D_2)}Q^{\D_1+\D_2+\ell}_{\ell,0}(t)+\cdots\,,\\
\equiv &\sum_{\D,\ell}c^{(s)}_{\D,\ell}q^{(s)}_{\D,\ell}Q^{\D_1+\D_2+\ell}_{\ell,0}(t)+\cdots\,.
\end{split}
\end{align}
Here we have defined
\be\label{Cs}
q^{(s)}_{\D,\ell}=-4^{1-\ell} \frac{\G(\frac{1}{2}(\D_3+\D_4+\D_1+\D_2)-h+\ell)}{(\D_1+\D_2+\ell-h)^2 -(\D-h)^2}\,,
\ee
and the $\cdots$ denotes contribution from the physical pole $\D$ as well as the other spurious pole at 
$s=\frac{1}{2}(\D_3+\D_4)$ which gets an identical contribution to above with $(1,2)$ replaced by $(3,4)$. The case of identical scalars will be discussed separately in the next section. 

In the next subsections we look at the $t,u$-channels and similar pole contributions that lead to anomalous $u^{\frac{1}{2}(\D_1+\D_2)}$ behaviour in the amplitude. Demanding that these cancel against the above $s$-channel contribution will give us constraints. Note that the cancellation conditions involve a whole function of $t$.
To facilitate the comparison, we have expanded the functional dependence on $t$ in terms of the orthogonal polynomials $Q^{\D_1+\D_2+\ell}_{\ell,0}(t)$. We will expand the amplitudes in the other channels in terms of the same polynomials and use the orthogonality to set the nett coefficients of each $Q^{\D_1+\D_2+\ell}_{\ell,0}(t)$ to zero\footnote{At the other spurious pole $s=\frac{1}{2}(\D_3+\D_4)$, it will be natural to expand in the set of orthogonal polynomials $Q^{\D_3+\D_4+\ell}_{\ell,0}(t)$.}. We also note another important utility of this particular orthogonal decomposition -- a spin $\ell$ Witten diagram contributes {\it only} to the orthogonal polynomial labelled by the {\it same} $\ell$. This property makes this decomposition the analogue of the usual partial wave decompositions for flat space scattering amplitudes. The contributions from a specific spin is, however, a special feature of the $s$-channel decomposition and will not hold in the $t$-channel (see \eqref{torth}).

\subsection{The $t$-channel}

The unitary block in the $t$-channel is given by (\ref{tunitrymell}) 
\begin{align}
\begin{split}\label{tunitrymell1}
M_{\Delta,\ell'}^{(t)}(s,t) = \int_{-i\infty}^{i\infty} d\nu\, \mu^{(t)}_{\Delta,\ell'}(\nu)\Omega_{\nu, \ell'}^{(t)}(t)
P^{(t)}_{\nu, \ell'}(s-\frac{1}{2}(\D_3+\D_4),t+\frac{1}{2}(\D_2+\D_3)) \, .
\end{split}
\end{align}
where
\be\label{specunitryt}
\mu^{(t)}_{\Delta,\ell}(\nu)=\frac{\G(\frac{\D_1+\D_4}{2}-\l_2)\G(\frac{\D_1+\D_4}{2}- \bar{\l}_2)\G(\frac{\D_3+\D_2}{2}-\l_2)\G(\frac{\D_3+\D_2}{2}- \bar{\l}_2)}{2\pi i ((\D-h)^2-\nu^2)\G(\nu)\G(-\nu)(h+\nu-1)_\ell(h-\nu-1)_\ell}\,
\ee
and
\be\label{omt}
\Omega^{(t)}_{\nu, \ell}(t)= \frac{\G(\l_2-t-\frac{1}{2}(\D_2+\D_3))\G(\bar{\l}_2-t-\frac{1}{2}(\D_2+\D_3))}{\G(-t)\G(-a_s-b_s-t)} \, .
\ee
Here $\l_2, \bar{\l}_2$ are as defined before in \eqref{lam2}.

We want to look at the contribution to the amplitude from terms which go like $u^{\frac{1}{2}(\D_1+\D_2)}$. In the $t$-channel this can be attributed entirely to the  $\G(\frac{1}{2}(\D_1+\D_2)-s)$ factor in the Mellin measure 
$\rho_{\{\D_i\}}(s,t)$ of \eqref{nonidmelldef} -- there are no compensating denominator factors like in the $s$-channel.  Hence we can directly evaluate 
$M_{\D,\ell'}^{(t)}(s = \frac{1}{2}(\D_1+\D_2) ,t)$ to get the residue at this pole. We will then expand the resulting function of $t$ in terms of the orthogonal polynomials of the previous subsection
\be\label{torth}
M_{\D,\ell'}^{(t)}(s = \frac{1}{2}(\D_1+\D_2) ,t) \equiv \sum_{\ell}q^{(t)}_{\D,\ell | \ell'}Q^{\D_1+\D_2+\ell}_{\ell,0}(t) \,.
\ee
Here the $q^{(t)}$ have an extra subscript, compared to the $q^{(s)}$, reflecting the fact that an operator of  fixed spin $\ell'$ in the $t$-channel can give rise to contributions to all $Q^{\D_1+\D_2+\ell}_{\ell,0}$. This is unlike the $s-$channel where the contribution to a given $Q^{\D_1+\D_2+\ell}_{\ell,0}$ comes from a single spin $\ell$.\footnote{To put the notation in the different channels on the same footing, it is sometimes convenient to define the partial sum coefficients $c_{\D\ell}q^{(t)}_{\D,\ell}\equiv \sum_{\ell'} c_{\D\ell'}q^{(t)}_{\D,\ell |\ell'}$. This is the notation that was employed for compactness in \cite{usprl}.}

To extract the coefficients $q^{(t)}_{\D,\ell | \ell'}$ we use the orthogonality relations (see Appendix \ref{A})
\begin{align}\label{Qorth}
\begin{split}
\frac{1}{2\pi i}\int_{-i\infty}^{i\infty}dt \ \G(\frac{\tau}{2}+a_s+t)\G(\frac{\tau}{2}+b_s+t)\G(-t)\G(-a_s-b_s-t) Q^{\tau+\ell}_{\ell,0}(t)Q^{\tau+\ell'}_{\ell',0}(t)=\kappa_{\ell}(\frac{\tau}{2})\d_{\ell,\ell'}\,,
\end{split}
\end{align}
where the normalisation factor $\kappa_{\ell}(\frac{\tau}{2})$ is given in \eqref{knorm}.
Using this orthonormality condition, we obtain
\begin{align}\label{qformt}
\begin{split}
q_{\D,\ell |\ell'}^{(t)}=& \k_\ell(\frac{\D_1+\D_2}{2})^{-1}\int \frac{dt}{2\pi i} \ \G(\D_2+t)\G(\frac{1}{2}(\D_1+\D_2+\D_3-\D_4)+t)
\G(-t)\G(-a_s-b_s-t)  \\
&\times M_{\D,\ell'}^{(t)}(s=\frac{1}{2}(\D_1+\D_2),t) Q^{\D_1+\D_2+\ell}_{\ell,0}(t) \\
=& \k_\ell(\frac{\D_1+\D_2}{2})^{-1}  \int \frac{dt}{2\pi i} d\n \  \G(\D_2+t)\G(\frac{\D_1+\D_2+\D_3-\D_4}{2}+t) \\
\times & \G(\l_2-t-\frac{\D_2+\D_3}{2})\G(\bar{\l}_2-t-\frac{\D_2+\D_3}{2})\mu^{(t)}_{\D,\ell'}(\n) P_{\nu, \ell'}^{(t)}(\frac{\D_1+\D_2-\D_3-\D_4}{2},t+\frac{\D_2+\D_3}{2}) \\
\times & Q^{\D_1+\D_2+\ell}_{\ell,0}(t) \, .
\end{split}
\end{align}
We alert the reader that the $\ell$ dependence is explicit in the above while $\ell'$ enters in $\l_2, \bar{\l}_2$. 
Below we will see how to simplify this expression for the case of identical scalars.

\subsection{The $u-$channel}

Similarly, for the $u-$channel, the Mellin amplitude for the unitary block is given by (\ref{uunitrymell})
\begin{align}
\begin{split}\label{uunitrymell1}
M_{\Delta,\ell'}^{(u)}(s,t) = \int_{-i\infty}^{i\infty} d\nu\, \mu^{(u)}_{\Delta,\ell'}(\nu)\Omega_{\nu, \ell'}^{(u)}(s+t) P^{(u)}_{\nu, \ell'}(s-\frac{1}{2}(\D_1+\D_4),t) \,,
\end{split}
\end{align}
where
\be\label{specunitryu}
\mu^{(u)}_{\Delta,\ell}(\nu)=\frac{\G(\frac{\D_1+\D_3}{2}-\l_2)\G(\frac{\D_1+\D_3}{2}- \bar{\l}_2)\G(\frac{\D_2+\D_4}{2}-\l_2)\G(\frac{\D_2+\D_4}{2}- \bar{\l}_2)}{2\pi i ((\D-h)^2-\nu^2)\G(\nu)\G(-\nu)(h+\nu-1)_\ell(h-\nu-1)_\ell}\,
\ee
and 
\be\label{omu}
\Omega^{(u)}_{\nu, \ell'}(s+t)= \frac{\G(\l_2+s+t-\frac{1}{2}(\D_1+\D_4))\G(\bar{\l}_2+s+t-\frac{1}{2}(\D_1+\D_4))}
{\G(b_s+s+t)\G(a_s+s+t)} \, .
\ee

As in the $t$-channel, the behaviour $u^{\frac{1}{2}(\D_1+\D_2)}$ can be attributed entirely to the  $\G(\frac{1}{2}(\D_1+\D_2)-s)$ factor in the Mellin measure. We thus can expand in terms of the {\it same} set of orthogonal polynomials we had in the $s$-channel 
\be
M_{\Delta,\ell'}^{(u)}(s = \frac{1}{2}(\D_1+\D_2) ,t) \equiv \sum_{\ell}q^{(u)}_{\D,\ell |\ell'}Q^{\D_1+\D_2+\ell}_{\ell,0}(t) \, ,
\ee
where, using orthonormality, we can write for the coefficients,
\begin{align}\label{qformu}
\begin{split}
q_{\D,\ell |\ell'}^{(u)}=& \k_\ell(\frac{\D_1+\D_2}{2})^{-1}\int \frac{dt}{2\pi i}\ \G(\frac{\D_1+\D_2}{2}+a_s+t)\G(\frac{\D_1+\D_2}{2}+b_s+t) \G(-t) \\
&\times\G(\frac{1}{2}(\D_1-\D_2-\D_3+\D_4)-t) M_{\D,\ell'}^{(u)}(s=\frac{1}{2}(\D_1+\D_2),t) Q^{\D_1+\D_2+\ell}_{\ell,0}(t) \\
=& \k_\ell(\frac{\D_1+\D_2}{2})^{-1} \int \frac{dt}{2\pi i} d\n \   \G(-t)\G(\frac{1}{2}(\D_1-\D_2-\D_3+\D_4)-t)\G(\l_2+t+\frac{\D_2-\D_4}{2}) \\
&\times \G(\bar{\l}_2+t+\frac{\D_2-\D_4}{2}) \mu^{(u)}_{\D,\ell'}(\n)P_{\nu, \ell'}^{(u)}(\frac{\D_2-\D_4}{2},t)  Q^{\D_1+\D_2+\ell}_{\ell,0}(t) \, .
\end{split}
\end{align}
As in the $t$-channel, note that the $\l_2, \bar{\l}_2$ depend on $\ell'$.

\subsection{The bootstrap constraints}

Putting together the results from the previous three subsections, we demand
\be
\sum_{\D, \ell}\bigg(c^{(s)}_{\D,\ell}M_{\D,\ell}^{(s)}(s,t)+c^{(t)}_{\D,\ell}M_{\D,\ell}^{(t)}(s,t)+c^{(u)}_{\D,\ell}M_{\D,\ell}^{(u)}(s,t)\bigg)\bigg|_{s=\frac{\D_1+\D_2}{2}} =0 \ .
\ee
Or equivalently
\be
\sum_{\D, \ell} \bigg(c^{(s)}_{\D, \ell} q^{(s)}_{\D,\ell}+\sum_{\ell'}(c^{(t)}_{\D, \ell'}q^{(t)}_{\D,\ell |\ell'}+c^{(u)}_{\D, \ell'}q^{(u)}_{\D,\ell |\ell'}) \bigg)Q^{\D_1+\D_2+\ell}_{\ell,0}(t) =0
\ee
which implies from the orthogonality of the polynomials $Q^{\D_1+\D_2+\ell}_{\ell,0}(t)$, that
\be
\sum_{\D}  \bigg( c^{(s)}_{\D, \ell} q^{(s)}_{\D,\ell}+\sum_{\ell'}(c^{(t)}_{\D, \ell'}q^{(t)}_{\D,\ell |\ell'}+c^{(u)}_{\D, \ell'}q^{(u)}_{\D,\ell |\ell'}) \bigg) =0 \ .
\ee
Here the individual $q_{\D,\ell}$ are given in (\ref{Cs}, \ref{qformt}, \ref{qformu}). 
We can also write parallel expressions for the cancellation of the spurious poles at $s=\frac{\D_3+\D_4}{2}$. 
We note that while the $s$-channel expression is very explicit, the $t$,$u$-channel expressions are given as integrals. We will simplify these expressions bringing them into useful form, in the case of identical scalars, in Appendix D and E.


\section{The case of identical scalars}

Thus far we have considered the case of generic external scalar operators for our four point function and set up the bootstrap constraints to illustrate the generality of the method. In our simplest set of applications we will confine ourselves to the particular instance of identical external scalar operators. There are a couple of features that we have to keep in mind when we specialise. The first is that there will be a contribution of the identity operator (with $\D=\ell=0$) in all channels. This `disconnected' piece can be explicitly taken into account and we will find it convenient to separate it out. The second is that when we take all the $\D_i$ to the common value 
$\D_{\phi}$, the two separate spurious poles at $s=\frac{\D_1+\D_2}{2}$ and $s=\frac{\D_3+\D_4}{2}$ coalesce into a double and single pole at $s=\D_{\phi}$. It is the double pole that gives rise to the behaviour 
$u^{\D_{\phi}}\ln{u}$ in position space. In this section, we discuss these two features in turn and the resulting bootstrap constraints.  

\subsection{Double and single poles}

Let us first consider what happens to the general bootstrap conditions of the previous section when we take 
$\D_i \rightarrow \D_{\phi}$. The definition of the Mellin amplitude becomes as in (\ref{idmelldef})
\be\label{idmelldef2}
{\cal A}(u,v)= \int_{-i\infty}^{i\infty}\frac{ds}{2\pi i} \, \frac{dt}{2\pi i} \, u^{s}v^t \rho_{\D_\phi}(s,t) {\cal M}(s,t)
\ee
where $\rho_{\{\D_i\}}(s,t)\rightarrow \rho_{\D_\phi}(s,t)=\Gamma^2(-t)\Gamma^2(s+t)\Gamma^2(\Delta_{\phi}-s)$. 
The Mellin amplitude ${\cal M}(s,t)$ in the spectral representation of the previous section now reads as 
\begin{align}\label{idmellsum}
{\cal M}(s,t)=&  \sum_{\Delta, \ell} c_{\D, \ell} \bigg(M_{\Delta,\ell}^{(s)}(s,t)+M_{\Delta,\ell}^{(t)}(s,t)
+M_{\Delta,\ell}^{(t)}(s,t)\bigg) \\ \nonumber
=& \sum_{\Delta, \ell} c_{\D, \ell} \int_{-i\infty}^{i\infty} d\nu\, \mu_{\Delta,\ell}(\nu)
\bigg(\Omega_{\nu, \ell}^{(s)}(s)P^{(s)}_{\nu, \ell}(s,t) \\ \nonumber
&+\Omega_{\nu, \ell}^{(t)}(t)P^{(t)}_{\nu, \ell}(s -\D_\phi,t+\D_\phi)+\Omega_{\nu, \ell}^{(u)}(s+t)P^{(u)}_{\nu, \ell}(s-\D_\phi,t)\bigg)\,.
\end{align}
Here the common spectral function 
\be\label{specunitry}
\mu_{\Delta,\ell}(\nu) =\frac{\G^2(\frac{2\D_{\phi}-h+\ell+\nu}{2})\G^2(\frac{2\D_{\phi}-h+\ell-\nu}{2})}{2\pi i ((\D-h)^2-\nu^2)\G(\nu)\G(-\nu)(h+\nu-1)_\ell(h-\nu-1)_\ell}\,,
\ee
and
\be\label{oms}
\Omega_{\nu, \ell}^{(s)}(s) = \frac{\G(\frac{h+\nu-\ell}{2}-s)\G(\frac{h-\nu-\ell}{2}-s)}{\G^2(\Delta_{\phi}-s)}\,,
\ee  
while $\Omega_{\nu, \ell}^{(t)}(t)$ and $\Omega_{\nu, \ell}^{(u)}(s+t)$ are obtained from (\ref{oms}) by the 
replacements $ s\rightarrow t+\D_\phi$ and $s\rightarrow \D_\phi-s-t$ respectively. The Mack Polynomials are as before with the specialisation of parameters $\D_i \rightarrow \D_\phi$. 
 
We now need to look at the individual contributions to the spurious double and single poles of 
$\rho_{\D_\phi}(s,t) {\cal M}(s,t)$ at $s=\D_\phi$ from each of the terms on the RHS of  (\ref{idmellsum}). We do this channel by channel. 

\subsubsection{$s-$channel}

Consider the $s$ channel first. Since we now need the double as well as single poles, we have to expand  (\ref{Mst_nuint}) to one higher order in $s$. We only need to consider the coefficients $q_{\D,\ell}^{(s)}(s)$ of the orthogonal polynomials $Q^{2\D_\phi+\ell}_{\ell,0}(t)$ in the $s-$channel 
\be
q_{\D,\ell}^{(s)}(s)=-\frac{4^{1-\ell}\G(\D_\phi+s+\ell-h)^2}{(\ell+2s-\D)(\ell+2s+\D-2h)\G(2s+\ell-h)}\,
\ee
and expand these to linear order about $s=\D_\phi$.\footnote{One may worry whether one needs to consider contributions from the derivatives of the orthogonal polynomials themselves when  $Q^{\D_1+\D_2+\ell}_{\ell,0}(t)$ and $Q^{\D_3+\D_4+ \ell}_{\ell,0}(t)$ approach the common limit $Q^{2\D_\phi+ \ell}_{\ell,0}(t)$. However, it is easy to convince oneself that such additional contributions are proportional to the ones at the double pole. This is true channel by channel. When one imposes the vanishing constraints at the double pole, combining all channels,  then these pieces are also automatically set to zero and hence we will not consider them. }
\begin{align}\label{qtaylor}
q_{\D,\ell}^{(s)}(s)\equiv & \, \, q^{(2,s)}_{\D, \ell} + (s-\D_\phi)q^{(1,s)}_{\D, \ell} + \ldots \\ \nonumber
=& -\frac{4^{1-\ell}\G(2\D_\phi+\ell-h)}{(\ell-\D+2\D_\phi)(\ell+\D+2\D_\phi-2h)}+(s-\D_\phi)\frac{4^{2-\ell}\G(2\D_\phi+\ell-h+1)}{(\ell-\D+2\D_\phi)^2(\ell+\D+2\D_\phi-2h)^2}\,.
\end{align}
The first term $q^{(2,s)}_{\D, \ell}$ in the above expression gives the contribution to the residue at the double pole (and thus the $\log$ term in position space) while the second term $q^{(1,s)}_{\D, \ell}$ is the contribution to the residue at the single pole (and thus the power law term). 

\subsubsection{$t-$channel}
As we saw, the $t-$channel analysis is less straightforward, since an infinite number of spins ($\ell'$) can contribute to a single $\ell$ term. Redoing the steps that led to expression \eqref{qformt}, for identical scalars but without setting $s$ to a particular value leads to the general expression
\begin{align}\label{qtidn}
\begin{split}
q_{\D,\ell |\ell'}^{(t)}(s)=& \k_\ell(s)^{-1}\int \frac{dt}{2\pi i} d\n \   \G^2(s+t)\G(\l_2-t-\D_\phi)\G(\bar{\l}_2-t-\D_\phi)\\
&\times \mu^{(t)}_{\Delta,\ell'}(\nu) 
P^{(t)}_{\nu, \ell'}(s-\D_\phi,t+\D_\phi) Q^{2s+\ell}_{\ell,0}(t)\,.
\end{split}
\end{align}
The expression has integrals over $t$ and $\nu$. In Appendix \ref{tints} we show how to evaluate the $t$ integral for general $\ell$ and $\ell'$. We write $q_{\D,\ell}^{(t)}$  in the form,
\be\label{qtf1f2}
q_{\D,\ell |\ell'}^{(t)}(s)=q^{(2,t)}_{\D, \ell |\ell'}+(s-\D_\phi)q^{(1,t)}_{\D, \ell |\ell'} +O((s-\D_\phi)^2) \,.
\ee
The expression \eqref{qtform} gives the coefficients $q^{(a,t)}_{\D, \ell |\ell'}$ after performing the $t$-integral. 
In this paper, we will mainly employ the case with $\ell'=0$ in which case the expressions are simpler and we have 
\begin{align}\label{qtid}
q^{(2,t)}_{\D, \ell |\ell'=0}=&\int d\n  \frac{\m^{(t)}_{\D,0}(\n)\G(\l)\G(\bar{\l})2^\ell((\D_\phi)_\ell)^2}{\k_\ell(\D_\phi)(2\D_\phi+\ell-1)_\ell} \sum_{q=0}^\ell \frac{(-\ell)_q(2\D_\phi+\ell-1)_q\G(q+\l)\G(q+\bar{\l})}{((\D_\phi)_q)^2\ q!\G(q-2k+\l+\bar{\l})} \, ,
\end{align}
and
\begin{align}\label{qt1}
\begin{split}
&q_{\D,\ell |\ell'=0}^{(1,t)}= \int d\n   \m^{(t)}_{\D,0}(\n)\partial_\ta\Bigg[\frac{2^{\ell+1}((\frac{\ta}{2})_\ell)^2}{\k_\ell(\frac{\ta}{2})(\ta+\ell-1)_\ell}\G\big(\frac{\ta}{2}-\D_\phi+\l\big) \G\big(\frac{\ta}{2}-\D_\phi+\bar{\l}\big)\\
&\times \sum_{q=0}^\ell\frac{(-\ell)_q(\ta+\ell-1)_q}{((\frac{\tau}{2})_q)^2\ q!} \frac{\G(q+\frac{\ta}{2}+\l-\D_\phi)\G(q+\frac{\ta}{2}+\bar{\l}-\D_\phi)}{\G(q+\ta+\l+\bar{\l}-2\D_\phi)} \Bigg]_{\ta=2\D_\phi} \ \ \,.
\end{split}
\end{align}
Here $\l=(h+\nu)/2$ and $\bar{\l}=(h-\nu)/2$. 

Although the $\nu$ integral in both \eqref{qtid} and \eqref{qt1} makes them look complicated, they turn out to be relatively simple in practice. The $\nu$ integral can be evaluated using the residue theorem at  the poles. Though there are an infinite number of these poles, in all the cases that we consider, only a small finite number of poles actually contribute. The others are subleading (in the perturbative parameter such as $\e$). For example, in the 
$\phi^4$ theory all poles except two are subleading in $\e$. This is explained in more detail later and in \cite{mathematica}.

\subsubsection{$u-$channel}

The $u-$channel need not be handled all separately. Hence we will not write down the expressions for the $u-$channel anymore. To see this more clearly, note from \eqref{tunitrymell} and \eqref{uunitrymell}, that for identical scalars,
\be
M_{\D,\ell'}^{(u)}(s,t)=M_{\D,\ell'}^{(t)}(s,-s-t)\,.
\ee
In terms of the $Q_\ell$ polynomials, this condition translates into the form (near $s=\D_\phi$),
\be
\sum_{\ell} Q^{2\D_\phi+\ell}_\ell(-\D_\phi-t)q_{\D,\ell |\ell'}^{(t)}(s)=\sum_{\ell} Q^{2\D_\phi+\ell}_\ell(t)q_{\D,\ell |\ell'}^{(u)}(s)\,.
\ee
In the appendix \ref{A} it is shown using the properties of the $Q_\ell$ polynomials, that,
\be
Q^{2\D_\phi+\ell}_\ell(-\D_\phi-t)=Q^{2\D_\phi+\ell}_\ell(t)\,.
\ee
Given the orthogonality property of the $Q^{2\D_\phi+\ell}_\ell(t)$ this translates into the fact that,
\be
q_{\D,\ell |\ell'}^{(t)}(s)=q_{\D,\ell |\ell'}^{(u)}(s)\,.
\ee
Hence we need not calculate the coefficients $q_{\D,\ell |\ell'}^{(u)}(s)$ separately. We simply multiply the answer for the $t$-channel by a factor of two.

\subsection{Identity operator contribution}

There is always a contribution of the identity operator in all the three channels when the external operators are identical. This disconnected contribution takes a simple form and hence this piece can be separated out and explicitly written out. In position space, this contribution (sum of all the three channels) takes the form 
\be\label{idpos}
{\cal A}_{\D=0,\ell=0}(u,v) = 1+\bigg(\frac{u}{v}\bigg)^{\D_\phi}+u^{\D_\phi} \, .
\ee
The Mellin space representation of the individual channels can be written down as simple pole contributions which give the above power laws on closing the Mellin contour integral. 
\begin{align}\label{disconn}
\begin{split}
M^{(s)}_{\D=0,\ell=0}(s,t)=&\frac{1}{\rho_{\D_\phi}(s,t)}\frac{1}{s t}\,,\\
M^{(t)}_{\D=0,\ell'=0}(s,t)=&\frac{1}{\rho_{\D_\phi}(s,t)}\frac{1}{(s-\D_\phi)(t+\D_\phi)}\,,\\
M^{(u)}_{\D=0,\ell'=0}(s,t)=&\frac{1}{\rho_{\D_\phi}(s,t)}\frac{1}{(-s-t+\D_\phi))t}\, .\\
\end{split}
\end{align}
We see from (\ref{disconn}) that ${\rho_{\D_\phi}(s,t)}M_{\D=0,\ell=0}(s,t)$ has at most a single pole at $s=\D_{\phi}$. Moreover, these can only arise from the $t$ and $u$ channel contributions. We can expand in the orthogonal basis of the $Q^{2\D_\phi+\ell}_{\ell,0}(t)$
\be\label{disconnexp}
M^{(t)}_{\D=0,\ell'=0}(s,t) =\sum_\ell Q_{\ell,0}^{2\D_\phi+\ell}(t)q^{(t)}_{\D=0,\ell| \ell'=0}(s)\,,
\ee
with 
\begin{align}\label{qdisdef}
\begin{split}
q^{(t)}_{\D=0,\ell |0}(s)&=\frac{\k_\ell(s)^{-1}}{\G(\D_\phi-s)^2}\int \frac{dt}{2\pi i}\ \frac{Q^{2\D_\phi+\ell}_{\ell,0}(t)}{(s-\D_\phi)(t+\D_\phi)}\\
&=-\frac{\k_\ell(s)^{-1}(s-\D_\phi)}{\G(\D_\phi-s+1)^2}Q^{2\D_\phi+\ell}_{\ell,0}(-\D_\phi)\,.
\end{split}
\end{align}
The second line makes clear that this contributes to the single pole, like the second terms of \eqref{qtaylor} and \eqref{qtf1f2}. Thus we have 
\be\label{idpole}
q^{(1,t)}_{\D=0,\ell |0} = -\kappa_{\ell}(\D_\phi)^{-1}Q^{2\D_\phi+\ell}_{\ell,0}(-\D_\phi)\,
\ee
with an identical contribution to $q^{(1,u)}_{\D=0,\ell |0}$ as explained above.

\subsection{The bootstrap constraints}

We are now ready to put together the expressions of the last couple of sections together to write down more explicitly the bootstrap constraints in this case of identical scalars. The constraint from the vanishing of the residue of the double pole is 
\be\label{dblres}
\sum_{\D \neq 0} (c_{\D,\ell} q^{(2,s)}_{\D, \ell} +2\sum_{\ell'}c_{\D,\ell'}q^{(2,t)}_{\D, \ell |\ell'})=0
\ee
where the factor of two comes from the equality of the $t$ and $u$ channels. Note this is true individually for each $\ell=0,1,2 \ldots$. Here $q^{(2,s)}_{\D, \ell}$ and $q^{(2,t)}_{\D, \ell}$ are given respectively by (the first term in)
(\ref{qtaylor}) and (\ref{qtid}). 

The constraint from the residue single pole gets a contribution from the identity operator as well, as we saw in the previous subsection. 
\be\label{snglres}
2q^{(1,t)}_{\D=0, \ell |0}+\sum_{\D \neq 0}(c_{\D,\ell} q^{(1,s)}_{\D, \ell}+2\sum_{\ell'}c_{\D,\ell'}q^{(1,t)}_{\D, \ell |\ell'}) =0
\ee
with $q^{(1,s)}_{\D, \ell}, q^{(1,t)}_{\D, \ell |\ell'}, q^{(1,t)}_{\D=0, \ell |0}$ given respectively by 
(the second term in) (\ref{qtaylor}), (\ref{qt1}), (\ref{idpole}).
With the above normalisation of the identity operator contribution, the coefficients $c_{\D,\ell}$ are related to the square of the OPE coefficients $C_{\D,\ell}$ by a normalisation factor -- $c_{\D,\ell} = C_{\D,\ell}\mathfrak{N}_{\D,\ell}$. The factor $\mathfrak{N}_{\D,\ell}$ is worked out in Appendix \ref{AppC}.
\begin{align}\label{norm}
&\mathfrak{N}_{\D,\ell}^{-1} = \frac{ \Gamma (\Delta -1) \Gamma ^4\left(\frac{\ell +\Delta }{2}\right) }{(-2)^{\ell }  (\ell +\Delta-1 ) \Gamma (1-h+\Delta ) \Gamma ^2(\ell +\Delta-1 )} \Gamma \left(\frac{\ell -\Delta +\Delta _1+\Delta _2}{2}\right) \nonumber\\
& \Gamma  \left(\frac{\Delta +\Delta _1+\Delta _2-2 h+\ell }{2}\right) \Gamma  \left(\frac{\ell -\Delta +\Delta _3+\Delta _4}{2}\right)\Gamma  \left(\frac{\Delta +\Delta _3+\Delta _4-2 h+\ell }{2}\right)\,.
\end{align}
This will be important in what follows to extract out the physical OPE coefficients from our equations.

\section{The $\e$ expansion}

One of the physically very important nontrivial fixed points of the renormalisation group is the Wilson-Fisher fixed point in three dimensions, which governs the critical behaviour of the 3d Ising model (and the broader universality class that governs the liquid-vapour transition e.g. in water). We have little analytic control on this fixed point, as yet, directly in three dimensions other than in a large N limit of the $O(N)$ version. The $\e$ expansion in $d=4-\e$ dimensions is a way of arriving at a reasonable estimate of physical quantities (operator dimensions - which translate to critical exponents - and OPE coefficients) through a series expansion in $\e$, though eventually setting $\e$ to one. This is not a convergent expansion but the idea is that the first few terms might give rise to a reasonable approximation. 

The reason to believe so is because the Wilson-Fisher fixed point is perturbative in $\e$. Thus to leading order in $\e$, the beta function for the coupling $g$ in 
\be
S=\int d^d x\left[ \frac{(\partial \phi)^2}{2}+\frac{g \mu^{\e}\phi^4 }{4!}\right]\,
\ee
has a zero at $g=g^*=\frac{16\pi^2\e}{3}+O(\e^2)$. Thus the $\e$ expansion is essentially a perturbative expansion in the coupling and can be done order by order through evaluation of Feynman diagrams. To go to even a couple of nontrivial orders in $\e$, one thus needs to go to two, three loops etc. This rapidly becomes tedious and also subtle because of divergences and their regularisation. We will mention below some of the existing Feynman diagram results while comparing with results from our approach.   

Our approach via the conformal bootstrap, as described till now, is independent of the specific form of the Lagrangian and will not need to know the perturbative location of the zero of the beta function. We will proceed using only the following assumptions as our input:
\begin{itemize}
\item{There is a conserved stress tensor with $\D=d=2h$ and $\ell=2$.}

\item{There is only one fundamental scalar, $\phi$ of dimension $\D_\phi=\frac{d-2}{2}+O(\e)$.}

\item{A $Z_2$ symmetry ($\phi \leftrightarrow -\phi$) is present.}


\end{itemize}
We will also make some mild assumptions about the leading behaviour of OPE coefficients which are obvious from a perturbative point of view. 
We focus on the 4-point function $\la \phi \phi \phi \phi \ra$ of four fundamental scalars where we can apply the considerations of the previous section. We will first look at the $\ell=0$ and $\ell=2$ exchanges in the $s$-channel which contribute to only the corresponding partial waves $Q^{2\D_\phi+\ell}_{\ell}(t)|_{\ell=0,2}$. However, the contribution to these two partial waves from the crossed ($t,u$)-channels are typically from all spins. What will enable us to solve the bootstrap equation to low orders in $\e$ is that all nontrivial operators start contributing to $q^{(a,t)}_{\D>0, \ell>0 |\ell'}$ only from $O(\e^2)$. In fact, to both $O(\e^2)$ and $O(\e^3)$, only the lowest scalar $\phi^2$ contributes. In the $s$-channel too, only the lowest dimension operators with $\ell=0,2$ contribute to $O(\e^2)$ and $O(\e^3)$ respectively. Thus we can systematically solve the equations in terms of a {\it finite} number of unknowns to these orders. We describe this iterative procedure of solving the constraint equations below.  We then go on to apply a similar strategy to the spin $\ell$ exchange. 
 


\subsection{Scalar dimensions and OPE coefficients}

Our first goal will be to determine the dimensions  $\D_\phi$ and $\D_0$ of the scalar operators $\phi$ and $\phi^2$ respectively, together with the OPE coefficient $C_0\equiv C_{\phi\phi\phi^2}$. We will do so using the $\ell=0$ and $\ell=2$ constraint equations. The latter will involve the conserved stress tensor $T$ (whose dimension is protected to be $d=2h$ as well as the OPE coefficient $C_2\equiv C_{\D=d,\ell=2}=C_{\phi\phi T}$. We will assume there is a series expansion in $\e$ for each of these unknown quantities above, expanding about the free field value in $d=4$. In other words, we will take 
\be\label{dimphi} 
\D_\phi=1+\d_\phi^{(1)} \e+\d_\phi^{(2)} \e^2+\d_\phi^{(3)} \e^3+O(\e^4) \, \, ;\, \, \,  \D_0=2+\d_{0}^{(1)} \e+\d_{0}^{(2)} \e^2+O(\e^3)
\ee
and 
\be\label{Clow}
C_{0}=C_{0}^{(0)}+C_{0}^{(1)}\e+C_{0}^{(2)} \e^2+C_{0}^{(3)} \e^3 +O(\e^4) \,; \, \, \, \,C_{2}=C_{2}^{(0)}+C_{2}^{(1)}\e+C_{2}^{(2)} \e^2+C_{2}^{(3)} \e^3 +O(\e^4).
\ee

Our strategy will be as follows. We begin with the $\ell=2$ channel. We will see that our bootstrap constraint to leading order in $\e$ will allow us to determine $\d_\phi^{(1)}$ and $C_{2}^{(0)}, C_{2}^{(1)}$. This will be taken as input into the $\ell=0$ equation. The leading order equations here will turn out to determine $\d_{0}^{(1)}$ and 
$C_{0}^{(0)}, C_{0}^{(1)}$. We can then return to the $\ell=2$ equation and go further to $O(\e^2)$. This will determine $\d_\phi^{(2)}$ and $C_{2}^{(2)}$. Finally, we return to the  $O(\e^2)$ terms for $\ell=0$ and $O(\e^3)$ for $\ell=2$  and obtain $\d_{0}^{(2)}, \d_\phi^{(3)}$ as well as $C_{0}^{(2)}, C_{2}^{(3)}$.


{\bf A. Bootstrap Constraints for $\ell=2$ (First Pass):} 

We start with the $s$-channel expression \eqref{qtaylor}. For the stress tensor contribution with $\D=2h$ and $\ell=2$ we have,
\be
c_{2h,2}q^{(2,s)}_{\D=2h,\ell=2}=-C_{2} \frac{(1+2h) \Gamma^2(1+2h )  \Gamma(2-h+2 \Delta _{\phi })}{4(1-h + \Delta _{\phi }) (1 + \Delta _{\phi }) \Gamma^3(1+h) \Gamma(2h-1) \Gamma^2(1-h+\Delta _{\phi }) \Gamma^2(1+\Delta _{\phi })}\,,
\ee
and the derivative,
\be
c_{2h,2}q^{(1,s)}_{\D=2h,\ell=2}=C_{2} \frac{(1+2h ) [2 (1+\Delta _{\phi })-h] \Gamma^2(1+2h )  \Gamma(2-h+2 \Delta_{\phi })}{2 \Gamma^3(1+h) \Gamma(2h-1 ) \Gamma^2(2-h+\Delta _{\phi }) \Gamma^2(2+\Delta _{\phi })}\,.
\ee
The $c_{\D,\ell}$ are related to the OPE coefficients $C_{\D,\ell}$ through the normalization given in \eqref{norm}. Expanding this in a power series expansion in $\e$ to leading order, we have
\be\label{spin2-ep}
c_{2h,2}q^{(2,s)}_{\D=2h,\ell=2}=-\frac{45}{4}C_{2}^{(0)} (1+2 \delta^{(1)}_{\phi }) \epsilon  +O(\e^2)\,,
\ee
and
\be\label{spin2der-ep}
c_{2h,2}q^{(1,s)}_{\D=2h,\ell=2}=\frac{45 C_{2}^{(0)}}{2}+\frac{3}{2}  \left(2 C_{2}^{(0)}+15 C_{2}^{(1)}+30 \g_E C_{2}^{(0)}  \delta^{(1)}_{\phi }\right)\epsilon +O(\e^2)\,.
\ee
Here $\g_E$ is the Euler gamma constant. 

We will argue later that all other contributions in the $s$-channel from spin $\ell >0$ fields will start at higher order (precisely $O(\e^4)$) in $\e$. As mentioned above, and as will also be justified later, all contributions to $q^{(a,t)}_{\D>0, \ell=2 |\ell'}$ start at $O(\e^2)$. Thus, to the order we are considering, we only need to keep 
the identity operator contribution $q^{(1,t)}_{\D=0, \ell=2 |\ell'=0}$ given in \eqref{idpole}.
Expanding this in $\e$ we have,
\be\label{kappa-ep}
2q^{(1,t)}_{\D=0, \ell=2 |0}=-\frac{15}{2}-\frac{47+60 \g_E}{4}  \delta^{(1)}_{\phi }\epsilon +O(\e^2)\,.
\ee


We are now ready to impose the bootstrap constraints $\eqref{dblres}, \eqref{snglres}$ to $O(\e)$ keeping in mind that the above are the only non-zero contributions to this order. 
Thus we demand  that \eqref{kappa-ep} cancel with \eqref{spin2der-ep} and that \eqref{spin2-ep} cancels on its own. this immediately implies
\be
\d^{(1)}_\phi=-\frac{1}{2}\,,\hspace{1cm}C_{2}^{(0)}=\frac{1}{3}\hspace{1cm}\text{and}\hspace{1cm}C_{2}^{(1)}=-\frac{11}{36}\,.
\ee

{\bf B. Bootstrap constraints for $\ell=0$ (First Pass) :}

We now examine the various contributions to the bootstrap conditions for $\ell=0$. We first expand 
$q^{(a,s)}_{\D_0, \ell=0}$ in \eqref{qtaylor}, upto $O(\e^2)$,
\begin{align}\label{phi21}
c_{\D_0,0}q^{(2,s)}_{\D_0,\ell=0}=& -C^{(0)}_{0} \d^{(1)}_{0}(1+\delta^{(1)}_{0}) \frac{\epsilon}{2} - \bigg(\delta^{(1)}_{0} (1+\delta^{(1)}_{0}) \left(C_{0}^{(1)}+C_{0}^{(0)} (-\gamma_E +\delta^{(1)}_{0})\right)  \nonumber\\&  +
C_{0}^{(0)} \left(1+2 \delta^{(1)}_{0}\right) \delta^{(2)}_{0}+ 2 C_{0}^{(0)} \left(1+2 \delta^{(1)}_{0} (1+\delta^{(1)}_{0})\right) \delta^{(2)}_{\phi }\bigg)\frac{\epsilon^2}{2}\,,
\end{align}
and 
\begin{align}\label{phi22}
c_{\D_0,0}q^{(1,s)}_{\D_0,\ell=0}=& C_{0}^{(0)} + \left(C_{0}^{(1)} +C_{0}^{(0)}  (-\gamma_E +\delta^{(1)}_{0})\right) \epsilon \nonumber\\&
+ \left( C_{0}^{(2)} + C_{0}^{(1)}  \left(-\gamma_E +\delta^{(1)}_{0}\right)+\frac{C_{0}^{(0)}}{2}\left(\gamma_E^2-2 \gamma_E  \delta^{(1)}_{0}+2 \delta^{(2)}_{0}\right)+2  \gamma_E C_{0}^{(0)}  \delta^{(2)}_{\phi }\right)\e^2\,.
\end{align}

In the $t$-channel we now get contributions both from the identity operator as well as the $\phi^2$ operator to leading order in $\e$. The former contribution is 
\be\label{phi23}
2q^{(1,t)}_{\D=0, \ell=0 |\ell'=0}=-2+2 (1+\gamma_E )   \epsilon -  \left(\gamma_E  (2+\gamma_E ) +4(1+\gamma_E ) \delta_\phi^{(2)}\right)\epsilon ^2 \,.
\ee
while the latter's double pole contribution is given from \eqref{qtid}, after making an expansion in $\e$
\begin{align}\label{phi24}
&c_{\D_0,0}q^{(2,t)}_{\D_0, \ell=0 |\ell'=0}=-C_{0}^{(0)}\left(1+\delta^{(1)}_{0}\right)^2 \frac{\epsilon}{2} \nonumber
\\ & -\frac{1}{4} \epsilon ^2 \left(1+\delta _0^{\text{(1)}}\right) \left(2 C_0^{\text{(1)}} \left(1+\delta _0^{\text{(1)}}\right)+C_0^{\text{(0)}} \left(-1+2 (\delta _0^{ \text{(1)}})^2+4 \delta _0^{\text{(2)}}-2 \gamma _{\text{E}} \left(1+\delta _0^{\text{(1)}}\right)+\delta _0^{\text{(1)}} \left(1+8 \delta _{\phi }^{\text{(2)}}\right)\right)\right) \nonumber\\ & +O(\e^3) \,,
\end{align}
while the single pole contribution \eqref{qt1}, is given by,
\be\label{phi25}
c_{\D_0,0}q^{(1,t)}_{\D_0, \ell=0 |\ell'=0}=- \frac{C_{0}^{(0)}(1+\d_0^{(1)})^2 \e^2}{2}+O(\e^3) \,.
\ee
Setting the constant and $O(\e)$ terms of \eqref{phi22}, \eqref{phi23} and \eqref{phi25} as well as $O(\e)$ term of \eqref{phi21} and \eqref{phi24} to zero, we get,
\be
C^{(0)}_{0}=2\,,\hspace{1cm}C^{(1)}_{0}=-\frac{2}{3}\hspace{1cm} \text{and}\hspace{1cm}\d^{(1)}_{0}=-\frac{2}{3}\,.
\ee

\vskip 1cm

{\bf C. Bootstrap constraints for $\ell=2$ (Second Pass):}

We can now feed this information back to the $\ell=2$ constraints going now to $O(\e^2)$. We have in the $t$-channel, the contribution from the $\phi^2$ operator to be
\be\label{spin2t1}
c_{\D,0}q^{(2,t)}_{\D, \ell=2 |\ell'=0}=\frac{5 \epsilon^2}{144}+\frac{\epsilon^3 }{864} \left(-1+15 C^{(1)}_{\D_0,0}-30 \gamma_E +180 \delta_{0}^{(2)}-360 \delta_\phi^{(2)}\right)+O(\e^4) \,,
\ee
and 
\be\label{spin2t2}
c_{\D,0}q^{(1,t)}_{\ell=2 |\ell'=0}=-\frac{7 \epsilon^2}{96}-\frac{7 \epsilon ^3 \left(23+54 C_{\D_0,0}^{(1)}-108 \gamma_E +648 \delta^{(2)}_{0}-1296 \delta_\phi^{(2)}\right)}{10368}+O(\e^4) \,.
\ee

As will be argued later, there are no other contributions to the $t$-channel, in fact, even till $O(\e^3)$. 
Thus we combine $O(\e^2)$ terms of \eqref{spin2-ep} and \eqref{spin2t1} and set them to zero. We also do the same for the $O(\e^2)$ terms of \eqref{spin2der-ep}, \eqref{kappa-ep} and \eqref{spin2t2}. This gives us
\be
\d^{(2)}_\phi=\frac{1}{108}\,,\mbox{\ \ \ \ \ \   and \ \ \ \ \ \   } C^{(2)}_{2}=\frac{37}{486}\,.
\ee

{\bf D. Bootstrap constraints for $\ell=0$ (Second Pass):}

We go back to the expressions \eqref{phi21} and \eqref{phi24} for the double pole contributions and go to $O(\e^2)$. And then similarly with \eqref{phi22}, \eqref{phi23} and \eqref{phi25}. These relations give,
\be
C_{0}^{(2)}=-\frac{34}{81}\hspace{1cm} \text{and}\hspace{1cm}\d^{(2)}_{0}=\frac{19}{162}\,.
\ee

{\bf E. Bootstrap constraints for $\ell=2$ (Third Pass):}

The above results can now be used to get the $O(\e^3)$ pieces of \eqref{spin2-ep} and \eqref{spin2t1} and the corresponding terms of \eqref{spin2der-ep}, \eqref{kappa-ep} and \eqref{spin2t2}. On setting the constraints to zero we obtain\footnote{This can be computed in an $O(N)$ theory (with $N$ identical scalars). It matches with the large $N$ result, given in \cite{klebanov2, petkou}. This calculation will be detailed in a work in progress \cite{dks}.},
\be
\d^{(3)}_\phi=\frac{109}{11664}\,,\mbox{\ \ \ \ \ \   and \ \ \ \ \ \   } C_{2}^{(3)}=\frac{451}{52488}\,.
\ee

\subsection{Higher spin anomalous dimensions and OPE coefficients} \label{higherspinsec}

Now we use the results obtained above to determine the anomalous dimensions and OPE coefficients of higher spin operators $J^\ell$, schematically of the form $\phi\partial^\ell \phi$ with spin $\ell>2$. 
Note that only even spins $\ell$ are allowed, because of $Z_2$ invariance. We expect the expansion around $d=4$ for the dimension $\D_\ell$ to take the form
\be\label{hspindim}
\D_\ell \equiv d-2+\ell +\gamma_{\ell} = 2+\ell+\d_{\ell}^{(1)}\e+\d_{\ell}^{(2)}\e^2+\d_{\ell}^{(3)}\e^3+O(\e^4)
\ee
 and for the OPE coefficient 
\be\label{hsope}
C_{\phi\phi J^\ell} \equiv C_{\ell}=C_{\ell}^{(0)}+C_{\ell}^{(1)}\e+C_{\ell}^{(2)}\e^2+C_{\ell}^{(3)} \e^3+O(\e^4).
\ee

We proceed in a similar manner to before. In the $s$-channel we can write the contributions,  from $J^\ell$ with dimension $\D_\ell$ and spin $\ell$, to the bootstrap constraints by using \eqref{qtaylor} to $O(\e)$
\begin{align}\label{genell}
&c_{\D_\ell,\ell}q^{(2,s)}_{\D_\ell,\ell}=2^{-1-\ell } C_{\ell}^{(0)}\frac{(1+2 \ell )   \left(1+\delta_{\ell}^{(1)}\right) \Gamma ^2(1+2 \ell )}{\Gamma ^5(1+\ell )} \epsilon+O(\e^2)\, 
\end{align}
and
\begin{align}\label{genellder}
c_{\D_\ell,\ell}q^{(1,s)}_{\D_\ell,\ell}&=2^{- \ell } C_{\ell}^{(0)}\frac{(1+2 \ell )! (2 \ell )!}{(\ell! )^5}\nonumber\\& +
\epsilon\frac{2^{- \ell }   ((2 \ell )!)^2 \left((1+2 \ell ) \left(C_{\ell}^{(1)}-C_{\ell}^{(0)} \gamma_E \right)-C_{\ell}^{(0)} \delta_{\ell}^{(1)} (3(1+2 \ell ) H_{\ell }-2 (1+2 \ell ) H_{2 \ell }-1)\right)}{(\ell ! )^5}\nonumber\\&+O(\e^2)\,.
\end{align}

This contribution cancels against the identity contribution in the crossed channel
\begin{align}\label{genellkappa}
2q^{(1,t)}_{\D=0, \ell |\ell'=0}= -\frac{2^{1-\ell } (1+2 \ell )!}{ (\ell ! )^3}+\epsilon\frac{2^{1-\ell }   (1+2 \ell )! (\gamma_E -H_{\ell}+H_{2\ell-1})}{ (\ell ! )^3}+O(\e^2)\,.
\end{align}
The cancellation of the single pole contributions (constant term as well as $O(\e)$) in \eqref{genellder} and \eqref{genellkappa} as well as that of the $O(\e)$ term of  \eqref{genell} by itself (since there is no double pole contribution from the crossed channels) gives us 
\be
\d_{{\ell}}^{(1)}=-1\text{\,, \ \ \ }C_{\ell}^{(0)}=\frac{2(\ell !)^2}{ (2 \ell)!}\text{\, \ \ \ and \ \ \ \ }C_{\ell}^{(1)}=-\frac{2\text{  }  (\ell !)^2 (2H_{\ell}-H_{2\ell})}{ (2\ell)!}\,.
\ee
Once again, all other operators give higher order in $\e$ contributions and hence could be neglected. 

Moreover, even when we go to $O(\e^2)$, we need to only additionally include the scalar $\phi^2$ in the crossed channel. Thus what we need are \eqref{qtid} and \eqref{qt1} for general $\ell$. We will not explicitly show here the $O(\e^2)$ expressions for $q^{(a,t)}_{\D_0,\ell | \ell'=0}$, since they are cumbersome. We refer the reader to Appendix \ref{qsum}. 
The bootstrap conditions to $O(\e^2)$ after combining with the corresponding cumbersome piece from \eqref{genell} read as 
\begin{align}
\Bigg[ c_{\D_\ell,\ell}q^{(2,s)}_{\D_\ell,\ell}+2  c_{\D_0,0}q^{(2,t)}_{\D_0,\ell|\ell'=0} \Bigg]_{\e^2}=0
\implies\d^{(2)}_{\ell}=\left(1-\frac{6}{\ell (\ell +1)}\right)\frac{\epsilon ^2}{54}\,.
\end{align}
The corresponding constraint from the single pole then determines the OPE coefficient. 
\begin{align}
& \Bigg[  c_{\D_\ell,\ell}q^{(1,s)}_{\ell}+2  c_{\D_0,0}q^{(1,t)}_{\D_0,\ell|\ell'=0}+2q^{(1,t)}_{\D=0, \ell |0} \Bigg]_{\e^2}=0\nonumber\\
&\implies C_{\ell }^{(2)}=\frac{\ell\Gamma^2 (\ell )}{54 (\ell+1 )^2 \Gamma \left(2\ell+1 \right)} \Big(12+216 \ell  (1+\ell )^2 H_{\ell }^2+4 (1+\ell ) H_{\ell } \left(-3+\ell +\ell ^2-54 \ell  (1+\ell ) H_{2 \ell }\right)\nonumber\\&
\left.+(1+\ell ) \left(-2 \left(-6+\ell +\ell ^2\right) H_{2 \ell }+54 \ell  (1+\ell ) H_{2 \ell }^2+27 \ell  (1+\ell ) \left(2 H^{(2)}_{2 \ell }-3 H^{(2)}_{\ell }\right)\right)\right) \Big)\,.
\end{align}

Here $H_n^{(2)}=\sum_{k=1}^n (1/k^2)$ is the generalized harmonic number of power 2. One can proceed to $O(\e^3)$ and obtain $\d_\ell^{(3)}$ since we know $\D_\phi$ to $O(\e^3)$ and $C_\ell$ to $O(\e^2)$. Once again we just give the results.
\begin{align}
&\Bigg[ c_{\D_\ell,\ell}q^{(2,s)}_{\D_\ell,\ell}+2  c_{\D_0,0}q^{(2,t)}_{\D_0,\ell|\ell'=0} \Bigg]_{\e^3}=0\nonumber\\
&\implies\d^{(3)}_{\ell}=\frac{373\ell^2-384\ell-324+109\ell^3(\ell+2)-432\ell(\ell+1)H_\ell}{5832 \ell ^2 (\ell+1 )^2}\,.
\end{align}
Furthermore we can also obtain $C_\ell^{(3)}$ for any $\ell$ by looking at the single pole equation to this order in $\e$. In our present approach, a closed form expression is difficult to find; we can explicitly solve for various values of $\ell$. However, using a different approach in \cite{rajanind} one can find a closed form expression for any $\ell$ which is given below.

\subsection{Justification for truncating operator sums}\label{simplfction}

In our analysis we considered only a few operators in the $s$ and $t$- channels, whereas we had a sum over an infinite number of operators in both channels. We could get away with that because all other operators in both channels start contributing to our constraint equations, from a higher order in the $\e$ expansion. To be precise, for nonzero $\ell$ in the $s$  channel, operators with dimensions greater than $J^\ell$ contribute from $O(\e^4)$ order. For $\ell=0$, their contribution begins from $O(\e^3)$, due to which we were only able to determine the anomalous dimension of $\phi^2$ up to $O(\e^2)$. This is demonstrated in appendix \ref{simplification}. Also in crossed channels the residues of poles (of the $\n$ integral) in $q^{(t)}_{\D,\ell|\ell'}$ for all $\D$ and $\ell'>0$ undergo mutual cancellations, such that their nett contributions also start from $O(\e^4)$. For $\ell'=0$ only $\phi^2$ contributes at $O(\e^2)$ as well as $O(\e^3)$, for both of which only two residues (again in $\n$) are sufficient and the rest cancel one another. These cancellations are discussed in more detail in appendix \ref{simplification}.

These justify our being able to solve the bootstrap constraints reliably to the order in $\e$ that we have done above. At the same time they also show that going beyond these low orders in $\e$ is difficult using only these constraint equations since we would need to introduce an infinite number of other operators (both in the $s$ and $t$-channels). Thus we are unable, as of now, to compute the $O(\e^3)$ anomalous dimension of $\phi^2$ since many higher dimensional scalars will contribute to $O(\e^3)$ of $q^{(a,s)}_{\D,\ell=0}$ and $q^{(a,t)}_{\D,\ell=0| 0}$. Similarly, to determine the order $O(\e^4)$ anomalous dimension of $J^\ell$, we will have to enumerate and resum all the infinite number of poles of the spectral function. 

However, lest this sound dispiriting, we should hasten to remind the reader that we have looked at the very simplest constraint (of four identical scalars) and that too for the simplest spurious poles (at $s=\D_\phi$). As mentioned above, there are spurious poles (including primaries) at $s=\D_\phi+m$. We expect there to be powerful constraints from these additional poles as well as correlators, which can give a more systematic way to perform an $\e$ expansion (or indeed for any other small parameter) systematically. We hope to report on some of these aspects in \cite{rajanind}.  

\subsection{A summary and comparison of results}

We first pull together the term wise results obtained above and summarize, highlighting the new results obtained for the OPE coefficients, and then compare with some existing calculations.  

\subsubsection{Anomalous dimensions}

We find for the basic scalar field
\be\label{Dphi}
\Delta_\phi =1-\frac{\epsilon }{2}+\frac{1}{108} \epsilon^2+\frac{109}{11664} \epsilon^3+O(\e^4)\,.
\ee
The dimension of $\phi^2$ is given by,
\be
\D_0=2-\frac{2}{3}\e+\frac{19}{162}\e^2+O(\e^3).
\ee
This agrees with known results to this order \cite{vicari}.

For general spin $\ell$ we obtain
\be\label{genspin}
\D_\ell=d-2+\ell+\left(1-\frac{6}{\ell (\ell +1)}\right)\frac{\epsilon^2}{54}++\d^{(3)}_{\ell}\e^3+O(\e^4)\,,
\ee
where $\d^{(3)}_{\ell}$ is given by\footnote{It can be checked that this function vanishes for $\ell=2$ \ !},
\be\label{dl3}
\d^{(3)}_{\ell}=\frac{373\ell^2-384\ell-324+109\ell^3(\ell+2)-432\ell(\ell+1)H_\ell}{5832 \ell ^2 (\ell+1 )^2}\,.
\ee
This matches precisely with the results of \cite{gracey}. What is noteworthy is the relative ease of obtaining these results compared to the formidable Feynman integrals over a growing number of diagrams that need to be carried out. 

\subsubsection{OPE coefficients}

While we have not yet been able to go beyond Feynman diagram computations for anomalous dimensions (though we hope to eventually!), the results for OPE coefficients are essentially all new. Feynman diagram calculations for three point functions are much harder than for two point functions. But in our approach, the two appear more or less on the same footing and the bootstrap constraints enable us to solve for both simultaneously. 
   
The simplest OPE coefficient $C_{\phi\phi\phi^2}=C_0$ is given by,
\be
C_{0}=2-\frac{2}{3}\e-\frac{34}{81}\e^2+C_0^{(3)}\e^3\,.
\ee
The order $\e^2$ piece is new.
We note that if we take as external input the order $\e^3$ anomalous dimension of $\phi^2$ i.e. $\delta^{(3)}_0=\frac{937}{17496} - \frac{4 \zeta(3)}{27}$, computed using Feynman diagrams \cite{kleinert}, we can use the equations of Sec. 5.1 to obtain the prediction
\be\label{C0new}
C_0^{(3)} = - \frac{611}{4374} + \frac{23 \zeta(3)}{54}.
\ee

For the higher spin OPE coefficients $C_{\ell}$ it is best to write the expressions in terms of the free theory or alternatively, mean field theory values.  Thus
\be\label{clcfr}
\frac{C_{\ell}}{C_{\ell}^{free}}=1+\frac{\epsilon ^2 \left(6(\ell+1)^{-1}+2(\ell^2+\ell-3)H_{\ell}-(\ell-2)(\ell+3)H_{2\ell}\right)}{54 \ell  (\ell +1)} \ + \ c^{(3)}_\ell\e^3\,.
\ee
Here, in our conventions
\be\label{Cfree}
C_{\ell}^{free}=\frac{2 \Gamma\left(\ell+h-1\right)^2\Gamma\left(\ell+2h-3\right)}{\ell!\Gamma\left(h-1\right)^2\Gamma\left(2h+2\ell-3\right)}\,.
\ee
The $c^{(3)}_\ell$ can be computed individually for specific $\ell$. The first few are given by,
\begin{align}\label{cl3}
&c^{(3)}_2=\frac{955 }{17496}\hspace{1cm} c^{(3)}_4=\frac{33071 }{699840} \hspace{1cm}c^{(3)}_6=\frac{3665024719}{72613648800}\nonumber\\& c^{(3)}_8=\frac{45230019647 }{834534005760} \hspace{1cm}c^{(3)}_{10}=\frac{16447155548067179}{285712467504710400}\,.
\end{align}
This process can be automated to calculate $c^{(3)}_{\ell}$ for any arbitrary $\ell$. Both the $O(\e^2)$ as well as $O(\e^3)$ terms are new results (except for $\ell=2$ which was known to $O(\e^2)$). The $O(\e^3)$ values are found to obey a closed form expression in $\ell$ given by,
\begin{align}
c_\ell^{(3)}&=\frac{1}{5832 \ell ^2 (\ell+1 )^3}\Big(324+12 (59-22 \ell ) \ell -432 \ell  (\ell+1 )^2 H_{\ell }^2+2 (\ell+1 ) H_{\ell } [-162\nonumber\\&
\left.+\ell  (24+\ell  (241+109 \ell  (\ell+2 )))+216 \ell  (\ell+1 ) H_{2 \ell }\right]+(\ell+1 ) \Big[[324-\ell  (-384+\ell  (373\nonumber\\&
+109 \ell  (\ell+2 )))] H_{2 \ell }+54 \ell  (\ell+1 ) \left((-8+3 \ell  (\ell+1 )) H^{(2)}_{\ell }-2 (\ell-2 ) (\ell+3 ) H^{(2)}_{2 \ell }\right)\Big]\Big)\,.
\end{align}
It will be shown  in \cite{rajanind} how this general expression can be obtained. 

Interestingly, the OPE coefficients are even simpler when compared to the mean field OPE coefficients.
\begin{align}\label{Cl3}
\frac{C_{\ell}}{C_{\ell}^{MFT}}&=1+\frac{\epsilon ^2 \left(\ell  (1+\ell ) \left(H_{2 \ell }-H_{\ell-1 }\right)-1\right)}{9 \ell ^2 (1+\ell )^2} \nonumber \\ &
+\frac{\epsilon ^3}{486 \ell ^2 (1+\ell )^3}\Big[27+(59-22 \ell ) \ell -36 \ell  (1+\ell )^2 H_{\ell }^2+(1+\ell ) H_{\ell } \left(22 \ell^2+4\ell-27+36 \ell  (1+\ell ) H_{2 \ell }\right)\nonumber\\&
+(1+\ell ) \big(\left(27+32 \ell -22 \ell ^2\right) H_{2 \ell }+18 \ell  (1+\ell ) \big(3 H_{2 \ell }^{(2)}-2 H_{\ell }^{(2)}\big)\big)\Big] \ + \ O(\e^4)\,.
\end{align}
 
Here $C_{\ell}^{MFT}$ are the OPE coefficients in a theory where we have only disconnected (or identity operator) contributions but the dimension of the external scalar is given as $\D_\phi$,  as in the interacting theory \eqref{Dphi}. They are given by 
\be\label{CMFT}
C_{\ell}^{MFT}=\frac{4 \Gamma ^2\left(\ell +\Delta _{\phi }\right) \Gamma \left(\ell +2 \Delta _{\phi }-1\right)}{\ell ! \Gamma ^2\left(\Delta _{\phi }\right) \Gamma \left(2 \ell +2 \Delta _{\phi }-1\right)}\, .
\ee

\subsubsection{Comparisons with numerics in the 3d Ising model}\label{ising}

The OPE coefficient $C_2$ involving the stress tensor has a special status. It is related to the so-called central charge $c_T$ by the formula 
\be\label{cTformula}
c_T=\frac{d^2 \Delta_\phi ^2}{(d-1)^2 C_{2}}.
\ee
Using our expressions \eqref{clcfr}, \eqref{cl3}, for $\ell=2$, we have 
\be\label{ct3}
\frac{c_T}{c_{free}}=1-\frac{5 \epsilon ^2}{324}-\frac{233 \epsilon ^3}{8748}+O(\e^4)\,.
\ee
This matches with the $O(\e^2)$ term obtained previously \cite{hathrell, jack-osborn, petkou} but the $O(\e^3)$ order is new. We can put $\e=1$ here and obtain 
\be
\Bigg[\frac{c_T}{c_{free}}\Bigg]_{\e=1}=0.9579\,.
\ee

\begin{figure}
  \centering
  \includegraphics[width=0.5\textwidth]{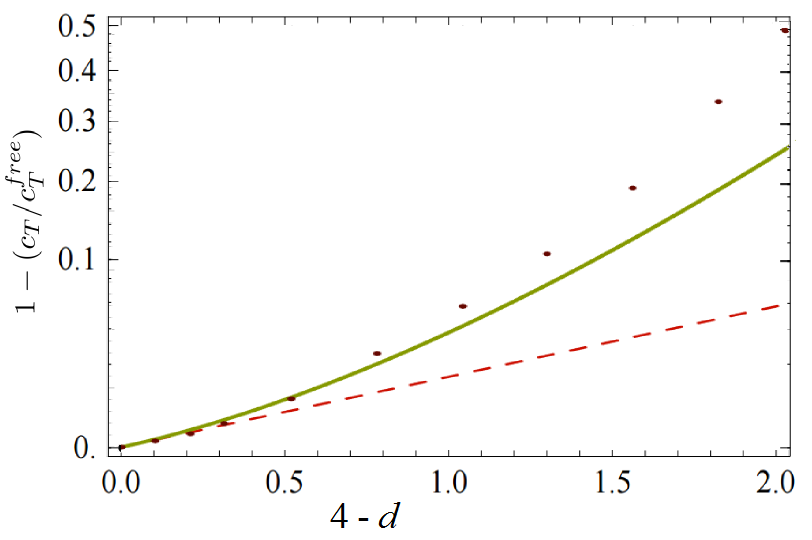}
  \caption{Plot of $\frac{c_T}{c_{free}}$ against $\e$, showing its variation in $2<d<4$. The dots indicate numerical bootstrap results \cite{epsnum}. The dashed red line indicates the analytic result with only the $O(\e^2)$ term. The smooth green line is the one including the $O(\e^3)$ and gives a more closer match.}
  \label{fig:CTalldim}
\end{figure}

This can be compared with precision values for the 3d Ising model obtained by numerical bootstrap \cite{3dising}, $c_T/c_{free} |_{numerics}=0.946534(11)$.   
We see that adding the $O(\e^3)$ terms gives a better estimate than what one gets from only the $O(\e^2)$ term (which gives $\sim 0.98$ for the same ratio). 
In fact, we can compare our analytical expression \eqref{ct3} with numerical results in $2<d<4$ dimensions. This is shown in Figure \ref{fig:CTalldim}.

While controlled numerical results are not yet available for the higher spin OPE coefficients, there are some estimates with not very clear error bars for the spin four coefficient which we can compare our answers to. 
Thus putting $\e=1$ in \eqref{clcfr}, \eqref{cl3}, for $\ell=4$, our value for $C_{4}=0.00489$. From \cite{zohar}, using our normalization\footnote{In \cite{zohar} the normalization is such that $C^{free}_\ell=2^{2-2\ell}$.}, we get $C_{4}=0.00476$ which is in good agreement.


With increasing spin we need to keep higher orders of $\e$ in order to get a better estimate with $\e=1$. A rough estimate can be made for what order of $\e$ we need to keep with increasing spin, by looking at the free theory OPE coefficients. Since we know $C^{free}$ in any dimension, we can compare $C^{free}_{d=4-\e}$ in an $\e$ expansion with $\e=1$, with $C^{free}_{d=3}$. The table below indicates the minimum power of $\e$, expansion up to which, the $C^{free}_{d=4-\e}$ gives more than 99 \% agreement with $C^{free}_{d=3}$.
\begin{center}
\begin{tabular}{|c|c|l|l|l|l|l|l|}
 \hline
 \text{power of $\e$} & 3 & 4 & 5 & 6 & 7 & 7 & 8\\ \hline
 \text{$\ell$} & 2 & 4 & 6 & 8 & 10 & 12 & 14\\
  \hline
\end{tabular}
\end{center} 
From the above table we can get an idea of the order in $\e$  required to get a reasonable estimate for the Ising model. For example, for the spin 10 OPE coefficient, barring any numerical coincidence, we may need to know till $O(\e^7)$ before getting a good match between $\e$ expansion and future numerical estimates. For larger spin the OPE coefficients actually get close to the mean field theory OPE coefficients $C^{MFT}_{\ell}$.

Coming back to anomalous dimensions to compare with numerics, we have for the higher spin currents, e.g. for $\ell=4$
\be
\Delta_{\ell=4}=6-\epsilon +\frac{7 \epsilon ^2}{540}+\frac{6991 \epsilon ^3}{583200} \ \stackrel{\e=1}{\approx} \  5.02495\,.
\ee
Without the $O(\e^3)$ term the value is $\approx 5.013$. With the $O(\e^3)$ term we are in better agreement with numerical bootstrap \cite{3dising,zohar} which gives $\D_{4}\approx 5.0227$.

Another cross-check we can perform is by expanding the $O(\e^3)$ anomalous dimension in \eqref{dl3} at large $\ell$. this gives
\be
{\d^{(3)}_{\ell\gg 1}}-2\d^{(3)}_{\phi}=\frac{11-18 \gamma_E -18 \text{log}(\ell)}{243 \ell ^2}\,.
\ee
Note the appearance of $\log \ell$ at $O(\e^3)$. This behaviour matches exactly with the prediction of \cite{alday} and the precise coefficient agrees with what we will find in the next section. 

\subsection{$\epsilon$ -expansion in other dimensions}

The method introduced above can be used for theories in other dimensions too. Here we will consider two: i) $\phi^3$ theory in $6-\epsilon$ dimensions; and ii)  $\phi^6$ theory in $3+\epsilon$ dimensions. Once again we will assume no prior knowledge of the Lagrangian. Our starting assumptions are same as for the $\phi^4$ theory, except that for $\phi^3$ in $6-\epsilon$ dimension, we will not assume any $Z_2$ invariance.

\subsubsection{$\phi^3$ in $6-\epsilon$ dimensions - a non-unitary example}

Let us start with $q^{(a,s)}_{\D_\phi,\ell=0}$. The fundamental field has the free theory dimension $\frac{d-2}{2}$. So, in the interacting theory let us write the dimension as,
\be
\D_\phi=2+\d^{(1)}_\phi \e+O(\e^2)\,.
\ee
In this theory the exchange operator can be $\phi$ itself. With $\phi$ as the exchange operator, we have using the general expression \eqref{qtaylor},
\be
c_{\D_\phi,0}q^{(2,s)}_{\D_\phi,\ell=0}=-C^{(0)}_0\frac{1+3\delta_\phi^{(1)}}{1+2\d_\phi^{(1)}}+O(\e)\,.
\ee
Here $C^{(0)}_0$ is the zeroth order of the OPE coeffcient for the exchange $C_{\phi\phi\phi}=C^{(0)}_0+C^{(1)}_0\e+O(\e^2)$.

Now we will look at the $t$- and $u$-channel. As before, we observe that only the lowest dimension scalar gives the leading contribution in the $t$ ($u$) channel. In this channel we find from \eqref{qtid},
\be
c_{\D_\phi,0}q^{(2,t)}_{\D_\phi,\ell=0|\ell'=0}=c_{\D_\phi,0}q^{(2,u)}_{\D_\phi,\ell=0|\ell'=0}=3 C_0^{(0)}  + O(\e)\,.
\ee
Now summing up the $s$, $t$ and $u$ channels, we get 
\begin{align}
\Bigg[ c_{\D_\phi,\ell=0}q^{(2,s)}_{\D_\phi,\ell=0}+2 c_{\D_\phi,\ell=0}q^{(2,t)}_{\D_\phi,\ell=0|\ell'=0} \Bigg]_{\e^0}=0
\implies\d^{(1)}_{\phi}=-\frac{5}{9}\,.
\end{align}
This agrees with the results of \cite{gracey2}.

We can also look at the single pole contribution of $\eqref{qtaylor}$ and \eqref{idpole}, which give,
\be
c_{\D_\phi,0}q^{(1,s)}_{\D_\phi,\ell=0}=\frac{C_0^{(0)}}{\epsilon  \left(\delta _{\phi }^{(1)}+\frac{1}{2}\right)}+O(\e^0)\,, \text{\ \ \ \ \ \ }2q^{(1,t)}_{\D=0,\ell=0|\ell'=0}=-12\,.
\ee
From this we get,
\begin{align}
\Bigg[c_{\D_\phi,0}q^{(1,s)}_{\D_\phi,0}+ 2q^{(1,t)}_{0,0|0} \Bigg]_{\e^0}=0
\implies C_{0}^{(0)}=-\frac{2\e}{3}\,.
\end{align}

Note that $C_0^{(0)}\sim O(\e)$ is negative\footnote{Our results are in agreement with the recent work \cite{naka}.}. This is a reflection of the fact that $\phi^3$ in $d=6-\e$ is a non-unitary theory. It is well known that the square of the coupling $\frac{(\l^*)^2}{(4\pi)^{d/2}}=-\frac{2\e}{3}$ is negative at the fixed point, if $\e>0$. Our above result is consistent with this since $C_0$ is proportional to the square of the 3-point function $\la \phi\phi\phi\ra \sim \lambda$. Note, once again that this result could be obtained only because all other scalars start contributing from a higher order in $\e$. In the $t$ channel, we find a similar cancellation, as described above for $\phi^4$ theory in $4-\e$ dimension, for heavy operators. A more careful and systematic analysis can extract more information. We leave this for future study.

\subsubsection{$\phi^6$ in $3+\e$ dimension}
 $Z_2$ invariance is preserved in this theory, and the external operator cannot appear in the OPE. So let us start with the $\ell=2$ conserved stress tensor expression, in order to get the dimension of $\phi$. Again, using \eqref{qtaylor} and \eqref{idpole},
\begin{align}
c_{2h,0}q^{(2,s)}_{2h,\ell=2}&=C_{2}^{(0)}\frac{512 (1-2\d^{(1)}_\phi )}{9\pi^2}\e \ + \ O(\e^2)\nonumber\\
c_{2h,0}q^{(1,s)}_{2h,\ell=2}&=\frac{1024 C_{2}^{(0)}}{9\pi^2}+O(\e)\nonumber\\
q^{(1,t)}_{0,2|0}&=-\frac{32}{3\pi^2}\,.
\end{align}
Here as before, $\d_\phi^{(1)}$ is the first subleading correction in $\D_\phi=\frac{1}{2}+\d_\phi^{(1)} \e+O(\e^2)$ and $C_{2}=C_{2}^{(0)}+O(\e)$ is the OPE coefficient $C_{\phi\phi J^2}$. In the $t$, $u$- channel all operators are found to contribute from a higher order in $\e$. So we immediately get the expected (free field) answers
\begin{align}
\d_\phi^{(1)}=\frac{1}{2}\,,\text{\ \ \ \ \ \ }C_{2}^{(0)}=\frac{3}{32}\,.
\end{align}
Now let us use this and look at $\ell=0$. Here we get,
\begin{align}
c_{\D_{\phi^2},0}q^{(2,s)}_{\D_{\phi^2},\ell=0}&=C_{0}^{(0)}\frac{\d^{(1)}_0-1}{2\pi^2}\e \ + \ O(\e^2)\,,\nonumber\\
c_{\D_{\phi^2},0}q^{(1,s)}_{\D_{\phi^2},\ell=0}&=\frac{C_{0}^{(0)}}{\pi^2}+O(\e)\,,\nonumber\\
q^{(1,t)}_{0,0|0}&=-\frac{2}{\pi^2}\,,
\end{align}
where we have $\D_{\phi^2}=1+\d_0^{(1)}\e+O(\e^2)$ and the OPE coefficient $C_{\phi\phi\phi^2}=C_0^{(0)}+O(\e)$. At $O(\e^2)$ for $q^{(2,s)}_{\D,0}$ and $O(\e)$ for  $q^{(1,s)}_{\D,0}$, there are an infinite number of scalars that can contribute. Also in the $t$ channel $q^{(2,t)}_{\D,\ell|\ell'}$ generically begins from $O(\e^2)$ and $q^{(1,t)}_{\D,\ell|\ell'}$ begins from $O(\e)$. Hence we get,
\be
\d^{(1)}_0=1\,, \text{\ \ \ \ \ \ \ }C_{0}^{(0)}=2 \,.
\ee
These results are consistent with the known results of $\phi^6$ in $3+\e$ dimensions \cite{3deps}.  It will be interesting to take the analysis beyond these orders, and find the anomalous dimensions and OPE coefficients systematically. In this paper, we will not pursue this problem any further.

\section{Large spin asymptotics}

\subsection{Strongly coupled theories}

Our formulation in Mellin space can be used to obtain results for large spin operators for a general scalar CFT in any dimension. As before, we shall consider a correlator with four identical external scalars 
($\D_i=\D_\phi$). We could then consider exchange of bilinear operators of the form ${\cal O}_{n,\ell} \sim \phi\square^n\partial^\ell \phi$ with large spin ($\ell \gg 1$) and assume there exists an operator of minimal twist in the OPE. This is the context studied in \cite{anboot,Komargodski}. We will limit our study to operators with $n=0$ for reasons mentioned below. We will show that the Mellin formalism easily reproduces the results \cite{anboot,Komargodski} for the anomalous dimensions $\g_\ell$ and OPE coefficients $C_\ell$ at leading order, of the large spin operators ${\cal O}_{0,\ell} =J^{\ell}$.


In the $s$-channel, we will be employing our workhorse  \eqref{qtaylor} which we reproduce below
\be\label{dq}
c_{\D,\ell}q_{\D,\ell}^{(s)}(s)=-\frac{C_{\ell} \mathfrak{N}_{\D,\ell}4^{1-\ell}\G(2\D_\phi+\ell-h)}{(\ell-\D+2\D_\phi)(\ell+\D+2\D_\phi-2h)}+(s-\D_\phi)\frac{C_{\ell}\mathfrak{N}_{\D,\ell} 4^{2-\ell}\G(2\D_\phi+\ell-h+1)}{(\ell-\D+2\D_\phi)^2(\ell+\D+2\D_\phi-2h)^2}\,.
\ee
We have included the normalization $ \mathfrak{N}_{\D,\ell}$ from \eqref{norm}, which for identical scalars, is given by,
\be
\mathfrak{N}_{\D,\ell}=\frac{(-2)^{\ell }  (\ell +\Delta-1 ) \Gamma (1-h+\Delta ) \Gamma ^2(\ell +\Delta-1 )}{ \Gamma (\Delta -1) \Gamma ^4\left(\frac{\ell +\Delta }{2}\right)  \Gamma^{2} \left(\frac{\ell +2\Delta_\phi -\Delta}{2}\right)\Gamma^{2}  \left(\frac{\ell+2\Delta_\phi+ \Delta - 2 h}{2}\right)}\,.
\ee\
We will presently take the large $\ell$ limit of these expressions.

Now let us go to the $t$-channel. First, there is the disconnected part of the Mellin amplitude given in \eqref{idpole}. The main trick that we will employ for the large spin analysis is to use an approximate form for the hypergeometric function that $Q^{\ta+\ell}_{\ell,0}$ is given in terms of (see Appendix G),
\begin{align}
{}_3 F_2\bigg[\begin{matrix} -n,k_1,k_2\\
k_3, k_4
\end{matrix};1 \bigg]& \rightarrow \frac{n^{-k_2} \Gamma \left(k_1-k_2\right) \Gamma \left(k_3\right) \Gamma \left(k_4\right)}{\Gamma \left(k_1\right) \Gamma \left(k_3-k_2\right) \Gamma \left(k_4-k_2\right)}\, ,
\end{align}
when $n \to \infty$. When applied to the $Q^{\ta+\ell}_{\ell,0}$ polynomials this is the same as requiring $\ell \gg \ta,t$. Since one is applying the bootstrap equations for finite values of $(\ta,t)$ in comparison to $\ell$, this is justified. One thus has
\be\label{Qapprox}
Q^{\ta+\ell}_{\ell,0}(t)=\frac{2^{\ell } \ell ^{-\frac{\ta}{2}-t} \Gamma^2 (\frac{\ta}{2}+\ell ) \Gamma (-1+\frac{\ta}{2}-t+\ell )}{\Gamma^2 (-t) \Gamma (-1+\ta+2 \ell )}\,.
\ee
Using the above approximation in \eqref{idpole} we get for the disconnected part,
\be\label{qdisfinal}
q_{\D=0,\ell |0}^{(1,t)}=\frac{2^{-\frac{3}{2}+\ell +2 \Delta_\phi } e^{\ell } \ell ^{-\ell }}{\pi  \G(\Delta_\phi )^2}+O(1/\ell)\,.
\ee

Now let us examine the rest of the $t$ and $u$ channel amplitudes. Here we will need the assumption that there is a {\it single} operator of minimum twist and that all other operators are separated from it by a (large) twist gap.
This typically happens in strongly coupled CFTs and hence the title of this subsection. We will denote the twist and spin of this minimum twist operator as $\ta_m$ and $\ell_m$ respectively. From our analysis it will become apparent that the operators with higher twists will contribute at subleading order. 

We begin with \eqref{qtf1f2}, 
\be\label{qtf1f22}
q^{(t)}_{\D,\ell|\ell'}(s)=q^{(2,t)}_{\D,\ell|\ell'}+(s-\D_\phi)q^{(1,t)}_{\D,\ell|\ell'}\,,
\ee
where we have from \eqref{qtidn}
\begin{align}\label{qt}
\begin{split}
q_{\D,\ell |\ell'}^{(2,t)}=& \Big[ \k_\ell\big(\frac{\ta}{2}\big)^{-1}\int \frac{dt}{2\pi i} d\n \   \G^2\big(\frac{\ta}{2}+t\big)\G(\l_2-t-\D_\phi)\G(\bar{\l}_2-t-\D_\phi)\\
&\times \mu^{(t)}_{\Delta,\ell'}(\nu) 
P^{(t)}_{\nu, \ell'}\big(\frac{\ta}{2}-\D_\phi,t+\D_\phi\big) Q^{\ta+\ell}_{\ell,0}(t)\Big]_{\ta=2\D_\phi}\,.
\end{split}
\end{align}
We remind the reader of the notation $\l_2=\frac{h+\nu-\ell'}{2}\text{ \  and \  }\bar{\l}_2=\frac{h-\nu-\ell'}{2}$. We will now use the approximation \eqref{Qapprox} for $Q^{\ta+\ell}_{\ell,0}$. Then the only poles for the $t$ integral are given by $t=\l_2-\D_\phi$ and $t=\bar{\l}_2-\D_\phi$. 
The residue of the other pole in $t$ from \eqref{Qapprox} is suppressed at large $\ell$. The power of $\ell$ in \eqref{Qapprox} requires that we close the contour on right.  At these poles the Mack polynomial simplifies and we get for general $\ta_m$ and $\ell_m$,
\begin{align}
& c_{\D_m,\ell_m}q^{(2,t)}_{\D_m,\ell|\ell_m}=  C_m\mathfrak{N}_{\ta_m+\ell_m,\ell_m} \k_\ell(\D_\phi)^{-1} \int d\nu  e^{2 \ell } 2^{\ell -2 \ell _m+2 \Delta _{\phi }-1} \text{  }\ell^{-2\ell -2\Delta _{\phi }+1-\frac{h}{2}-\frac{\nu }{2}+\frac{\ell _m}{2}}\nonumber\\& \times \frac{ \Gamma \left(\frac{h+\nu }{2}\right) \Gamma \left(\frac{h+\nu +\ell _m}{2}\right) \Gamma \left(\ell +\Delta _{\phi }\right) \Gamma \left(2\D_\phi-1-\frac{h}{2}+\ell -\frac{\nu }{2}+\frac{\ell _m}{2}\right) \Gamma \left(\frac{2\D_\phi-h+\nu +\ell _m}{2}\right)^2}{\pi ^{3/2} (h+\nu-1 ) \left(\frac{1+h+\nu }{2}\right)_{\frac{\ell _m}{2}-1} \left(\left(h-\ell _m-\tau _m\right)^2-\nu^2\right) \Gamma (\nu ) \Gamma \left(\D_\phi+\ell-\frac{1}{2}\right)}\,.
\end{align}
Here $C_m \equiv C_{\ell_m}$ is the OPE coefficient of the minimal twist operator. Since there is a factor $\ell^{-\nu/2}$ we must also close the $\nu$-contour on the right. The leading power in $\ell$ will come from the smallest positive value of the $\nu$ pole. Thus we pick the denominator pole at  $\nu=\D_m-h=\ta_m+\ell_m-h$. The only other possible pole, lying inside the contour, is from $\Gamma \left(-1-\frac{h}{2}+\ell -\frac{\nu }{2}+\frac{\ell _m}{2}+2 \Delta _{\phi }\right)$ and its residue contributes at subleading order in $\ell$. 
Thus, to leading order, for large $\ell$, we get a simple result,
\be\label{afternuint}
q^{(2,t)}_{\D_m,\ell|\ell_m}= -\frac{2^{-\frac{1}{2}+\ell -3 \ell _m+2 \Delta _{\phi }-\tau _m} e^{\ell }\ell^{-\ell -\tau _m} \Gamma \left(\ell _m+\frac{\tau _m}{2}\right) \Gamma \left(\ell _m+\Delta _{\phi }+\frac{\tau _m}{2}-h\right)^2 \Gamma \left(\ell _m+\tau _m-1\right)}{\sqrt{\pi } \Gamma \left(1-h+\ell _m+\tau _m\right) \Gamma \left(\frac{2 \ell _m+\tau _m-1}{2}\right)}\,.
\ee
This is the double pole contribution. Associated with the single poles is the second term in \eqref{qtf1f22}, 
\be
q^{(1,t)}_{\D_m,\ell|\ell_m}= \int\frac{ dt d\n}{2\pi i} \  \partial_\ta \Big[2\k_\ell\big(\frac{\ta}{2}\big)^{-1} \G^2(\frac{\ta}{2}+t)\G(\l_2-t)\G(\bar{\l}_2-t)  \mu^{(t)}_{\Delta_m,\ell_m}(\nu) \Omega_{\nu, \ell_m}^{(t)}(t)
P^{(t)}_{\nu, \ell_m}(\frac{\ta}{2}-\D_\phi,t+\D_\phi) Q^{\ta+\ell}_{\ell,0}(t)\Big]_{\ta=2\D_\phi}\,.
\ee
For large $\ell$ this integral can be done in a way similar to above. We simply quote the result,
\be\label{f2final}
c_{\D_m,\ell_m}q^{(1,t)}_{\D_m,\ell|\ell_m}=\frac{C_m 2^{\ell +2 \Delta_\phi -\frac{3}{2}} e^{\ell } \ell ^{-\ell-\tau _m} \left(\log\left(\frac{\ell }{2}\right)-\psi \left(\ell _m+\frac{\tau _m}{2}\right)\right) \Gamma \left(2 \ell _m+\tau _m\right)}{\pi  \Gamma \left(\Delta_\phi -\frac{\tau _m}{2}\right)^2 \Gamma \left(\ell _m+\frac{\tau _m}{2}\right)^2}\,.
\ee
From \eqref{qdisfinal}, \eqref{afternuint} and \eqref{f2final}, we find $q_{\D_m,\ell |\ell_m}^{(t)}$ to be suppressed by an additional factor of $\ell^{-\ta_m}$ compared to the identity or disconnected piece. Since $\ta_m$ was assumed to be the minimal twist, any other operator (contributing to the same $\ell$) in the $t$ or $u$ channel will have an even more subleading contribution. 

The fact that the identity operator dominates implies that at large $\ell$ the theory behaves asymptotically like a free theory. In other words, the operators ${\cal O}_{n,\ell}$ at large spin, have their dimension of the form, $\D_{n,\ell}=2\D_\phi+2n+\ell +\g(n,\ell)$, where $\g(n,\ell)$ must be small at large $\ell$. As mentioned before we will limit ourselves to only the $n=0$ operators, and denote $\g(n=0,\ell)=\g_\ell$. This is possible because the contributions of operators with $n>0$ to both the terms of \eqref{dq}  are suppressed with extra factors of $\g(n,\ell)$. This is because of the denominators, which are small only for $n=0$. 

In \eqref{dq} we insert $\D=\D_{0,\ell}=2\D_\phi+\ell+\g_\ell$ and $C_{\ell}=C^{(0)}_\ell(1+\d C)$, where $\d C$ is OPE coefficient correction due to the minimal twist operator. The single pole term on the rhs of \eqref{dq} becomes,
\be\label{Pmft}
C_\ell^{(0)}\frac{2^{3 \ell +4 \Delta_\phi -\frac{9}{2}} e^{\ell } \left(\frac{1}{\ell }\right)^{\ell +2 \Delta_\phi -\frac{3}{2}}}{\pi ^{3/2}} \ + \ O(1/\ell , \d C, \g_\ell)\,.
\ee
We have indicated the various subleading terms as ones coming from an explicit $\frac{1}{\ell}$ expansion as well as those proportional to $\d C$ and $\g_\ell$ which are also down by powers of  $\frac{1}{\ell}$.
Now the leading term in \eqref{Pmft} must cancel the leading term of the $Q_{\ell,0}$ component of the $t$ and $u$ channel. As we saw, this comes from the disconnected part, which is given by \eqref{qdisfinal}. We therefore get 
\be\label{mftlargeell}
C_\ell^{(0)} = \frac{2^{3-2\ell -2 \Delta _{\phi }} \sqrt{\pi } \left(\frac{1}{\ell }\right)^{\frac{3}{2}-2 \Delta _{\phi }}}{\Gamma \left(\Delta _{\phi }\right)^2}\,.
\ee
This is, in fact, nothing but the leading large $\ell$ behaviour of mean field theory OPE coefficients \eqref{CMFT}.

To find the anomalous dimension $\g_\ell$, we use the constraint from the double pole term in \eqref{dq} which simplifies at large $\ell$
\be\label{gammalhs}
-\frac{C_\ell^{(0)} \mathfrak{N}_{\D,\ell}4^{1-\ell}\G(2\D_\phi+\ell-h)}{(\ell-\D+2\D_\phi)(\ell+\D+2\D_\phi-2h)}={C_\ell^{(0)} \pi^{-3/2}2^{-\frac{9}{2}+3 \ell +4 \Delta _{\phi }} e^{\ell } \g_\ell \left(\frac{1}{\ell }\right)^{-\frac{3}{2}+\ell +2 \Delta _{\phi }}  }\,.
\ee
Demanding its cancellation with double pole terms of the $t$ and $u$ channel amplitudes (given by \eqref{afternuint}), we get\footnote{Here the OPE coefficients are normalized such that $\la \phi \phi\phi\phi \ra =(x_{12}^2 x_{34}^2)^{\D_\phi}(1+C_m u^{\ta_m/2}(1-v)^{\ell_m}+\cdots)$},
\be\label{gammastrng}
\g_\ell=-C_m\frac{2\Gamma ^2\left(\Delta _{\phi }\right)\Gamma \left(2\ell _m+\tau _m\right)}{\Gamma ^2\left(\Delta _{\phi }-\frac{\tau _m}{2}\right)\Gamma ^2\left(\ell _m+\frac{\tau _m}{2}\right)}\left(\frac{1}{\ell }\right)^{\tau _m}\,.
\ee
This result agrees with those obtained in  \cite{anboot,Komargodski} by very different techniques. 

Finally let us compute the leading correction $\d C$ to the free field OPE coefficient. For this we expand the single pole term in \eqref{dq} to its subleading orders. 
\be\label{OPElhs}
\frac{2^{-\frac{3}{2}+\ell +2 \Delta \phi } e^{\ell } \ell ^{-\ell }}{\pi  \Gamma (\Delta_\phi )^2}+\frac{2^{-\frac{3}{2}+\ell +2 \Delta _{\phi }} e^{\ell } \ell ^{-\ell } \left(\text{$\delta C$}/2-\gamma_E  \gamma _{\ell }+2\log 2 \gamma _{\ell }-\log (\ell ) \gamma _{\ell }\right)}{\pi  \Gamma \left(\Delta _{\phi }\right)^2}\,.
\ee
Now this must cancel with \eqref{f2final}. Note that in the above expression the $O(1/\ell)$ correction term has not been considered since it does not involve the minimal twist operator. It cancels with the subleading $O(1/\ell)$ term of the identity operator piece, just like the first term of \eqref{OPElhs} cancels with the leading term. Both the anomalous dimension $\g_\ell$ and $q^{(1,s)}_{\D_\ell,\ell}$ are suppressed by factors of $\ell^{\ta_m}$ from the leading terms, and hence we expect $\d C$ to receive contribution only from them. In this way we find
\be\label{opestrng}
\d C=-\frac{2 C_m  \ell ^{-\tau _m} \left(\gamma_E -\log 2+\psi \left(\ell _m+\frac{\tau _m}{2}\right)\right) \Gamma^2 \left(\Delta _{\phi }\right) \Gamma \left(2 \ell _m+\tau _m\right)}{\Gamma^2 \left(\Delta _{\phi }-\frac{\tau _m}{2}\right) \Gamma^2 \left(\ell _m+\frac{\tau _m}{2}\right)}\,.
\ee
This again agrees with the results of \cite{anboot,Komargodski}.

\subsection{Weakly coupled theories}
In this subsection we will carry out a similar large spin analysis but for ``weakly coupled" theories. This will be a somewhat broader notion in that we merely require that the anomalous dimension of the fundamental scalar as well as that of the higher spin operators $J^{\ell}$ be small. Thus we will have a near continuum of higher spin operators, instead of just one as in the previous section, with minimal twist. To make this more precise we write the dimension of $\phi$ to be 
\be\label{phidim}
\D_\phi=\frac{d-2}{2}+\g_\phi  \text{  \ \ \ \ where \ \ \ \  } \g_\phi=g \d_\phi^{(1)}+g^2 \d_\phi^{(2)}+O(g^3)\,.
\ee  
Here $g$ is a small parameter. We keep the precise definition of $g$ ambiguous, so that we can fix it to be any convenient small parameter available in the theory we have in mind. In many cases, we can take $\gamma_\phi$ to be the small parameter. But in other cases like in the $\e$ expansion (where $\gamma_\phi \propto \e^2$) the expansion parameter is more naturally the anomalous dimension $\g_0$ of $\phi^2$. 
In anycase, we will assume we can expand the dimensions of the higher spin operators $J^\ell$ also in $g$. Hence we have,
\be\label{genellexpand}
\D_\ell \ = \ 2\D_\phi+\ell+\g_\ell \ = \ d-2+\ell+g \d_{\ell}^{(1)}+g^2 \d_{\ell}^{(2)}+O(g^3)\,.
\ee
This notion of weakly coupled theories includes not only perturbative CFTs (like the Wilson-Fisher point in $d=4-\e$ dimension) but also others such as the 3d Ising model which has a sector of operators with small anomalous dimensions. 

Now we have two perturbative parameters at our disposal: $1/\ell$ and $g$. We will work in the regime where $\ell^{-1}\ll g$. Thus we will expand in large $\ell$ first and then expand in small $g$. 
In the $s$-channel the leading contribution is given as in the previous subsection by taking the large $\ell$ limit of the LHS  of \eqref{gammalhs} 
\be\label{weakgammalhs}
c_{\D_\ell,\ell}q^{(2,s)}_{\ell\gg 1}=\frac{2^{2\ell +d-\frac{5}{2}} e^{\ell } \ell ^{-\ell } \g_\ell}{\pi  \Gamma \left(\frac{d-2}{2}\right)^2}\,.
\ee

Coming to the $t$-channel we now have to take into account the contributions of the infinite number of operators $J^{\ell'}$ whose twists lie very close to the minimal twist. Let us denote their OPE coefficients as $C_{\ell'}$. To get the contribution of each of these, we put $\tau_m=2\D_\phi+\g_{\ell'}$ and 
$\ell_m=\ell'$ in \eqref{afternuint}. This gives,
\begin{align}\label{weakafternuint}
&c_{\D_{\ell'},\ell'}q^{(2,t)}_{\D_{\ell'},\ell|\ell'}=  -C_{\ell'} \mathfrak{N}_{d-2+\ell'+2\g_\phi+\g_{\ell'},\ell'} 2^{-\frac{1}{2}+\ell -3 \ell'-\g_{\ell'}} e^{\ell }\ell^{-\ell -2h+2-\g_{\ell'}-2\g_\phi}\nonumber\\ & \hspace{2cm}\times\frac{ \Gamma \left(\ell'+h+\frac{\g_\ell'+2\g_\phi-2}{2}\right) \Gamma \left(h+\ell'-2+\frac{\g_{\ell'}}{2}+2\g_\phi\right)^2 \Gamma \left(\ell'+2h-3+\g_{\ell'}+2\g_\phi\right)}{\sqrt{\pi } \Gamma \left(h+\ell'-1+\g_{\ell'}+2\g_\phi\right) \Gamma \left(\frac{2 \ell'+\g_{\ell'}+2\g_\phi-3}{2}+h\right)}\,.
\end{align}
We now have to sum over $\ell'$ and this should cancel against \eqref{weakgammalhs}. 
Now let us use \eqref{genellexpand} and expand the above in $g$. In addition to an overall power law, we get an expansion in $\log{\ell}$ from expansing the piece $\ell^{-\g_\ell'}$ in a power series in $g$. Thus we find an expansion of the form,
\be
c_{\D,\ell'}q^{(2,t)}_{\D,\ell|\ell'}=g^2\hat{f}_0(\log\ell)+g^3\hat{f}_1(\log\ell)+\cdots
\ee
In the above expansion, $\hat{f}_p(x)$ is a polynomial of degree $p$. This means the $O(g^2)$ term has no $\log \ell$ term. The $\log \ell$ dependence enters at $O(g^3)$, $(\log \ell)^2$ at $O(g^4)$ and so on. Using this to evaluate $\g_\ell$, we get,
\be\label{gammaellweak}
\g_\ell=\frac{\a_0(g) +\a_1(g) \log \ell+\a_2(g) 
(\log \ell)^2+\cdots}{\ell^{d-2}}
\ee
where\footnote{Here the OPE coefficients are normalized such that the large spin mean field theory OPE coefficients are given by \eqref{mftlargeell} },
\be\label{alphai}
\alpha_p=- \sum _{\ell '}C_{\ell '} \ g^{2+p}\frac{2^{2h'-3}   \left(-\delta^{(1)}_{\ell'}\right)^p\left(\delta ^{(1)}_{\ell'}-2\d_\phi^{(1)} \right)^2 \Gamma (h-1)^2 \Gamma \left(h-\frac{1}{2}+\ell '\right)}{2 \sqrt{\pi } \Gamma \left(h+\ell '-1\right)}+O\left(g^{3+p}\right)\,.
\ee
So we have a precise form of $\a_i$ up to the leading order in $g$. The order of $g$ was also predicted in \cite{alday}. Any CFT satisfying \eqref{phidim} and \eqref{genellexpand} will have large spin anomalous dimensions given by the above. However there are situations where this formula can be made more compact. Below we discuss two such special cases. 

\subsubsection{CFTs close to a free theory}
We can apply the above considerations to theories which are perturbatively near the free theory limit. The 
$\phi^4$ theory in $d=4-\e$, or $\phi^6$ theory in $d=3+\e$, etc fall under this category. From the expression \eqref{alphai} we see that $\g_\ell$ already starts from $O(g^2)$. The absence of a term of $O(g)$ implies  we must have $\delta^{(1)}_{\ell'}=2\d_\phi^{(1)}$ for sufficiently large $\ell'$. Now every $\a_i$ has a factor of $\big(\delta^{(1)}_{\ell'}-2\d_\phi^{(1)} \big)^2$, so the sum over $\ell'$ in $\a_i$ gets contribution from finitely many $\ell'$. We examine low-lying $\ell$'s to pinpoint which terms contribute to the above sum over $\ell'$. For this we will use the  expansion \eqref{genellexpand} in equation \eqref{dq}. For $\ell>0$ we get quite generally,
\be\label{qswk}
q^{(2,s)}_{\D_\ell,\ell>0}=g\frac{ (2h+2 \ell-3 ) \left(\delta^{(1)}_{\ell}-2\d_\phi^{(1)}\right)  \Gamma (2h+2 \ell-3 )^2}{4^{\ell }\Gamma (h+\ell-1 )^4 \Gamma (2h+\ell-3 )}+O(g^2)\,.
\ee
In this expression, we are only considering only the higher spin operators $J^\ell$. There can be other higher spin operators as well, but  taking a cue from perturbation theory, we assume their OPE coefficients would be $C_{\D,\ell}\sim O(g^2)$. Thus they would not contribute to leading order and the $s$-channel answer to $O(g)$ is simply given by the above expression.

Now it can also  be verified that for any operator in the $t$-channel, we have,
\be
c_{\D,\ell'}q^{(2,t)}_{\D,\ell|\ell'}=O(g^2)\,.
\ee
This is true for any operator in $t$-channel with the form $\D=m+O(g)$ with $m$ being an integer $\ge d-2+\ell'$. The $t$-channel thus starts from $O(g^2)$ for all $\D$ and $\ell'$.
For the $s$-channel double pole in \eqref{qswk} to cancel to $O(g)$ we must have 
$\delta^{(1)}_{\ell}=2\d_\phi^{(1)}$. 

However this is not the case when $\ell=0$ in the $s$-channel. It can be checked that for $d$ close or equal to either two or four dimensions we can have $q^{(s)}_{\D_0,\ell=0}$ and $q^{(t)}_{\D_0,\ell=0|\ell'}$ starting from the same order in $g$. For this reason, generically $\d^{(1)}_{\ell=0}\ne 2\d_\phi^{(1)}$ in these cases. Then we see that the sum over $\ell'$ in \eqref{alphai} gets contribution only from scalars
\be\label{alphap}
\alpha _p=- C_{0} \ g^{2+p}\frac{(-\delta^{(1)}_{0})^p}{{2p!}}   \left(\delta^{(1)}_{0}-2\d_\phi^{(1)}\right)^2 \Gamma (d-2) +O\left(g^{3+p}\right)\,.
\ee
As a check we can cansider the the $\phi^4$ theory in $d=4-\e$ and use the above expression in \eqref{gammaellweak} to evaluate the coefficient of $(\log \ell)^p$ terms in the anomalous dimension of the higher spin operators. The result matches with what one gets from expanding \eqref{genspin} at large $\ell$.

Note that in cases where $\delta^{(1)}_{0}=2\d_\phi^{(1)}$ we will have $\a_p \sim O(g^{3+p})$ for which there is no compact expression like above. However even in cases where such compact expressions are not possible, for any weakly coupled theory if the assumptions \eqref{genellexpand} hold, then one can easily repeat the above analysis and systematically evaluate $\alpha_p$.

\subsection{Theories in $4\le d \le 6$}
Now in weakly coupled theories (in $d<6$) there is often a scalar whose dimension starts with $2$. Examples of this can be the fundamental field in $\phi^3$ theory in $d=6-\e$ or the shadow operator corresponding to $\vec{\phi}\cdot \vec{\phi}$ in large $N$ critical theories. Let us call this operator $\sigma$. 
\be
\ta_\s=2+\d^{(1)}_\s g+O(g^2)\,,
\ee
In such theories it is this $\s$ operator that has the minimal twist and decides the leading large $\ell$ behavior. It is separated from other operators by a finite twist gap. So we can directly use the result \eqref{gammastrng}  from the strongly coupled section. We can put $\tau_m=\ta_\s$ and $\ell_m=\ell_\s=0$ to obtain the anomalous dimension,
\be
\g_\ell=-C_{\s}\frac{2\Gamma ^2\left(\Delta _{\phi }\right)\Gamma \left(\tau _\s\right)}{\Gamma ^2\left(\Delta _{\phi }-\frac{\tau _\s}{2}\right)\Gamma ^2\left(\frac{\tau _\s}{2}\right)}\left(\frac{1}{\ell }\right)^{\tau _\s}\,,
\ee
where $C_{\phi\phi\s}\equiv C_\s$ is the OPE coefficient. Expanding this in large $\ell$, we get an equivalent of \eqref{gammaellweak} with a different spin dependence, 
\be\label{gammaellweak1}
\g_\ell=\frac{\a_0(g) +\a_1(g) \log \ell+\a_2(g) 
(\log \ell)^2+\cdots}{\ell^{2}}\,.
\ee
Here $\a_i(g)$ are given by,
\begin{align}\label{alphaidim2}
\a_p(g)=C_{\s} &  \frac{(-1)^{p+1}}{2 p!}  \left(\delta _{\s}^{(1)}\right)^{p}  (d-4)^{2}g^{p} \ + \ O(g^{p+1})\,.
\end{align}

We should point out that both the results \eqref{gammaellweak} and \eqref{gammaellweak1} are correct, and  depending on the CFT and the dimension one or the other formula will be relevant. Thus in $d>4$, the $\s$ operator having the leading twist gives the leading large spin dependence. Note that in $d=4-\e$, with $\e=g$, we can choose either of the two equations. It is the scalar $\phi^2$ that plays the role of $\s$ and has the dimension $\D=d-2+O(\e)=2+O(\e)$. So both expressions can reproduce the correct large spin form of \eqref{genspin}. However, in the second expression \eqref{gammaellweak1}, the $\a_p$ must be evaluated to  subleading orders of $g$ to get the correct result\footnote{Another verification is possible for a large $N$ critical model, where the large spin behavior of singlet, tracless symmetric and antisymmetric higher spin operators \cite{giombi} can be correctly reproduced. This will be elaborated in \cite{dks}.}. Note that in this example, with $d$ close to 4, it is still only $\s \equiv \phi^2$ that contributes in the $t$ channel, because other operators are suppressed when we expand in $g$.

\section{Discussion}
In this paper we have described in detail a new approach to the conformal bootstrap, formulated in Mellin space, which was outlined in \cite{usprl}. There are several directions to investigate in the future. Below we discuss a few of them:

\begin{itemize}
\item {\bf Numerics:}

In light of the recent success in constraining conformal field theories using numerics, it is natural to ask if we could investigate our bootstrap equations numerically. 
We give a preliminary discussion of a possible numerical approach using the Mellin space Witten diagram blocks. The way we have set up the calculation, for the $\e$-expansion, is not very well suited for numerics: firstly, there is a residual integral over the spectral parameter which is left to do and secondly it appears cumbersome to look at the constraints arising from the additional spurious poles at $s=\D_\phi+n$ with $n > 0$. In order to potentially use the already existing powerful numerical algorithms that build on the original work of \cite{rrtv}, we should try to look at our equations around a specific point in $(s,t)$. As a function of $(s,t)$, the natural choice to expand around, in our approach is $s=\D_\phi+n$. Let us then consider expanding around a particular $t=t_0$ in powers of $(t-t_0)$ (with $t_0<0$). This can be viewed as an alternative to expanding in the continuous Hahn polynomials. The constraints obtained by setting the coefficients of the various powers of  $(t-t_0)$ to zero are infinite linear combinations of our previous constraints\footnote{It may well be that for this combination, the sum over $\D, \ell$  converges faster and is therefore better suited for numerics.}.  It can be verified that in this way or organising the equations, the leading order in $\epsilon$ result for the anomalous dimension of $\phi^2$ is easily reproduced (higher spin operators start contributing from the next order itself unlike for the partial wave approach). For the following we will choose to expand around $t=-\D_\phi$. We will focus on the double pole or log term only. The leading order 
constraint will be given by the vanishing of the double pole residue
\be\label{logcons}
\sum_{\Delta, \ell} c_{\D, \ell}  \bigg(M_{\Delta,\ell}^{(s)}(s,t)+M_{\Delta,\ell}^{(t)}(s,t)+M_{\Delta,\ell}^{(u)}(s,t)\bigg)\bigg|_{s=\D_\phi,t=-\D_\phi}=0\,,
\ee
where $c_{\D,\ell}$ are the same coefficients of expansion as in \eqref{idmellsum}. Below we plot the individual spin contributions to the Mellin space Witten diagram blocks  in Fig. 3. This will demonstrate how one may hope to see a bound arising from these numerics. We will restrict to $d=3$ and take as input (purely for illustrative purposes) $\D_\phi=0.518$ which is the value for the 3d-Ising model.

\begin{figure}
  \centering
  \includegraphics[width=0.5\textwidth]{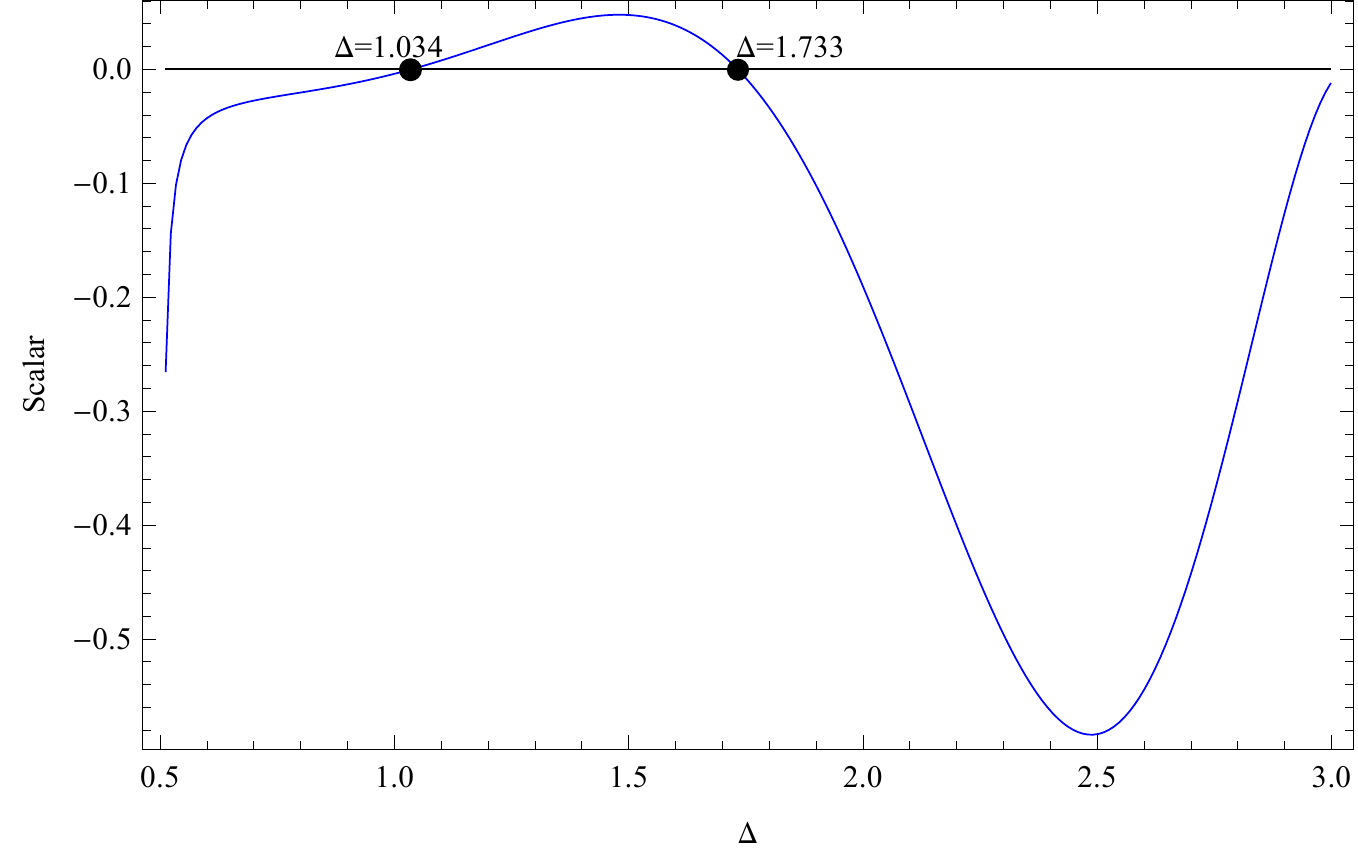}
  \caption{Plot of the sum of all three channels for a scalar exchange as a function of the scalar dimension. This is mostly negative except for $\D \in (1.034-1.733)$ where this flips sign and becomes positive.}
  \label{fig:scalar}
\end{figure}

\begin{figure}[!htpb]
\begin{subfigure}{0.5\textwidth}
  \centering
  \includegraphics[width=0.9\textwidth]{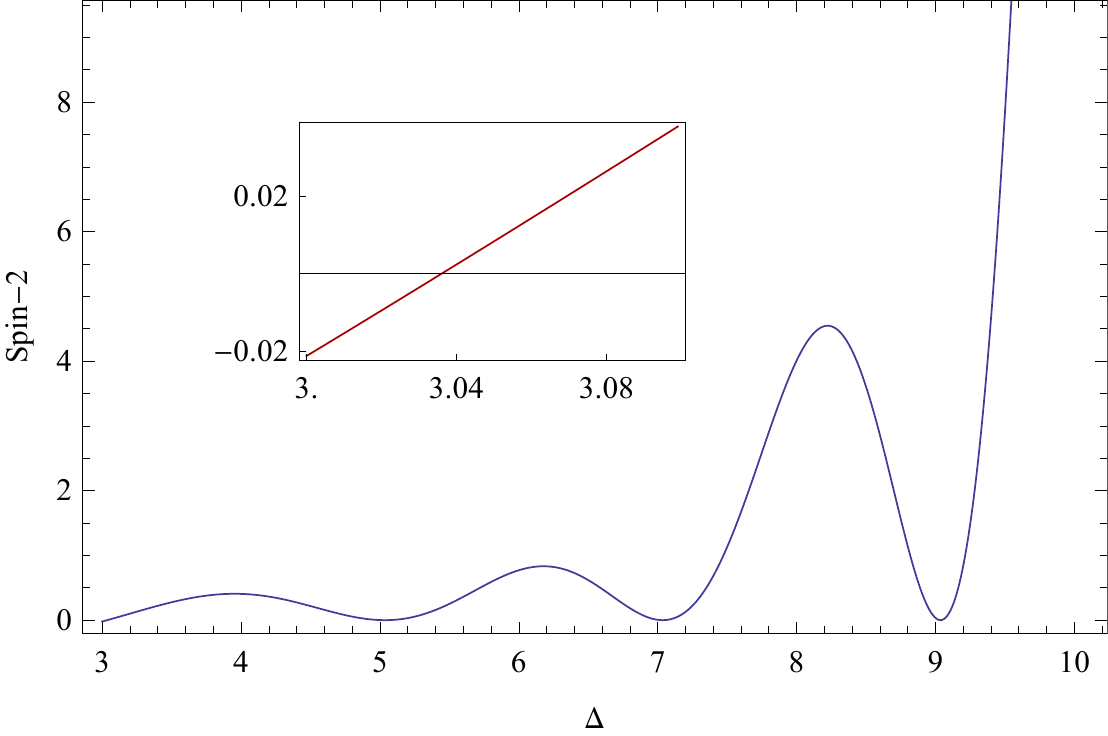}
  \caption{spin-2}
  \label{fig:sub1}
\end{subfigure}\hfill%
\centering
\begin{subfigure}{0.5\textwidth}
  \centering
  \includegraphics[width=0.9\textwidth]{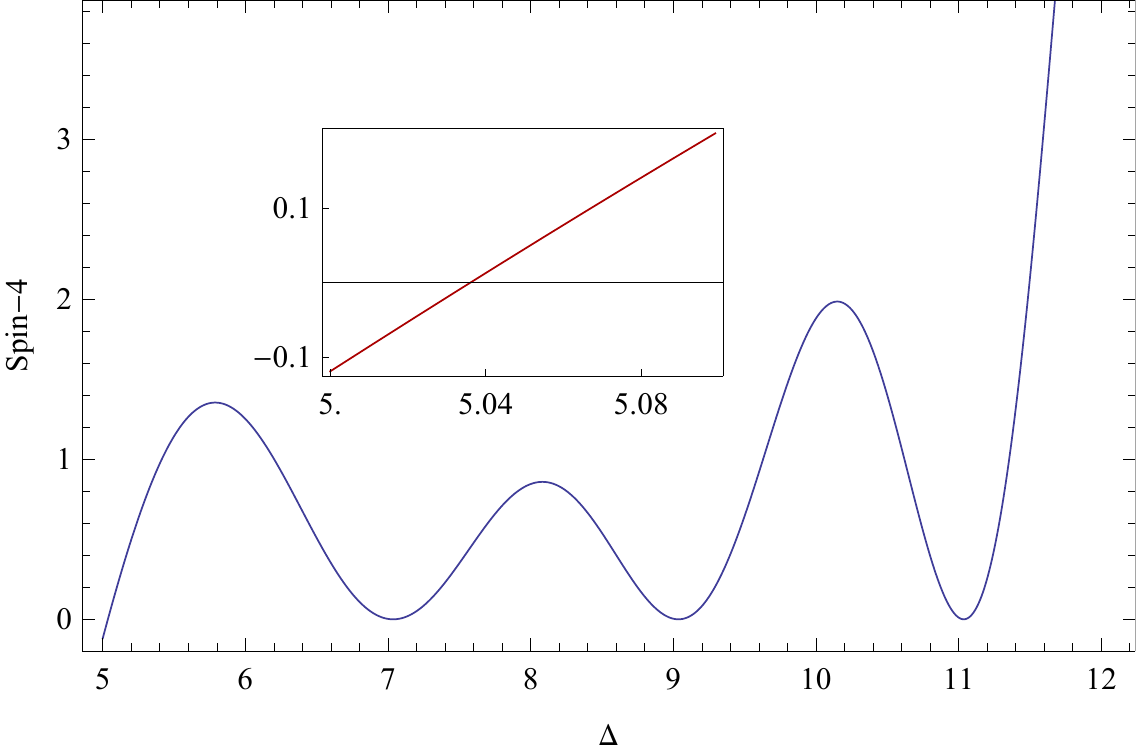}
  \caption{spin-4}
  \label{fig:sub2}
\end{subfigure}\hfill%
\begin{subfigure}{0.5\textwidth}
  \centering
  \includegraphics[width=0.9\textwidth]{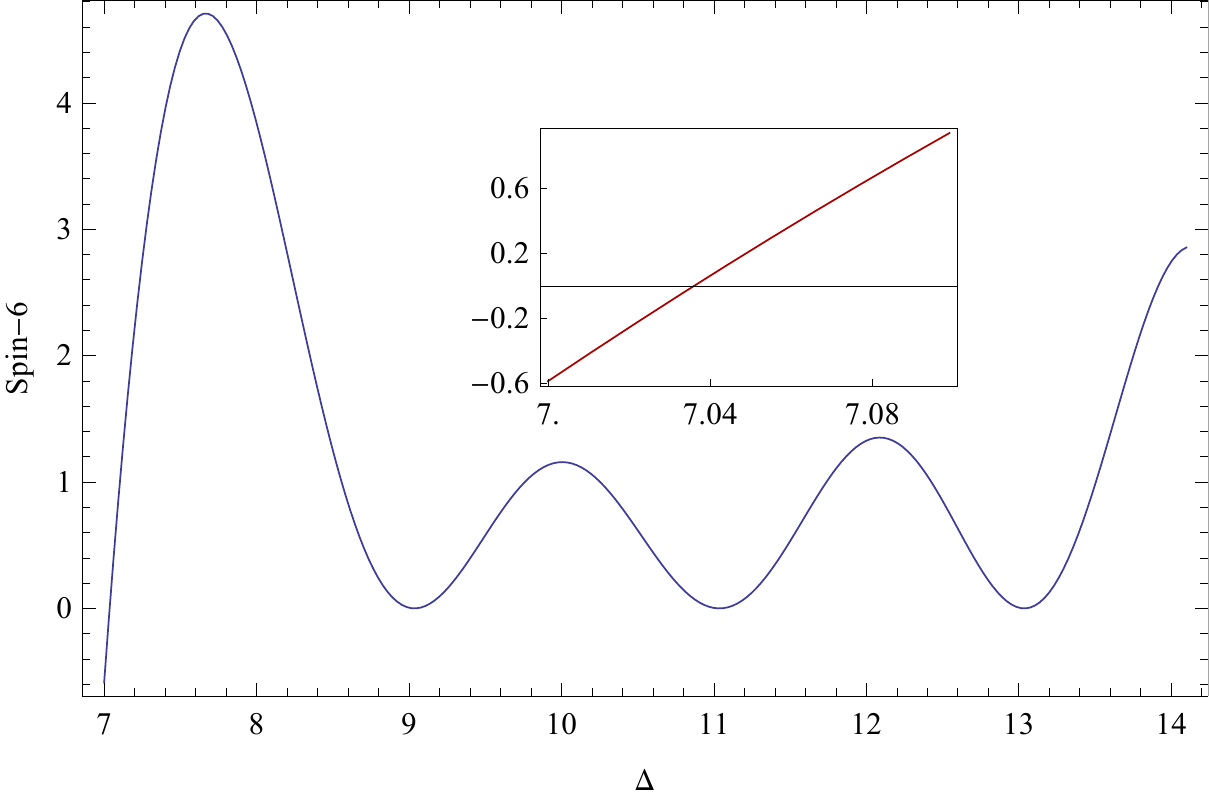}
  \caption{spin-6}\label{n8n9}
  \label{fig:sub3}
\end{subfigure}\hfill
\caption{Plots showing the variation of spin-2,4 and 6 blocks (sum of all three channels) as a function of the dimension of the exchanged operator dimensions $(\D)$. All of these plots are mostly positive except for near the unitarity bound given by $\D\approx d-2+\ell$. The sign flip near the unitarity bound is given by the inset plots. }
\label{otherns}
\end{figure}
As is clear from the spin block plots, close to the unitarity bound $(\D \approx d-2+\ell)$, the spin blocks are negative but positive elsewhere.
The scalar block on the other hand  is positive only for a small region of $\D =\D_0  \in (1.034-1.733)$ (see Fig. 2). In the rest of the range of $\D_0$, the scalar block is negative. If we {\it assume} that the non-zero spin operators, which contribute most to the constraints, are close to the unitarity bound (as happens for the 3d Ising model), then their contribution to the constraint equation will be negative. So the only way to satisfy the constraint equation would be if the scalar block were to give a positive contribution. This gives $\D_0 \in (1.034-1.733)$ which is indeed the case for the $\phi^2$ operator. Note that we used just one constraint for illustrative purposes to demonstrate that investigating numerics along these directions ought to be a promising future endeavour. Of course, one needs to demonstrate, since there is an infinite sum over the spectrum $\D,\ell$, that the resulting numerics converge. Our preliminary investigations, of theories living at the border of the known allowed regions, using presently available numerical methods, does suggest that the approach above will lead to convergent numerics\footnote{We thank Slava Rychkov for suggesting this check.}. A more thorough investigation of this issue should be carried out.

\item {\bf Higher orders in $\e$:}

We have had striking success in using our approach to obtain results to $O(\e^3)$ - therefore it is natural to ask how to go to the next order in the $\e$-expansion. Indeed one would like to know if there is a systematic approach that allows one to go to any arbitrary order in the expansion, if one so desired\footnote{Note that the 
$\e$-expansion, like any perturbative QFT expansion, is only an asymptotic one and needs to be Pade-resummed to obtain something useful. The question is whether there is an in-principle systematic method to obtain the $n$-th term in this expansion using our Mellin space approach.}. Once we set up the formalism obtaining the $O(\e^3)$ results was conceptually and mathematically straightforward (though, perhaps a bit tedious) and needed very few and rather mild assumptions. The two main inputs were the existence of a conserved stress tensor and the leading behaviour of OPE coefficients for higher order operators which we know from the perturbation expansion--for instance  $C_{\phi\phi\phi^2}$ begins at $O(\e^2)$ since it is the square of the OPE coefficient which is assumed to be $O(\e)$. We had pointed out in Sec. 5, that at $O(\e^4)$, the constraint equations at the spurious pole $(s=\D_\phi)$ involved an infinite number of operators. However, it is plausible that, by appropriately combining the enormous amount of information in the additional constraints at $s=\D_\phi+n$, one can give an algorithm to continue to higher orders in $\e$ in a controlled manner. Furthermore, we have only investigated the case of identical scalars. Thus we may need to combine the information from other spurious poles with correlators of other scalar operators. At present, our approach yielded information about operators which were bilinear in the elementary scalar $\phi$. It is possible to extend our results to operators with higher powers of $\phi$ \cite{rajanind}, for which some information is known in the $\e$-expansion \cite{kehrein}. At some stage we expect the non-unitary behaviour in $4-\e$ to show up for some large dimension operator \cite{rychnon}. Our approach, however, did not crucially rely on unitarity as the non-unitary example in $6-\e$ showed and should be able to capture this behaviour in $4-\e$ dimensions. It will also be interesting to use our approach to study the theories in higher dimensions considered in \cite{klebanov}.
 
\item {\bf Other small expansion parameters:}

We have studied our equations with two small expansion parameters, namely $\e$ and $\frac{1}{\ell}$, and found quite remarkable simplifications. It will therefore be interesting to investigate our equations when we have other small expansion parameters. These could be, for example, large dimensions for external operators, large spacetime dimension $d$ limit, strong or weak coupling limits and of course, large $N$. In these cases one might hope to have similar simplifications which organise the bootstrap conditions so that there is a controlled way of incorporating the contributions from different families of operators.  
Recently in \cite{Alday:2016njk}, a systematic procedure has been outlined to solve the conventional boostrap equations in the large spin limit using ``twist blocks''. These twist blocks resum the contribution of all operators of degenerate twist and different spins and appear to be a useful way to compute the anomalous dimension of large spin operators with arbitrary twists. One could attempt a similar procedure to approximate the Mack polynomials in this limit and set up the analogous equations in Mellin space.

\item{\bf Technical hurdles}

In order to use our constraint equations systematically, one bottleneck is the integration over the spectral parameter in the crossed channel and another is the lack of a compact expression for the Mack polynomials. In the way we have currently set-up the equations there were coincidences which led to remarkable (almost) cancellations between various $\nu$-poles. This fact enabled us to go to higher orders in $\e$ than what one may have naively expected from \cite{Polyakov, sensinha}. In fact, as we saw in our calculation, in the crossed channel only the $\phi^2$ operator contributed to yield the nontrivial results at $O(\e^3)$. In \cite{rajanind} we will show how to get rid of the integration over the spectral parameter leading to an enormous simplification in the form of the equations. This should enable a systematic investigation of a plethora of questions, some of which have been indicated above. It will also be desirable to have a better understanding of the Mack polynomials to see if more compact representations for them exist, compared to the present one. One could also perhaps try to see if there is a geometric way (in $AdS_{d+1}$) of understanding our consistency conditions\footnote{There is a geometric way of understanding the conventional conformal blocks in terms of geodesic Witten diagrams \cite{perlmutter}.}. This could lead to a new way of doing quantum field theory for critical phenomena which uses a different set of diagrams rather than Feynman diagrams to systematize general perturbative expansions.

\item{\bf Connection with AdS/CFT}

Our building blocks are Witten diagrams in Mellin space. This suggests the tantalizing possibility of AdS/CFT playing an important role to understand the Wilson-Fisher fixed point. Of course, any such dual string theory is likely to be in the quantum regime. In a companion paper \cite{dks}, the present method will be extended to $O(N)$ both in the $\e$-expansion as well as in the large-$N$ limit showing that it works analogously, yielding the first few subleading orders. A systematic study of our constraints in the large-$N$ limit will be an important question to investigate in the future in order to explicate the connection with a weakly coupled string theory/Vasiliev theory. We do not have further insights to offer at this stage, but clearly it will be fascinating to unearth a direct connection between string theory and the 3d Ising model.

\end{itemize}

\section*{Acknowledgments}
Special thanks to J. Penedones for collaboration during the initial stages of this work and for discussions. We acknowledge useful discussions with S. Giombi, T. Hartman, J. Kaplan, I. Klebanov, G. Mandal, S. Minwalla, D. Poland,  S. Pufu, Z. Komargodski, J. Maldacena, H. Osborn, E. Perlmutter, L. Rastelli, M. Serone, S. Wadia and especially S. Rychkov. We also thank all our other colleagues at IISc, ICTS and TIFR-Mumbai for numerous discussions and encouragement during various stages of this work. R.G. acknowledges the support of the J. C. Bose fellowship of the DST. A.S. acknowledges support from a DST Swarnajayanti Fellowship Award DST/SJF/PSA-01/2013-14. This work would not have been possible without the unstinting support for the basic sciences by the people of India.

\appendix

\section{The Mack polynomial}\label{mack}

The Mack polynomials $P^{(s)}_{\nu,\ell}(s,t)$ are explicitly known \cite{mack, joao, dolanosborn2}, albeit in terms of a multiple sum 
\begin{align}\label{sumt}
\begin{split}
& P^{(s)}_{\nu,\ell}(s,t)=\widetilde{\sum}\frac{\g_{\l_1,a_s}\g_{\bar{\l}_1,b_s}(\l_2-s)_k(\bar{\l}_2-s)_k(s+t+a_s)_\b(s+t+b_s)_\a(-t)_{m-\a}(-a_s-b_s-t)_{\ell-2k-m-\b}}{\prod_i\G(l_i)}\,,\\
& \text{where \ \  }\widetilde{\sum}\equiv \frac{\ell!}{2^{\ell}(h-1)_\ell}\sum_{k=0}^{[\frac{\ell}{2}]}\sum_{m=0}^{\ell-2k}\sum_{\a=0}^m\sum_{\b=0}^{\ell-2k-m}\frac{(-1)^{\ell-k-\a-\b}\G(\ell-k+h-1)}{\G(h-1)k!(\ell-2k)!}\binom{\ell-2k}{m}\binom{m}{\a}\\
& \ \ \ \ \ \ \ \times\binom{\ell-2k-m}{\b}\,.
\end{split}
\end{align}
Here we employ the notation in \cite{dolanosborn2}
\be
\l_1=\frac{h+\nu+\ell}{2} \,, \ \ \bar{\l}_1=\frac{h-\nu+\ell}{2}\,, \ \ \l_2=\frac{h+\nu-\ell}{2}\text{ \ \  and \ \  }\bar{\l}_2=\frac{h-\nu-\ell}{2}\, .
\ee
The $\g_{\l_1,a_s},\g_{\bar{\l}_1,b_s}$ are defined in \eqref{gammadef} with $a_s,b_s$ defined in \eqref{asbs} 
and the $l_i$-s are given by,
\be\label{li}
l_1=\l_2-a_s+\ell-k-m+\a-\b\,,\ l_2=\l_2+a_s+k+m-\a+\b\,, \ l_3=\bar{\l}_2+b_s+k+m\,,\ l_4=\bar{\l}_2-b_s+\ell-k-m\,.
\ee

\section{The continuous Hahn polynomials}\label{A}

In \eqref{hahndef} we specialised the two variable Mack polynomials defined in Appendix \ref{mack} to obtain polynomials of degree $\ell$ in one variable. 
\be\label{hahndef2}
Q^{\tau+\ell}_{\ell,0}(t)= \frac{4^\ell }{(\ta+\ell-1)_{\ell}(2h-\ta-\ell-1)_{\ell}}P^{(s)}_{\tau+\ell-h,\ell}(s=\frac{\tau}{2},t) .
\ee
Note that we have suppressed the dependence on the parameters $a_s,b_s$ 
These polynomials have a number of remarkable properties which enable us to simplify the bootstrap conditions. 

The first is that the multiple sum that defines the Mack polynomials in \eqref{sumt} collapses into a simple single sum which is, in fact, a familiar ${}_3F_2$ hypergeometric function.
\be\label{Qdefn}
Q^{\tau+\ell}_{\ell,0}(t)=\frac{2^\ell (\frac{\tau}{2}-a)_\ell(\frac{\tau}{2}+b)_\ell}{(\tau+\ell-1)_\ell}\ {}_3F_2\bigg[\begin{matrix} -\ell,\tau+\ell-1,\frac{\tau}{2}+b+t\\
\frac{\tau}{2}-a, \frac{\tau}{2}+b
\end{matrix};1\bigg] .
\ee
Here and in the rest of this appendix, for generality, we replace the $s$-channel parameters $(a_s, b_s)$ by arbitrary parameters $(a,b)$.  

The second remarkable fact is that the $Q^\D_{\ell,0}$ polynomials are orthogonal and known in the mathematics literature as the continuous Hahn polynomials \cite{AAR}. They obey the orthonormality condition,
\be
\frac{1}{2\pi i}\int_{-i\infty}^{i\infty}d t \G(a+\frac{\tau}{2}+t)\G(b+\frac{\tau}{2}+t)\G(-t)\G(-t-a-b)Q^{\tau+\ell}_{\ell,0}(t)Q^{\tau+\ell '}_{\ell',0}(t)=\kappa_{\ell}({\tau/2})\d_{\ell,\ell'}\,,
\ee
where,
\be\label{knorm}
\kappa_{\ell}(\tau/2)=
\frac{(-1)^\ell4^\ell \ell!}{(\tau+\ell-1)_\ell^2}\frac{\g_{\ell+\tau/2,a}\g_{\ell+\tau/2,b}}{(\tau+2\ell-1)\G(\tau+\ell-1)}\,,
\ee

We note another useful property of the $Q^\D_{\ell,0}$ polynomials (in the special case of identical scalars) which follows from properties of the hypergeometric function ${}_3F_2$. The transformation
\be
{}_3 F_2\bigg[\begin{matrix} -n,k_1,k_2\\
k_3, k_4
\end{matrix};1\bigg] = \frac{(k_3-k_2)_n}{(k_3)_n} {}_3 F_2\bigg[\begin{matrix} -n,k_1,k_4-k_2\\
k_4, k_2-k_3-n+1
\end{matrix};1\bigg]\,.
\ee
If we take  $n=\ell$, $k_1=\tau+\ell-1$, $k_2=b+\frac{\ta}{2}+t$, $k_3=\frac{\ta}{2}-a$ and $k_4=\frac{\ta}{2}+b$, we see that this gives an identity for the $Q^{\tau+\ell}_{\ell,0}$. For identical scalars, $a=0$ and $b=0$. Then the external factor $(k_3-k_2)_n/(k_3)_n$ becomes $(-1)^\ell$ using the reflection identity of $\G$ functions and we obtain the relation
\be\label{Qid}
Q^{\tau+\ell}_{\ell,0}(t)=(-1)^\ell Q^{\tau+\ell}_{\ell,0}(-\frac{\tau}{2}-t)\,.
\ee
Now let us use this to show that for identical scalars, and for an even spin $\ell$ exchange in the $s$-channel, we have the $t$-channel expansion coefficient in \eqref{qformt} equal to that in the $u$-channel \eqref{qformu}. In the $t$-channel, $q_\ell^{(t)}(s)$ for identical scalars is given by \eqref{qtidn},
\begin{align}
\begin{split}
q_{\D,\ell|\ell'}^{(t)}(\tau/2)=&\kappa_{\ell}(\tau/2)^{-1}\int\frac{ dt d\n}{2\pi i} \G^2(\ta/2+t)\G(\l_2-t-\D_\phi)\G(\bar{\l}_2-t-\D_\phi)\\
&\times \m^{(t)}_{\ell'}(\n)P_{\n,\ell'}^{(t)}(\ta/2-\D_\phi,t+\D_\phi)Q^{\ta+\ell}_{\ell,0}(t)\,,
\end{split}
\end{align}
while in the $u$-channel $q_{\D,\ell|\ell'}^{(u)}(s)$ is given by,
\begin{align}
\begin{split}
q_{\D,\ell|\ell'}^{(u)}(\ta/2)=&\kappa_{\ell}(\ta/2)^{-1}\int \frac{dt d\n}{2\pi i} \G^2(-t)\G(\ta/2+t+\l_2-\D_\phi)\G(\ta/2+t+\bar{\l}_2-\D_\phi)\\
&\times \m^{(u)}_{\ell'}(\n)P_{\n,\ell'}^{(u)}(\ta/2-\D_\phi,t)Q^{\ta+\ell}_{\ell,0}(t)\,.
\end{split}
\end{align}
Now if we use the identity \eqref{Qid}, the above two expressions become equal under the exchange $t \leftrightarrow -\frac{\ta}{2}-t$. Hence one can conclude in general for identical scalars and even spin exchange in $s$-channel, that the $t$-channel is equal to the $u$-channel.

Finally the more general polynomials $Q^{\tau+\ell}_{\ell,m}(t)$\,,
\be
Q^{\tau+\ell}_{\ell,m}(t)= \frac{4^\ell }{(\tau+\ell-1)_{\ell}(2h-\ta-\ell-1)_{\ell}}P_{\tau+\ell-h,\ell}(s=\frac{\tau}{2}+m,t) \,
\ee
appear at the descendant poles. However, analogues of the above nice properties of $Q^{\D}_{\ell,0}(t)$ are not known for the polynomials with $m>0$.

\section{A key normalisation}\label{AppC}

We have been expanding the amplitude in terms of Witten diagram blocks as in \eqref{polyapp} with to-be determined coefficients $c_{\D, \ell}$. These $c_{\D, \ell}$ are proportional to the (square of the) OPE coefficients 
$C_{\D, \ell}$ which appear in the conventional  conformal block expansion of \eqref{convapp}. We need to fix the relative normalisation between the two if we are to be able to compute the OPE coefficients $C_{\D, \ell}$. We will do so in this appendix. 

This is simplest to do in position space.
The conformal blocks $G_{\D,\ell}(u,v)$ are normalised such that in an ($s$-channel) expansion around $u=0$ and $v=1$ we have 
\be\label{confblnorm}
G^{(s)}_{\D,\ell}(u,v) \rightarrow u^{\frac{\D-\ell}{2}}(1-v)^\ell +\ldots \, .
\ee
Thus we need to ensure that as $(u \rightarrow 0, v\rightarrow 1)$
\be\label{ccomp}
c_{\D, \ell}W^{(s)}_{\D, \ell}(u,v) \bigg|_{u \rightarrow 0, v\rightarrow 1} = C_{\D, \ell}u^{\frac{\D-\ell}{2}}(1-v)^\ell +\ldots
\ee
By explicitly evaluating the LHS we will be able to obtain the relative normalisations. We will actually carry this out for non-identical scalars for generality. 

The Witten diagram in position space has the Mellin representation is obtained from the general expression  \eqref{nonidmelldef} by plugging in $M^{(s)}(s,t)$ as the reduced mellin amplitude. The latter is given in the spectral representation by \eqref{sunitrymell}. In the integral over the spectral parameter, we now focus on the contribution from the physical pole at $\nu=\D-h$. This gives
\begin{align}
\begin{split}
c_{\D,\ell}W^{(s)}_{\D,\ell}(u,v)=& c_{\D,\ell}\int \frac{ds }{2\pi i}\frac{dt}{2\pi i} u^s v^t  \G(-t)\G(-t-a_s-b_s)\G(s+t+a_s)\G(s+t+b_s)\\
& \times \frac{ \Gamma (2 h-\Delta -1) \Gamma \left(\frac{1}{2} (2h-\ell-2 s-\Delta )\right) \Gamma (-1+\Delta ) \Gamma \left(\frac{1}{2} (-\ell-2 s+\Delta )\right) }{2 (\Delta -h ) \Gamma (h-\Delta ) \Gamma (-1+2 h+\ell-\Delta ) \Gamma (-h+\Delta ) \Gamma (-1+\ell+\D) }\\
& \times\Gamma \left(\frac{1}{2} (\ell-\Delta +\text{$\Delta_1 $}+\text{$\Delta_2 $})\right) \Gamma \left(\frac{1}{2} (-2 h+\ell+\Delta +\text{$\Delta_1$}+\text{$\Delta_2$})\right) \\& \times \Gamma \left(\frac{1}{2} (\ell-\Delta +\text{$\Delta_3$}+\text{$\Delta_4$})\right) \Gamma \left(\frac{1}{2} (-2 h+\ell+\Delta +\text{$\Delta_3$}+\text{$\Delta_4$})\right)  P^{(s)}_{\D-h,\ell}(s,t)\,.
\end{split}
\end{align}
In doing the $s$-integral, we pick the pole at $s=\frac{\D-\ell}{2}$. Actually, there there is another contribution to the spectral integral from a $\nu$ pole at $\nu=2h-\ell-s$. Evaluating the residue at this pole, we find an additional $s$ pole at $s=\frac{\D-\ell}{2}$ which gives an identical contribution. Thus we will multiply the above expression by a factor of two.

To proceed we expand $v^t$ in $1-v$. This will give,
\begin{align}\label{Uphystintgrl}
\begin{split}
c_{\D,\ell}W^{(s)}_{\D,\ell}(u,v)=& c_{\D,\ell}\frac{2^{-2\ell} }{\Gamma (\Delta-h+1 )  }\Gamma \left(\frac{1}{2} (\ell-\Delta +\text{$\Delta_1 $}+\text{$\Delta_2 $})\right) \Gamma \left(\frac{1}{2} (-2 h+\ell+\Delta +\text{$\Delta_1$}+\text{$\Delta_2$})\right)\\
& \times \Gamma \left(\frac{1}{2} (\ell-\Delta +\text{$\Delta_3$}+\text{$\Delta_4$})\right) \Gamma \left(\frac{1}{2} (-2 h+\ell+\Delta +\text{$\Delta_3$}+\text{$\Delta_4$})\right)  u^{\frac{\D-\ell}{2}} \sum_{m}(-1)^{m}(1-v)^{m}\\& \times  \int \frac{dt}{2\pi i}\binom{t}{m}  \G(-t)\G(-t-a_s-b_s)\G(\frac{\D-\ell}{2}+t+a_s)\G(\frac{\D-\ell}{2}+t+b_s) Q^\D_{\ell,0}(t,a_s,b_s)\,.
\end{split}
\end{align}
Here we have used the relation 
\be
P^{(s)}_{\D-h,\ell}(s=\frac{\D-\ell}{2},t)= 2^{-2\ell}(\D-1)_\ell(2h-\D-1)_\ell Q^{\D}_{\ell,0}(t)\ .
\ee

The binomial expansion coefficient $(-1)^{m}\binom{t}{m} =(-1)^{m}\frac{\G(t+1)}{\G(t-m+1)m!}==\frac{(-1)^{m}}{m!}t^{m}+O(t^{m-1})$ is a polynomial in $t$ of degree $m$. Therefore we can rewrite it as a sum over the orthogonal polynomials $Q^\D_{\ell',0}(t)$ for $0\le \ell' \le m$. Since $Q_{\ell,0}^\D$ is normalized such that $Q_{\ell,0}^\D(t)=2^\ell t^\ell+O(t^{\ell-1})$ we must have $(-1)^{m}\binom{t}{m}=\frac{(-1)^{m}2^{-m}}{m!}Q_{m,0}^\D(t)+\cdots$. Using the orthonormality of the $Q_{\ell,0}^\D$ polynomials as given in Appendix \ref{A}, we can evaluate the contribution to \eqref{Uphystintgrl} which goes as $(1-v)^\ell$ purely from the $m=\ell$ term in the binomial expansion. Doing the $t$ integral gives the contribution
\begin{align}\label{Uphys}
\begin{split}
c_{\D,\ell}W^{(s)}_{\D,\ell}(u,v)=& u^{\frac{\D-\ell}{2}} (1-v)^{\ell} c_{\D,\ell}\k_\ell\bigg(\frac{\D-\ell}{2}\bigg)\frac{(-1)^{\ell}}{2^\ell\ell!}  \frac{ 2^{-2\ell}}{\Gamma(\Delta -h+1) }\Gamma\left(\frac{1}{2} (-2 h+\ell+\Delta +\text{$\Delta_1$}+\text{$\Delta_2$})\right)  \\& \times \Gamma\left(\frac{1}{2} (\ell-\Delta +\text{$\Delta_1 $}+\text{$\Delta_2 $})\right) \Gamma\left(\frac{1}{2} (\ell-\Delta +\text{$\Delta_3$}+\text{$\Delta_4$})\right) 
\Gamma\left(\frac{1}{2} (-2 h+\ell+\Delta +\text{$\Delta_3$}+\text{$\Delta_4$})\right)  +\cdots\,.
\end{split}
\end{align}
The omitted terms denote higher powers of $u$ and $1-v$. 

We can now combine \eqref{Uphys} and \eqref{ccomp} to obtain the relation 
$c_{\D,\ell} = C_{\D,\ell}\mathfrak{N}_{\D,\ell}$ where
\begin{align}\label{norm}
&\mathfrak{N}_{\D,\ell}^{-1} = \frac{ \Gamma (\Delta -1) \Gamma ^4\left(\frac{\ell +\Delta }{2}\right) }{(-2)^{\ell }  (\ell +\Delta-1 ) \Gamma (1-h+\Delta ) \Gamma ^2(\ell +\Delta-1 )} \Gamma \left(\frac{\ell -\Delta +\Delta _1+\Delta _2}{2}\right) \nonumber\\
& \Gamma  \left(\frac{\Delta +\Delta _1+\Delta _2-2 h+\ell }{2}\right) \Gamma  \left(\frac{\ell -\Delta +\Delta _3+\Delta _4}{2}\right)\Gamma  \left(\frac{\Delta +\Delta _3+\Delta _4-2 h+\ell }{2}\right)\,.
\end{align}

\section{$t$ integrals in the crossed channels}\label{tints}

This section will deal with performing the $t$ integrals in  \eqref{qtidn}.  We will carry out the integrals for generic 
$\ell, \ell'$.  The technique can be generalised to the non-identical scalar cases of \eqref{qformt} and \eqref{qformu} but we will not do so here since we only need the identical scalar results. Let us begin with the $Q_{\ell,0}^{\ta+\ell}$ expression, with $a=b=0$, according to \eqref{Qdefn},
\be
Q_{\ell,0}^{\ta+\ell} = \frac{2^{\ell } \Gamma ^2(\ta/2+\ell ) \Gamma (\ta+\ell -1)}{\Gamma ^2(\ta/2) \Gamma (\ta+2 \ell -1)} \ {}_3F_2\bigg[\begin{matrix} -\ell,\ta+\ell-1,\ta/2+t\\
\frac{\ta}{2} \ \ , \ \  \frac{\ta}{2}
\end{matrix};1\bigg]\,.
\ee
Now the hypergeometric function can be written as a sum as below,
\be
{}_3F_2\bigg[\begin{matrix} -\ell,\ta+\ell-1,\ta/2+t\\
\frac{\ta}{2} \ \ , \ \  \frac{\ta}{2}
\end{matrix};1\bigg] = \sum_{q=0}^\ell \frac{\big(\frac{\ta}{2}+t\big)_q (-\ell )_q (\ta+\ell -1)_q}{q! {\big(\frac{\ta}{2}\big)_q}^2}\,.
\ee
With this let us evaluate the $t$-channel  $q^{(t)}_{\D,\ell}$ as given in \eqref{qformt},
\begin{align}\label{qformula}
\begin{split}
& q_{\D,\ell |\ell'}^{(t)}(\ta/2)  = \k_\ell(\ta/2)^{-1}\int \frac{dt}{2\pi i} d\n \   \G^2\big(\frac{\ta}{2}+t\big)\G(\l_2-t-\D_\phi)\G(\bar{\l}_2-t-\D_\phi) \\
&\times \m^{(t)}_{\D,\ell'}(\n)P^{(t)}_{\nu,\ell}(\ta/2-\D_\phi,t+\D_\phi)Q^{\ta+\ell}_{\ell,0}(t)\\
=&\k_\ell(\ta/2)^{-1}\sum_{q=0}^\ell\widetilde{\sum}\int d\n   \m^{(t)}_{\D,\ell'}(\n)\g_{\l_1,0}\g_{\bar{\l}_1,0}\big(\D_\phi-\frac{\ta}{2}\big)_{m-\a}\big(\D_\phi-\frac{\ta}{2}\big)_{\ell'-2k-m-\b}\frac{1}{\prod_i\G(l_i)}\frac{2^\ell((\ta/2)_\ell)^2}{(\ta+\ell-1)_\ell}\\
&\times \frac{(-\ell)_q(\ta+\ell-1)_q}{((\ta/2)_q)^2\ q!} \int \frac{dt}{2\pi i}\ \G\big(\frac{\ta}{2}+t+\a\big)\G\big(\frac{\ta}{2}+t+\b\big)\G(\l_2-t-\D_\phi+k)\G(\bar{\l}_2-t-\D_\phi+k)\big(\frac{\ta}{2}+t\big)_q
\end{split}
\end{align}
In the second equality we have used the explicit forms of the Mack Polynomials in \eqref{sumt} and $P^{(t)}_{\nu,\ell}(s,t)=P^{(s)}_{\nu,\ell}(t,s)$. We want to compute the $t$ integral in the last line above. To do that, we will use the integral representation of the ${}_2F_1$ function, which is given by,
\be
{}_2F_1(a,b,c;z)=\frac{1}{2\pi i}\frac{\G(c)}{\G(a)\G(b)\G(c-a)\G(c-b)}\int_{\g-i\infty}^{\g+i\infty} ds \G(s)\G(c-a-b+s)\G(a-s)\G(b-s)(1-z)^{-s}\,,
\ee
where $\g$ denotes that we have shifted the contour along the real axis so that all the poles of the same sign lie on the same side of the contour. We can further use,
\be
(1-z)^{-\tau/2-t}=\sum_{m=0}^\infty \frac{\big(\frac{\ta}{2}+t\big)_m}{m!}z^m\,.
\ee
The idea is to use the above as a generating funtion for the part  $\big(\frac{\ta}{2}+t\big)_q$ in the $t$-integral above, by picking up the power of $z^q$. So we have,
\begin{align}
\begin{split}
&\sum_{m=0}^\infty\int\frac{ dt}{2\pi i}\ \G\big(\frac{\ta}{2}+t+\a\big)\G\big(\frac{\ta}{2}+t+\b\big)\G(\l_2-t-\D_\phi+k)\G(\bar{\l}_2-t-\D_\phi+k)\frac{\big(\frac{\ta}{2}+t\big)_m}{m!}z^m\\
&=\int \frac{dt}{2\pi i}\ \G\big(\frac{\ta}{2}+t+\a\big)\G\big(\frac{\ta}{2}+t+\b\big)\G(\l_2-t-\D_\phi+k)\G(\bar{\l}_2-t-\D_\phi+k)(1-z)^{-\ta/2-t}\,.
\end{split}
\end{align}
 Shifting the variables $t\rightarrow t-s-\a$ and using the mapping,
\be
a=\l_2+\frac{\ta}{2}+\a-\D_\phi+k\, ;  b=\bar{\l}_2+\frac{\ta}{2}+\a-\D_\phi+k\, ;  c=2k+\ta+\a+\b-2\D_\phi+\l_2+\bar{\l}_2\,,
\ee
we can write,
\begin{align}
\begin{split}
&\int \frac{dt}{2\pi i}\ \G\big(\frac{\ta}{2}+t+\a\big)\G\big(\frac{\ta}{2}+t+\b\big)\G(\l_2-t-\D_\phi+k)\times\G(\bar{\l}_2-t-\D_\phi+k)(1-z)^{-\ta/2-t}\\
=&(1-z)^\a \frac{\G\big(k+\frac{\ta}{2}+\b-\D_\phi+\l_2\big)\G\big(k+\frac{\ta}{2}+\a-\D_\phi+\l_2\big)}{\G\big(2k+\ta+\a+\b-2\D_\phi+\l_2+\bar{\l}_2\big)}\G\big(k+\frac{\ta}{2}+\b-\D_\phi+\bar{\l}_2\big)\G\big(k+\frac{\ta}{2}+\a-\D_\phi+\bar{\l}_2\big)\\
&\times{}_2F_1\bigg[\begin{matrix}k+\frac{\ta}{2}+\a-\D_\phi+\l_2,k+\frac{\ta}{2}+\a-\D_\phi+\bar{\l}_2\\2k+\ta+\a+\b-2\D_\phi+\l_2+\bar{\l}_2\end{matrix};z\bigg]\,.
\end{split}
\end{align}
Collecting the powers of $z^q$, we finally have,
\begin{align}
\begin{split}
&\int \frac{dt}{2\pi i}\ \G\big(\frac{\ta}{2}+t+\a\big)\G\big(\frac{\ta}{2}+t+\b\big)\G(\l_2-t-\D_\phi+k)\G(\bar{\l}_2-t-\D_\phi+k)(\frac{\ta}{2}+t)_q\\
=&\frac{\G\big(k+q+\frac{\ta}{2}+\a+\l_2-\D_\phi\big)\G\big(k+q+\frac{\ta}{2}+\a+\bar{\l}_2-\D_\phi\big)}{\G\big(q+\ta+2k+\a+\b+\l_2+\bar{\l}_2-2\D_\phi\big)}\G\big(k+\frac{\ta}{2}+\b-\D_\phi+\l_2\big)\G\big(k+\frac{\ta}{2}+\b-\D_\phi+\bar{\l}_2\big)\\
&\times{}_3F_2\bigg[\begin{matrix}-q,-\a,1-2k-q-\ta-\a-\b+2\D_\phi-\l_2-\bar{\l}_2\\
1-k-q-\frac{\ta}{2}-\a-\l_2+\D_\phi,1-k-q-\frac{\ta}{2}-\a-\bar{\l}_2+\D_\phi\end{matrix};1\bigg]\,.
\end{split}
\end{align}
Using this in \eqref{qformula} we get,
\begin{align}\label{qtform}
\begin{split}
\!\!\!\!\!\! \hspace{-1cm}&q_{\D,\ell |\ell'}^{(t)}(\ta/2)= \k_\ell(\ta/2)^{-1}\sum_{q=0}^\ell\widetilde{\sum}\int d\n   \m^{(t)}_{\D,\ell'}(\n)\g_{\l_1,0}\g_{\bar{\l}_1,0}\big(\D_\phi-\frac{\ta}{2}\big)_{m-\a}\big(\D_\phi-\frac{\ta}{2}\big)_{\ell'-2k-m-\b}\frac{1}{\prod_i\G(l_i)}\frac{2^\ell((\ta/2)_\ell)^2}{(\ta+\ell-1)_\ell}\\
&\times \frac{(-\ell)_q(\ta+\ell-1)_q}{((\ta/2)_q)^2\ q!} \frac{\G\big(k+q+\frac{\ta}{2}+\a+\l_2-\D_\phi\big)\G\big(k+q+\frac{\ta}{2}+\a+\bar{\l}_2-\D_\phi\big)}{\G\big(q+\ta+2k+\a+\b+\l_2+\bar{\l}_2-2\D_\phi\big)}\G\big(k+\frac{\ta}{2}+\b-\D_\phi+\l_2\big) \\&\times \G\big(k+\frac{\ta}{2}+\b-\D_\phi+\bar{\l}_2\big) {}_3F_2\bigg[\begin{matrix}-q,-\a,1-2k-q-\ta-\a-\b+2\D_\phi-\l_2-\bar{\l}_2\\
1-k-q-\frac{\ta}{2}-\a-\l_2+\D_\phi,1-k-q-\frac{\ta}{2}-\a-\bar{\l}_2+\D_\phi\end{matrix};1\bigg]\,.
\end{split}
\end{align}
The above expression also gives $q_{\D,\ell |\ell'}^{(u)}$ in the $u$-channel.

Now let us see, how the expression \eqref{qtform} reduces to \eqref{qtid} and \eqref{qt1}, when $\ell'=0$. In this case the sum $\widetilde{\sum}$ goes away. In the summand, we have $k=m=\a=\b=0$. Also $\lambda_1=\lambda_2=\lambda=(h+\nu)/2$ and $\bar{\lambda}_1=\bar{\lambda}_2=\bar{\lambda}=(h-\nu)/2$.
\begin{align}\label{qtformell0}
\begin{split}
\!\!\!\!\!\! \hspace{-1cm}&q_{\D,\ell |\ell'=0}^{(t)}(\ta/2)= \k_\ell(\ta/2)^{-1}\sum_{q=0}^\ell\int d\n   \m^{(t)}_{\D,0}(\n)\g_{\l_1,0}\g_{\bar{\l}_1,0}\frac{1}{\prod_i\G(l_i)}\frac{2^\ell((\ta/2)_\ell)^2}{(\ta+\ell-1)_\ell}\frac{(-\ell)_q(\ta+\ell-1)_q}{((\ta/2)_q)^2\ q!}\\
&\times  \frac{\G\big(q+\frac{\ta}{2}+\l-\D_\phi\big)\G\big(q+\frac{\ta}{2}+\bar{\l}-\D_\phi\big)}{\G\big(q+\ta+\l+\bar{\l}-2\D_\phi\big)}\G\big(\frac{\ta}{2}-\D_\phi+\l\big)\G\big(\frac{\ta}{2}-\D_\phi+\bar{\l}\big) \\&\times  {}_3F_2\bigg[\begin{matrix}-q,0,1-q-\ta+2\D_\phi-\l-\bar{\l}\\
1-q-\frac{\ta}{2}-\l+\D_\phi,1-q-\frac{\ta}{2}-\bar{\l}+\D_\phi\end{matrix};1\bigg]\,.
\end{split}
\end{align}
The hypergeometric ${}_3F_2$ simply reduces to 1. Also note that  $\g_{\l_1,0}\g_{\bar{\l}_1,0}/{\prod_i\G(l_i)}$ is  1. So we get,
\begin{align}\label{qtell0}
\begin{split}
&q_{\D,\ell |\ell'=0}^{(t)}(\ta/2)= \int d\n   \m^{(t)}_{\D,0}(\n)\frac{2^\ell((\ta/2)_\ell)^2}{\k_\ell(\ta/2)(\ta+\ell-1)_\ell}\G\big(\frac{\ta}{2}-\D_\phi+\l\big) \G\big(\frac{\ta}{2}-\D_\phi+\bar{\l}\big)\\
&\times \sum_{q=0}^\ell\frac{(-\ell)_q(\frac{\ta}{2}+\ell-1)_q}{((\ta/2)_q)^2\ q!} \frac{\G\big(q+\frac{\ta}{2}+\l-\D_\phi\big)\G\big(q+\frac{\ta}{2}+\bar{\l}-\D_\phi\big)}{\G\big(q+\ta+\l+\bar{\l}-2\D_\phi\big)}  \ \ \,.
\end{split}
\end{align}

\section{$q$ sums in the $t$ channel}\label{qsum}

In this section, we will demonstrate, how the expression in \eqref{qtf1f2} can lead to simple expressions, in an 
$\e$ expansion. Here we will show this for the  leading term in \eqref{qtid}. Let us write down the full expression, given by,
\begin{align}
\begin{split}
q_{\D,\ell|\ell'=0}^{(2,t)}=&\sum_{q=0}^\ell\int d\n \ \frac{2^{-\ell}(-\ell)_q\G(2\ell+2\D_\phi)\G(2\D_\phi+\ell+q-1)\G(\frac{2\D_\phi-h+\n}{2})^2\G(\frac{2\D_\phi-h-\n}{2})^2}{(\n^2-(\D-h)^2)\ell!q!\G(\ell+\D_\phi)^2\G(q+\D_\phi)^2\G(2\D_\phi+\ell-1)\G(-\n)\G(\n)}\\
&\times \frac{\G(\l)\G(\bar{\l})\G(\l+q)\G(\bar{\l}+q)}{\G(q+\l+\bar{\l})} \ \ + \ \cdots \,,
\end{split}
\end{align}
Here $\l=\frac{h+\nu}{2}$ and $\bar{\l}=\frac{h-\nu}{2}$. The expression can be expanded in $\e$. The $\nu$ integral can be carried out by evaluating residues at only the poles $\nu=\D-h$ and $\nu=2\D_\phi-h$. As discussed in section \ref{simplfction} and Appendix F, the other poles are subleading. The leading term is given by,
\begin{align}
\begin{split}
&-\e^{-2}\sum_{q=0}^\ell\frac{(-1)^q 2^{-\ell}\G(2\ell+1)\G(\ell+1+q)}{q!\G(\ell+1-q)\G(\ell+1)^3\G(2+q)}\times\\
&\bigg(-10+16\ell+(1+2\ell)\bigg(45\g_E+\frac{9}{1+q}-18H_{2\ell}+18H_q-18H_{\ell+q}+36\psi(\ell+1)\bigg)\bigg)\,,
\end{split}
\end{align}
where, note that the problematic terms are $H_{q}$ and $H_{\ell+q}$. Using the identity,
\be
H_{x+N-1}-H_{x-1}=\sum_{k=0}^{N-1}\frac{1}{x+k}\,,
\ee
we can pull out the $k$ sum and first perform the $q$ sum over these term. After the $q$ sum, we can perform the sum over $k$ to get,
\be
-\sum_{q=0}^\ell\frac{(-1)^q 2^{-\ell}\G(2\ell+1)\G(\ell+1+q)}{q!\G(\ell+1-q)\G(\ell+1)^3\G(2+q)}
\bigg((1+2\ell)(18H_q-18H_{\ell+q})\bigg)=9\frac{ 2^{2 + \ell}
  \Gamma(3/2 + \ell)}{\ell (1 + \ell) \sqrt{\pi} (\ell!)^2}\,.
\ee
The remaining terms can be handled with the usual sum over $q$ to obtain,
\begin{align}
\begin{split}
&-\sum_{q=0}^\ell\frac{(-1)^q 2^{-\ell}\G(2\ell+1)\G(\ell+1+q)}{q!\G(\ell+1-q)\G(\ell+1)^3\G(2+q)}\bigg(-10+16\ell+(1+2\ell)\bigg(45\g+\frac{9}{1+q}-18H_{2\ell}+36\psi(\ell+1)\bigg)\bigg)\\
&=-9\frac{ 2^{-\ell} (1 + 2 \ell) (2 \ell)! }{\ell (1 + \ell) (\ell!)^3}\,.
\end{split}
\end{align}
Adding these two separate contributions, we find that,
\begin{align}
\begin{split}
q^{(2,t)}_{\D,\ell|0}=&-\e^{-2}\sum_{q=0}^\ell\frac{(-1)^q 2^{-\ell}\G(2\ell+1)\G(\ell+1+q)}{q!\G(\ell+1-q)\G(\ell+1)^3\G(2+q)}\times\\
&\bigg(-10+16\ell+(1+2\ell)\bigg(45\g_E+\frac{9}{1+q}-18H_{2\ell}+18H_q-18H_{\ell+q}+36\psi(\ell+1)\bigg)\bigg)\\
=&\frac{2^{-\ell} 9 \e^{-2}\G(2 + 2 \ell) }{\ell (1 + \ell) \G(\ell+1)^3} +\cdots\,.
\end{split}
\end{align}
The $\cdots$ indicate subleading terms in $\e$. One must be careful while handling the above expression, since with the normalization inside $c_{\D,\ell}$ that multiplies this, the whole thing starts from $O(\e^2)$.

\section{Simplifications for the $\e$ expansion}\label{simplification}
Our analysis for the Wilson-Fisher point in $d=4-\e$ dimensions rested on several simplifications, which occur when we Taylor expand our equations in $\e$. In this appendix we will address all of them. 
\subsection{$s$-channel}\label{schcancel}
Recall that the $s$-channel has $\sum_{\D}c_{\D,\ell}q_{\D,\ell}^{(s)}$, which is a sum over all operators of spin $\ell$. The first simplification here is that only the lowest dimension operator of spin $\ell$ contributes to the sum to the order we consider. This is the operator with dimension $\D_\ell=2+\ell-\e+O(\e^2)$.  For the $\phi^4$ theory, we are considering, there are higher dimension operators with $\D_{2m,\ell}=\ell+2+2m +\d_m\e+O(\e^2)$. These operators have the generic form $O_{2m,\ell}\sim \phi (\partial^2)^m\partial^\ell \phi$. Using the equation of motion, $O_{2m,\ell} \sim \phi\partial^a\phi\partial^b\phi\partial^c\phi$, where among the $a+b+c \ ( = 2m-2+\ell)$ derivatives, $2m-2$ derivatives are contracted and $\ell$ derivatives carry indices. We will show that these operators are suppressed in an $\e$ expansion.  We will demonstrate this only for the $q^{(2,s)}_{\D,\ell}$ term. The $q^{(1,s)}_{\D,\ell}$ follows a similar logic. Using $\D_\phi=1-\e/2+O(\e^2)$, we have from \eqref{qtaylor} for $\D=\D_{2m,\ell}$,
\begin{align}
c_{\D,\ell}q_{\D,\ell}^{(2,s)}=&\frac{C_{2m,\ell }\epsilon ^2(-1)^{2 m+\ell } 2^{-2+4 m+3 \ell }\text{  }m (m+\ell ) (1+2 m+2 \ell ) \left(\delta _m+1\right){}^2  \Gamma ^2(m) \Gamma (\ell ) \Gamma ^2\left(\frac{1}{2}+m+\ell \right)}{\pi  \Gamma ^4(1+m+\ell )}\nonumber\\ &+O\left(\epsilon ^3\right)\,.
\end{align}
Here $C_{2m,\ell}$ is the OPE coefficient of this operator. Now these operators do not exist in the free theory. Since $O_{m,\ell}$ has four $\phi$-s it is easy to guess that the 3-point function starts from $\la \phi\phi O_{m,\ell}\ra \sim O(\e) (\sim \lambda^*)$. The OPE coefficient $C_{2m,\ell}$ goes as square of this quantity and hence $C_{2m,\ell} \sim O(\e^2)$. Accounting for this, we must have,
\be
c_{2m,\ell}q_{\D_{2m,\ell},\ell}^{(2,s)}=O(\e^4)\,.
\ee
There can be other ``heavier'' operators too, with spin $\ell$, also contributing to the sum  $\sum_{\D}c_{\D,\ell}q_{\D,\ell}^{(s)}$. However, such operators are composites with a higher number of $\phi$-s, for example $\phi\partial^\a\phi\partial^\b\phi\partial^\g\phi\partial^\r\phi\partial^\s\phi$. Now such operators will have the OPE coefficient $C_{\D,\ell}$ which are even further suppressed in $\e$. So the corresponding $q_{\D,\ell}^{(s)}$ will begin from $O(\e^6)$ or beyond. Because of this, when we considered only up to the $O(\e^3)$ term, keeping only the $J_\ell \sim \phi\partial^\ell\phi$ operator had sufficed.

For completeness, we give the spin 0 contribution, $q^{(2,s)}_{\D,\ell=0}$, from the $O_{2m,\ell=0}$ operator.
\be
c_{2m,0}q_{\D_{2m,0},0}^{(2,s)}=\frac{C_{2m,0} \ \epsilon \  (-1)^{2 m+1}\text{  }\left(\delta _m+1\right){}^2\text{   }\Gamma ^2(2+2 m)}{2\text{   }\Gamma ^4(1+m)}+O(\e^2)\,.
\ee
Then, again since $C_{2m,0} \sim O(\e^2)$, we have $c_{2m,0}q_{\D_{m,0},0}^{(s)} \sim O(\e^3)$. Thus, for 
$\ell=0$ we could only go to  $O(\e^2)$ order, by keeping just one operator.

\subsection{Crossed channels}
In the crossed channels, we have $\sum_{\D,\ell'}c_{\D,\ell}q^{(t)}_{\D,\ell|\ell'}$, which is a sum over all operators in the OPE. However we had considered only the $\ell'=0$ scalar operator $\phi^2$. Even for that operator, we had considered only a certain set of poles and neglected others under the assumption that the other set of poles would contribute at a higher order in the $\e-$expansion. Here we will put these assumptions on solid grounds. We will first go through the case when the exchanged operator is a spin$-0$ particle (scalar) and then move on to substantiate the same arguments for the case of spin$-\ell'$ exchanged operators. In the subsection below, we will first demonstrate the case of the scalar exchange (for identical external scalars, which is the case of interest). 

\subsubsection{Spin$-0$ exchange}
We begin by explicitly writing down the spin$-0$ contribution for the $t$ channel (since the crossed channels give identical contributions for the identical scalar case, we can just consider either one of them) in \eqref{qtform},
\begin{align}
\begin{split}
c_{\D,0}q_{\D,\ell|0}^{(t)}(s)=&\sum_{q=0}^\ell\int d\n\ \frac{2^{-\ell}c_{\D,0}\G(-\ell+q)\G(2\ell+2s)\G(2s+\ell+q-1)\G(\frac{2\D_\phi-h+\n}{2})^2\G(\frac{2\D_\phi-h-\n}{2})^2}{(\n^2-(\D-h)^2)\ell!q!\G(-\ell)\G(\ell+s)^2\G(q+s)^2\G(2s+\ell-1)\G(-\n)\G(\n)}\\
&\times \frac{\G(\l-\D_\phi+s)\G(\bar{\l}-\D_\phi+s)\G(\l-\D_\phi+q+s)\G(\bar{\l}-\D_\phi+q+s)}{\G(q+2s-2\D_\phi+\l+\bar{\l})}\,,
\end{split}
\end{align}
We will show that the leading $\e-$dependence comes from only the $\phi^2$ operator and also considering only two poles in the spectral integral is sufficient. This happens because of a nontrivial cancellation among residues of poles for every operator. We will demonstrate this with the log term, which is $q_{\D,\ell |0}^{(2,t)}$. The power law term which is $q_{\D,\ell |0}^{(1,t)}$ will have similar cancellations. Let us introduce the notation,
\be
f_\n(q,s)=\frac{2^{-\ell}(-\ell)_q(2s+\ell-1)_q\G(q+s-\D_\phi+\frac{h+\n}{2})\G(s-\D_\phi+\frac{h+\n}{2})\G(\frac{2\D_\phi-h+\n}{2})^2}{\ell!q!\G(\ell+s)^2\G(q+s)^2\G(h+q+2(s-\D_\phi))}\,.
\ee
We can write $q_{\D,\ell|0}^{(t)}(s)$ as,
\be
c_{\D,0}q_{\D,\ell|0}^{(t)}(s)=\sum_{q=0}^\ell\int d\n\ \frac{c_{\D,0}f_\n(q,s)}{\G(-\n)\G(\n)}\frac{\G(\frac{2\D_\phi-h-\n}{2})^2\G(\frac{h-2\D_\phi+2s-\n}{2})\G(\frac{h+2(q+s)-2\D_\phi-\n}{2})}{\nu^2-(\D-h)^2}\,.
\ee

The poles of the above integral, lying on the positive real part of the $\nu$- contour, occur at
\begin{itemize}
\item{I. \ \ $\nu= (\D-h)$ }

\item{II. \  $\nu= (2\D_\phi-h+2n_1)$ }

\item{III. $\nu= (h-2\D_\phi+2s+2n_2)$ }\,.

\end{itemize}

Note that for $n_2\ge q$ the poles III become double poles.

\begin{enumerate}

\item{{\bf Lowest dimension scalar}\\
\\
Consider the operator $\phi^2$ with dimension $\D=\D_0=2+\d_0^{(1)}\e+O(\e^2)$. This is the only operator that contributes in the crossed channels up to the $O(\e^3)$ order. In our computation we used only the poles $\nu=\Delta_0-h$ and $\nu=2\D_\phi-h$. Let us see why the other poles do not contribute. Let us assume $n_1<q$ and $n_2<q+1$. Then residues at the poles II and II give (where we have used $h=2-\e/2$ and the familiar dimension of $\phi$, $\D_\phi=1-\e/2+\d_\phi^{(2)} \e^2$),
\begin{align}
\begin{split}
Res_{\n=2\D_\phi-h+2n_1}&=-\frac{C_{2,0}(-1)^{-n_1+1} (\delta_0^{(1)}) ^2 (1+\delta_0^{(1)} )^2   f_{2n_1}(q,1) \Gamma \left(q-n_1+1\right)}{32 (\delta_\phi^{(2)})^2 \Gamma ^3\left(1+n_1\right)}\epsilon+O(\e^2)\,,\\
Res_{\n=h-2\D_\phi+2s+2n_2}&=-\frac{C_{2,0}(-1)^{-n_2} (\delta_0^{(1)}) ^2 (1+\delta_0^{(1)} )^2   f_{2+2n_2}(q,1) \Gamma \left(q-n_2\right)}{32 (\delta_\phi^{(2)})^2 \Gamma ^3\left(2+n_2\right)}\epsilon+O(\e^2)\,.
\end{split}
\end{align}
Here $C_{2,0}$ is the OPE coefficient of $\phi^2$. The subleading terms in $\e$ are cumbresome and not shown. The above residues cancel off for the replacement $n_2=n_1-1$, and quite nicely this cancellation is till the $O(\e^3)$ order. Hence we have,
\be\label{cancel1}
Res_{\n=2\D_\phi-h+2n}+Res_{\n=h-2\D_\phi+2s+2n-2}=O(\e^4)\,,
\ee
with $n$ being a positive integer. This cancellation is true for any function $f_{\nu}(q,s)$. When the poles III become a double pole for $n_2\ge q$, the individual residues start from a diffrent order, viz $Res_{\n=2\D_\phi-h+2n_1} \sim Res_{\n=h-2\D_\phi+2s+2n_2} \sim O(\e^{-1}) $. However the cancellation \eqref{cancel1} till $O(\e^3)$ would still hold.

}

\item{{\bf Heavier scalars}\\
\\

Now let us look at heavier scalars with dimensions of the form $\D_{2m,0}=2+2m+\d_m \e+O(\e)$. Here $m$ is a positive integer. It was argued in section \ref{schcancel} that OPE coeficients of such operators begin from $C_{2m,0}\sim O(\e^2)$ at least. With this taken into account one makes the following observations:
\begin{align}
Res_{\nu=2\D_\phi-h}&=O(\e^5)\nonumber\\
Res_{\nu=\D-h}+Res_{\nu=2\D_\phi-h+2m}+Res_{\nu=h-2\D_\phi+2s+2m-2}&=O(\e^4)\nonumber\\
Res_{\nu=2\D_\phi-h+n}+Res_{\nu=h-2\D_\phi+2s+2n-2}&=O(\e^4) \text{ \ \ with \ \ }n\ne m\,. 
\end{align}
Hence we see, none of the higher dimensional scalars can contribute to the crossed channels.

}

\end{enumerate}

\subsubsection{Spin$-\ell'>0$ exchange}

For spin $\ell'$ operators too, there are analogous cancellations just like what we have seen above. Let us take the general expression \eqref{qtform}. To avoid tedious expressions, we will not give the explicit forms of the individual residues.  We will just indicate the pairs of poles that cancel each other under $\e$- expansion. For both the log term $c_{\D,\ell'}q^{(2,s)}_{\D,\ell|\ell'}$ and the power law term $c_{\D,\ell'}q^{(1,s)}_{\D,\ell|\ell'}$ we can simply put $s=\D\phi$, which will give us the following poles,
\begin{itemize}
\item{I. \ \ $\nu= (\D-h)$ - from the denominator of the spectral weight  $\mu^{(t)}_{\Delta,\ell'}(\nu)$}

\item{II. \  $\nu= (2\D_\phi+\ell'-h+2n)$ - from the numerator $\G$ factors of the spectral weight}

\item{III. $\nu= (h+\ell'+2n)$ - From the other $\G$ factors in numerator}

\item{IV.   $\nu= (h-1), (h-2), \cdots , (h-2+\ell')$ from the denominator Pochhammer terms in the spectral weight\,.}

\end{itemize}

\begin{enumerate}

\item{{\bf Lowest dimension spin $\ell'$}\\ \\
The lowest dimension operators with spin $\ell'$ are the $J_{\ell'}$ operators, with dimensions $\D_{\ell'}=\ell'+2-\e+O(\e^2)$\,. For these operators we find the following cancellations,
\begin{align}
Res_{\nu=\D-h}+Res_{\nu=2\D_\phi+\ell'-h}+Res_{\nu=h-2+\ell'}&=O(\e^4)\nonumber\\
Res_{\nu=2\D_\phi-h+\ell'+2n+2}+Res_{\nu=h+\ell'+2n}&=O(\e^4) \text{\ \ where \ \ } n=0,1,2,\cdots\nonumber\\
Res_{\nu=h-1}\sim Res_{\nu=h-2}\sim \cdots Res_{\nu=h+\ell'-3}&=O(\e^4)\,.
\end{align}

}

\vskip 1cm
\item{{\bf Higher dimensional spin $\ell'$ operators}\\ \\
There are heavier operators with spin $\ell'$ as we discussed in \ref{schcancel}. These operators, labelled $O_{2m,\ell'}$ have the dimensions $\D_{2m,\ell}=\ell+2+2m +\d_m\e+O(\e^2)$\,. We had argued that since these operators must have the composition $\phi(\partial^2)^m\partial^\ell\phi\sim \phi\partial^a\phi\partial^b\phi\partial^c\phi$, or a higher number of $\phi$-s, their OPE coefficients go like $C_{2m,\ell'}\sim O(\e^2)$ or higher. Taking this into account for these operators, we have the cancellations,
\begin{align}
Res_{\nu=2\D_\phi+\ell'-h}+Res_{\nu=h-2+\ell'}&=O(\e^4)\nonumber\\
Res_{\nu=\D-h}+Res_{\nu=2\D_\phi-h+\ell'+2m}+Res_{\nu=h+\ell'+2m-2}&=O(\e^4) \nonumber\\
Res_{\nu=2\D_\phi-h+\ell'+2n+2}+Res_{\nu=h+\ell'+2n}&=O(\e^4) \text{\ \ where \ \ } n\ne m-1\nonumber\\
Res_{\nu=h-1}\sim Res_{\nu=h-2}\sim \cdots Res_{\nu=h+\ell'-3}&=O(\e^4)\,.
\end{align}

}

\end{enumerate}
Thus we conclude none of the operators in the crossed channels contribute except the $\phi^2$ operator for which only two poles are sufficient up to the $O(\e^3)$ order.

\section{Large $\ell$ behavior of $Q_{\D,\ell}$}
In this appendix we will derive the large $\ell$ approximation for $Q_{\ell,0}$. Let us start by deriving an approximation for the $_3F_2$ hypergeometric function. It has the integral representation,
\be\label{3F2}
{}_3 F_2\bigg[\begin{matrix} -n,k_1,k_2\\
k_3, k_4
\end{matrix};1\bigg]=\frac{\G(k_4)}{\G(k_1)\G(k_4-k_1)}\int_0^1 z^{k_1-1}(1-z)^{-k_1+k_4-1}{}_2F_1(-n,k_2,k_3,z) dz\,.
\ee
Now as $n \to \infty$, we have \cite{Luke},
\be
_2F_1(-n,k_2,k_3,z) \approx \frac{\G(k_3)}{\G(k_2)\G(k_3-k_2)}\Big[ \G(k_2)(n z)^{-k_2}+\G(k_3-k_2)(-n z)^{k_2-k_3}(1-z)^{k_3-k_2+n} \Big]\,.
\ee
For identical external scalars, the $Q_{\ell,0}^{\D}$ has a $_3F_2$ which is equal to the above under the map $n=\ell$, $k_1=\tau+\ell-1$, $k_2=s+t$, $k_3=\frac{\ta}{2}$ and $k_4=\frac{\ta}{2}$. Since $n \gg k_3-k_2$ we can neglect the second term in the parentheses above. Then \eqref{3F2} becomes,
\begin{align}\label{3F2next}
\!\!{}_3 F_2\bigg[\begin{matrix} -n,k_1,k_2\\
k_3, k_4
\end{matrix};1 & \bigg]=\frac{\G(k_3)\G(k_4)}{\G(k_2)\G(k_3-k_1)\G(k_3-k_2)}\frac{1}{n^{k_2}}\int_0^1 z^{k_1-k_2-1}(1-z)^{k_4-k_1-1}dz\,.
\end{align}
Carrying out the $z$ integral it gives,
\be
\frac{\G(k_3)\G(k_4)}{\G(k_2)\G(k_3-k_1)\G(k_3-k_2)}\frac{1}{n^{k_2}}\int_0^1 z^{k_1-k_2-1}(1-z)^{k_4-k_1-1}dz=\frac{\G(k_3)\G(k_4)\G(k_1-k_2)}{\G(k_2)\G(k_3-k_2)\G(k_4-k_2)}\frac{1}{n^{k_2}}\,.
\ee
So we get,
\begin{align}
{}_3 F_2\bigg[\begin{matrix} -n,k_1,k_2\\
k_3, k_4
\end{matrix};1 \bigg]& =\frac{n^{-k_2} \Gamma \left(k_1-k_2\right) \Gamma \left(k_3\right) \Gamma \left(k_4\right)}{\Gamma \left(k_1\right) \Gamma \left(-k_2+k_3\right) \Gamma \left(-k_2+k_4\right)}.
\end{align}
Finally using this in \eqref{Qdefn} and putting the values of $k_{1,2,3,4}$, we get the large $\ell$ approximation,
\be
Q_{\ell,0}^{\ta+\ell}(t) \approx \frac{2^\ell\ell ^{-\frac{\ta}{2}-t} \Gamma (\frac{\ta}{2}+\ell )^2 \Gamma (-1+\frac{\ta}{2}-t+\ell )}{\Gamma (-t)^2 \Gamma (-1+\ta+2 \ell )}\,.
\ee


\begin{thebibliography}{99}
\bibitem{mathematica}
An anciliary mathematica notebook is provided with this submission.
\end{thebibliography}
\end{document}